\documentclass[journal]{IEEEtran}

\usepackage{fontenc}
\usepackage{lmodern}
\usepackage{microtype}

\usepackage[noadjust]{cite}
\usepackage{graphicx}
\usepackage[caption=false,format=hang]{subfig}
\usepackage{physics}
\usepackage{amssymb}
\usepackage{hyperref}
\usepackage[capitalise,compress,nameinlink]{cleveref}
\usepackage[abbreviations,british]{foreign}
\usepackage{interval}
\usepackage{paralist}
\usepackage[separate-uncertainty=true]{siunitx}
\usepackage{tikz}
\usepackage{dblfloatfix}

\renewcommand{\abs}[1]{\big\lvert#1\big\rvert}

\newcommand{\almost}[1]{{\thicksim}#1}
\newcommand{\conj}[1]{#1^{\ast}}
\newcommand{\convolution}[2]{(#1 \circledcirc{} #2)}
\newcommand{\eigenMatrix}{K_{\ell{} m,\ell' m'}}
\newcommand{\euclideanNorm}[1]{{\big\lVert#1\big\rVert}_{2}}
\newcommand{\factor}[1][]{\frac{2\ell#1+1}{4\pi}}
\newcommand{\fwhm}[1]{\text{FWHM}=\SI{#1}{\degree}}
\newcommand{\harmonic}[2][]{{#2}_{\ell#1 m#1}}
\newcommand{\harmonicSum}[1][]{\displaystyle\sum\limits_{\ell#1 m#1}}
\newcommand{\hermitian}[1]{#1^{\dagger}}
\newcommand{\hilbert}[1]{L^{2}(#1)} 
\newcommand{\inducedNorm}[1]{\big\lVert#1\big\rVert}
\newcommand{\integrateRegion}[1]{\displaystyle\int\limits_{R} #1}
\newcommand{\integrateSphere}[1]{\displaystyle\int\limits_{\mathbb{S}^{2}} \dd{\Omega(#1)}}
\newcommand{\lmax}{\ell_{\text{max}}}
\newcommand{\phaseFactor}[1][]{{(-1)}^{m#1}}
\newcommand{\pixel}[2][]{#2(\omega#1)}
\newcommand{\powerSpectrum}{C_{\ell}}
\newcommand{\realPosParam}{\mathbb{R}_{\ast}^{+}}
\newcommand{\rotation}[1]{\mathcal{R}_{#1}}
\newcommand{\rotationGroup}{\text{SO(3)}} 
\newcommand{\rotationMatrix}[1][]{\vb*{R}_{\rho#1}}
\newcommand{\set}[1]{\{#1\}} 
\newcommand{\slepian}[2][]{{#2}_{p#1}}
\newcommand{\slepianSum}[1][]{\displaystyle\sum\limits_{p#1}}
\newcommand{\snr}[1]{\text{SNR}(#1)} 
\newcommand{\sphereVolume}[1][]{\dd{\Omega(\omega#1)}}
\newcommand{\translation}[1]{\mathcal{T}_{#1}}
\newcommand{\twoSphere}{\mathbb{S}^{2}}
\newcommand{\variance}[1]{{\big[\Delta#1\big]}^{2}}
\newcommand{\waveletSum}{\displaystyle\sum\limits_{j=J_{0}}^{J}}

\intervalconfig{soft open fences}

\graphicspath{{figures/}}

\usetikzlibrary{hobby,patterns}

\begin{document}

\title{Slepian Scale-Discretised Wavelets on the Sphere}

\author{
	\IEEEauthorblockN{
		Patrick J. Roddy\IEEEauthorrefmark{1} and
		Jason D. McEwen\IEEEauthorrefmark{1}
	}

	\IEEEauthorblockA{
		\IEEEauthorrefmark{1}
		University College London (UCL), Gower Street, London, WC1E 6BT, UK
		\newline
		\footnotesize \texttt{patrick.roddy@ucl.ac.uk}, \texttt{jason.mcewen@ucl.ac.uk}
	}
}

\maketitle

\begin{abstract}
	This work presents the construction of a novel spherical wavelet basis designed for incomplete spherical datasets, \ie{} datasets which are missing in a particular region of the sphere.
	The eigenfunctions of the Slepian spatial-spectral concentration problem (the Slepian functions) are a set of orthogonal basis functions which are more concentrated within a defined region.
	Slepian functions allow one to compute a convolution on the incomplete sphere by leveraging the recently proposed sifting convolution and extending it to any set of basis functions.
	Through a tiling of the Slepian harmonic line, one may construct scale-discretised wavelets.
	An illustration is presented based on an example region on the sphere defined by the topographic map of the Earth.
	The Slepian wavelets and corresponding wavelet coefficients are constructed from this region and are used in a straightforward denoising example.
\end{abstract}


\IEEEpeerreviewmaketitle{}

\section{Introduction}

\IEEEPARstart{M}{any} applications in science and engineering measure data on the sphere, such as in computer graphics~\cite{Ramamoorthi2004}, cosmology~\cite{Bennett1996}, geophysics~\cite{Simons2006}, and planetary science~\cite{Turcotte1981}.
Often these data are not observed over the whole sphere and are missing in some regions.
For example, in analyses of the cosmic microwave background, the region around the Galactic plane is often removed due to strong foreground microwave emissions~\cite{Mortlock2002}.
A common approach to deal with data of this form is to use wavelets, which allow one to probe spatially localised, scale-dependent features of signals on the sphere.
However, spherical wavelet methods still have problems when data are missing, as the boundaries of the region of missing data contaminate nearby wavelet coefficients.
A possible approach to solve this problem is to construct wavelets that are concentrated in the region itself.

Extracting non-trivial patterns and structures of interest is a common task in data analysis.
To overcome this problem the data may be projected onto an appropriate basis.
In contrast to Fourier analyses, where oscillatory features are considered, wavelets extract the contributions of scale-dependent features in both space and frequency.
Wavelets on the sphere have been effectively applied in fields such as astrophysics and cosmology~\cite{Pen1999,Barreiro2001,Rocha2004}, where datasets are increasingly large and require analysis at high resolutions for accurate theoretical predictions.
Wavelet theory is well established in the Euclidean domain.
Scale-discretised wavelets~\cite{Wiaux2008,McEwen2018,Leistedt2013,McEwen2013,McEwen2015} lean on a tiling of the harmonic line to produce an exact wavelet transform in both the continuous and discrete settings.

Functions cannot simultaneously have finite support in both the spatial (time) and spectral (frequency) domains~\cite{Slepian1961,Slepian1983}.
Slepian, Landau and Pollak solved the fundamental problem of finding and representing the functions that are optimally energy concentrated in both the time and frequency domains~\cite{Slepian1961,Landau1961,Landau1962}.
The Slepian spatial-spectral concentration problem, or Slepian concentration problem for short, produces the orthogonal functions optimally concentrated in the spatial (spectral) domain and exactly limited in the spectral (spatial) domain.
The Slepian functions and their multidimensional counterparts~\cite{Slepian1964,Simons2011a} have been used in many branches of science and engineering (\eg{} signal processing~\cite{Thomson1982}, geophysics~\cite{Thomson1976,Simons2006a,Simons2011}, and medical imaging~\cite{Jackson1991}).
In particular, these functions have been used in solving partial differential equations~\cite{Boyd2003,Chen2005}, inverse problems~\cite{Villiers2001,Abdelmoula2015}, interpolation~\cite{Moore2004,Shkolnisky2006} and extrapolation~\cite{Xu1983}.
In fields with spatially limited observations, the functions have become the dominant spatial or spectral windows for regularisation of quadratic inverse problems of power spectral estimation~\cite{Thomson1976}.

The initial formulation of the Slepian concentration problem was developed in the Euclidean domain; however, this has since been generalised to other geometries~\cite{Simons2006,Wieczorek2005,Albertella1999,Cohen1989,Meaney1984,Daubechies1988,Bolton2018}.
The Slepian concentration problem for functions defined on the 2-sphere has been extensively investigated~\cite{Simons2006,Wieczorek2005,Albertella1999}; often simplifying to the axisymmetric case~\cite{Simons2006a,Simons2007}.
The ensuing bandlimited, spatially concentrated functions have been applied for localised spectral analysis~\cite{Wieczorek2005}, and spectral estimation of signals~\cite{Wieczorek2007} defined on the sphere.

Various wavelet-like analysis techniques to simultaneously probe signal content localised in space and frequency have been considered before.
Wavelets constructed on Slepian functions in the Euclidean setting (\ie{} prolate spheroidal wave functions) have been constructed to yield a multiresolution analysis~\cite{Walter2004}.
Many techniques on the sphere have also been developed, often motivated by applications in geophysics or cosmology.
Sometimes the Slepian functions on the sphere themselves have been used~\cite{Simons2009}; whereas other times, spatially localised spherical harmonic transforms are used instead (\eg{}~\cite{Simons1997,Wieczorek2005,Khalid2013,Khalid2013a}).
Slepian frames on the sphere are constructed in~\cite{Simons2011}, which provide a wavelet-like representation but do not constitute a tight frame with an explicit inverse transform.
A Slepian-spatial transform is developed in~\cite{Aslam2021} based on the directional convolution between the signal and a Slepian function.
An approach to solving inverse problems with regional data on the sphere is presented in~\cite{Michel2017}, which results in Slepian functions that can be used to derive a singular value decomposition (SVD) approach.
Standard regularisation techniques based on a known SVD can then be applied to inverse problems where data are only defined on a region.

This work presents a novel spherical wavelet basis designed for incomplete spherical datasets.
Here, the scale-discretised wavelet construction on the sphere~\cite{Wiaux2008,McEwen2018,Leistedt2013,McEwen2013,McEwen2015} is extended to support situations where data are only defined over a partial region of the sphere.
Scale-discretised wavelets have the advantage that they constitute a tight frame, exhibit an explicit inversion formula, and have excellent localisation properties in both spatial and spectral domains.
The eigenfunctions of the Slepian concentration problem provide the orthogonal basis functions (the Slepian functions) for the region from which they are constructed.
A scale-discretised wavelet transform is built on these basis functions.
A tiling of the Slepian line allows one to define Slepian scale-discretised wavelets.
By generalising the sifting convolution~\cite{Roddy2021}, recently presented by the authors of the current article, one may perform convolutions over the incomplete sphere using the Slepian functions as a basis.
The Slepian wavelet transform then follows by performing convolutions of signals defined over the incomplete sphere with the Slepian wavelets.
The original function can then be reconstructed from its Slepian wavelet coefficients.
Slepian scale-discretised wavelets on the sphere have many possible applications wherever data are defined over a partial region of the sphere; for example, in many cosmological analyses.

The remainder of this article is as follows.
\cref{sec:mathematical_background_problem_formulation} presents some mathematical preliminaries of signals on the sphere.
The sifting convolution on the sphere is discussed, in particular where the translation is defined as a product of basis functions.
A review of the Slepian concentration problem on the sphere is presented.
The proposed Slepian wavelet theory is developed in \cref{sec:slepian_wavelets}, which first extends the sifting convolution to the Slepian basis.
A scale-discretised wavelet transform is introduced, along with the generating functions which define the Slepian wavelets.
An example region is constructed from the topographic map of the Earth in \cref{sec:numerical_illustration}.
The Slepian functions and the resulting Slepian wavelets of this region are presented, and the wavelet transform is performed.
A straightforward denoising example using hard-thresholding of the wavelet coefficients illustrates a potential use case of these wavelets.
Lastly, \cref{sec:conclusion} sets out some concluding remarks.

\section{Mathematical Background and Problem Formulation}\label{sec:mathematical_background_problem_formulation}

Some mathematical preliminaries are discussed in \cref{sec:mathematical_preliminaries} with an introduction to signals on the sphere, the spherical harmonic basis functions, and rotations on the sphere.
A directional convolution on the sphere, the sifting convolution, is reviewed in \cref{sec:sifting_convolution_sphere} which was recently developed by the authors of this work.
Lastly, in \cref{sec:slepian_concentration_problem} the Slepian spatial concentration problem is summarised.

\subsection{Mathematical Preliminaries}\label{sec:mathematical_preliminaries}

\subsubsection{Signals on the Sphere}

The 2-sphere \(\twoSphere{}\) is defined as such \(\twoSphere{} = \set{\omega \in \mathbb{R}^{3} : \euclideanNorm{\omega} = 1}\), where \(\euclideanNorm{\omega}\) denotes the Euclidean norm.
A point on the unit sphere is parameterised by \(\omega=(\theta,\phi)\), where the colatitude is \(\theta \in \interval{0}{\pi}\) and the longitude is \(\phi \in \interval[open right]{0}{2\pi}\).
The Hilbert space \(\hilbert{\twoSphere}\) is formed by the complex-valued square-integrable functions \(\pixel{f}\).
The inner product induces a norm \(\inducedNorm{f} = \sqrt{\braket*{f}}\).
Functions with a finite induced norm are signals on the sphere.

\subsubsection{Spherical Harmonics}

The complete set of orthonormal basis functions for the Hilbert space \(\hilbert{\twoSphere}\) are the spherical harmonics.
By the completeness of the spherical harmonics one may decompose any signal \(f \in \hilbert{\twoSphere}\) as
\begin{equation}
	\pixel{f}
	= \sum\limits_{\ell=0}^{\infty} \sum\limits_{m=-\ell}^{\ell} \harmonic{f} \pixel{\harmonic{Y}},
\end{equation}
where \(\harmonic{f}\) are the spherical harmonic coefficients given by the usual projection onto the basis functions \(\harmonic{f} = \braket*{f}{\harmonic{Y}}\).
The phase convention adopted here is \(\pixel{\conj{\harmonic{Y}}} = \phaseFactor \pixel{Y_{\ell(-m)}}\), such that \(\conj{\harmonic{f}} = \phaseFactor f_{\ell(-m)}\) for a real field.
One often considers functions bandlimited at \(\lmax{}\), \ie{} signals such that \(\harmonic{f} = 0,\ \forall \ell \geq \lmax{}\).
The shorthand convention \(\sum_{\ell m} = \sum_{\ell=0}^{\lmax-1} \sum_{m=-\ell}^{\ell}\) may be adopted.
The addition theorem of the spherical harmonics states
\begin{equation}\label{eq:addition_theorem}
	\sum\limits_{m=-\ell}^{\ell} \pixel{\harmonic{Y}} \pixel[']{\conj{\harmonic{Y}}}
	= \factor P_{\ell}(\omega \cdot \omega'),
\end{equation}
where \(P_{\ell}(x)\) are the Legendre polynomials.

\subsubsection{Rotation of a Signal on the 2-Sphere}

Three-dimensional rotations may be parametrised by the Euler angles with \(\rho = (\alpha,\beta,\gamma) \in \text{rotation group } \rotationGroup{}\), where \(\alpha \in \interval[open right]{0}{2\pi}\), \(\beta \in \interval{0}{\pi}\), and \(\gamma \in \interval[open right]{0}{2\pi}\).
A rotation \(\rotation{\rho}\) consists of the sequence of rotations:
\begin{inparaenum}[(i)]
	\item \({\gamma}\) rotation about the \(z\)-axis;
	\item \({\beta}\) rotation about the \(y\)-axis; and
	\item \({\alpha}\) rotation about the \(z\)-axis.
\end{inparaenum}
A function rotated on the sphere is defined by \(\pixel{(\rotation{\rho}f)} = f(\rotationMatrix^{-1} \omega)\) where the three-dimensional rotation matrix corresponding to \(\rotation{\rho}\) is \(\rotationMatrix{}\).

\subsection{Sifting Convolution on the Sphere}\label{sec:sifting_convolution_sphere}

Convolutions are a central part of a continuous wavelet transform.
To construct Slepian wavelets, a directional convolution (\ie{} a convolution which accepts inputs that are \emph{not} invariant under azimuthal rotation) with an output which remains on the sphere with minimal leakage outside the region is required.
The sifting convolution~\cite{Roddy2021} is such a convolution which was recently developed by the authors of this work.

The rotation operator on the sphere is the usual counterpart of the Euclidean translation operator in real space.
An alternative operator, \(\translation{\omega}\), may be defined which follows as the harmonic space complement of the Euclidean setting.
In contrast to the standard rotation, this translation considers two angles rather than three, which ensures its output remains on the sphere.
The complex exponentials \(\zeta_{u}(x) = \exp(iux)\), with \(x,\ u \in \mathbb{R}\) form the standard orthonormal basis of the Euclidean setting.
The translation of the basis functions is defined by a shift of coordinates where \(\zeta_{u}(x + x') = \zeta_{u}(x') \zeta_{u}(x)\), with \(x' \in \mathbb{R}\) and where the final equality follows by the standard rule for exponents.

In an analogous manner a translation operator on the sphere may be defined as a product of the basis functions (\ie{} the spherical harmonics)
\begin{equation}\label{eq:translation_basis_functions}
	\pixel{(\translation{\omega'}\harmonic{Y})}
	\equiv \pixel[']{\harmonic{Y}} \pixel{\harmonic{Y}},
\end{equation}
where \(\omega'=(\theta',\phi')\).
The translation of an arbitrary function \(f \in \hilbert{\twoSphere}\) is thus
\begin{equation}
	\pixel{(\translation{\omega'}f)} = \harmonicSum \harmonic{f} \pixel[']{\harmonic{Y}} \pixel{\harmonic{Y}},
\end{equation}
which implies
\begin{equation}
	\harmonic{(\translation{\omega'}f)} = \harmonic{f} \pixel[']{\harmonic{Y}}.
\end{equation}
The sifting convolution on the sphere of \(f,\ g \in \hilbert{\twoSphere}\) follows by the inner product
\begin{equation}\label{eq:sifting_convolution}
	\pixel{\convolution{f}{g}}
	\equiv \integrateSphere{\omega'} \pixel[']{(\translation{\omega}f)} \pixel[']{\conj{g}},
\end{equation}
where \(\sphereVolume=\sin{\theta}\dd{\theta}\dd{\phi}\) is the usual invariant measure on \(\twoSphere{}\).
In harmonic space this becomes
\begin{equation}
	\harmonic{\convolution{f}{g}}
	= \harmonic{f} \conj{\harmonic{g}},\ \forall \ell, m.
\end{equation}
This convolution permits directional kernels, whose output remains on the sphere, and is efficient to compute --- as it is a product in harmonic space.

The translation \cref{eq:translation_basis_functions} is a product of basis functions and hence follows as the analogue of the Euclidean setting.
This translation can be extended to any set of arbitrary basis functions.
Thus, to create wavelets restricted to a region of the sphere, one must seek the basis functions of a region on the sphere.
These basis functions are defined in \cref{sec:slepian_concentration_problem}.

\subsection{Slepian Concentration Problem on the Sphere}\label{sec:slepian_concentration_problem}

\begin{figure}
	\centering
	\begin{tikzpicture}
		\draw [use Hobby shortcut, 
			closed=true, 
			pattern=north east lines, 
			even odd rule] 
		(0,0) circle (2cm)
		(-0.2,0.3) .. (0.2,0.2) .. (1.1,0.6) .. (1.1,1) .. (0.3,1.4) .. (-0.4,1.2) .. (-0.3,0.8); 
		\draw (0,0) circle (2cm) 
		(-2,0) arc (180:360:2 and 0.6); 
		\draw[dashed] (2,0) arc (0:180:2 and 0.6); 
		\fill [fill=black] (0,0) circle (1pt); 
		\node at (0.4,0.8) {\(R\)}; 
	\end{tikzpicture}
	\caption{
		In many application domains, data are observed on a partial region of the sphere only, such as \(R\).
	}\label{fig:region}
\end{figure}
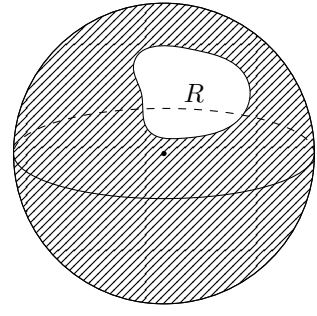

A function cannot be strictly spacelimited as well as strictly bandlimited~\cite{Slepian1961,Slepian1983}.
This work considers optimally concentrated functions within a region \(R\).
\cref{fig:region} presents such an example region on the sphere.

\subsubsection{Spatial Concentration of a Bandlimited Function}

To maximise the spatial concentration of a bandlimited function \(f \in \hilbert{\twoSphere}\) within a region \(R\) one may maximise the following ratio:
\begin{equation}\label{eq:spatial_concentration_ratio_pixel}
	\mu
	= \frac{\integrateRegion{\sphereVolume} \abs{\pixel{f}}^{2}}
	{\integrateSphere{\omega} \abs{\pixel{f}}^{2}},
\end{equation}
where \(0 < \mu < 1\) is a measure of the spatial concentration.
Using the spherical harmonic expansion of \(f\) this becomes
\begin{equation}\label{eq:spatial_concentration_ratio_harmonic}
	\mu
	= \frac{\harmonicSum \harmonic{f} \harmonicSum['] \eigenMatrix \conj{\harmonic[']{f}}}
	{\harmonicSum \abs{\harmonic{f}}^{2}},
\end{equation}
where
\begin{equation}\label{eq:slepian_matrix_calc}
	\eigenMatrix
	= \integrateRegion{\sphereVolume} \pixel{\harmonic{Y}} \pixel{\conj{\harmonic[']{Y}}}
\end{equation}
is an \(\lmax^{2} \times{} \lmax^{2}\) matrix if including all \(\ell{}\) and \(m\).
In practice, \cref{eq:slepian_matrix_calc} may be computed by discretising it using quadrature weights over the sphere.
Computing the integral is non-trivial for a general region \(R\) with a high \(\lmax{}\).
One may rewrite \cref{eq:spatial_concentration_ratio_harmonic} as a classical matrix variational problem
\begin{equation}
	\mu
	= \frac{\hermitian{\vb*{f}} \vb*{K} \vb*{f}}{\hermitian{\vb*{f}}\vb*{f}}.
\end{equation}
The harmonic coefficients \(\vb*{f}\) are the solutions to the \(\lmax^{2} \times{} \lmax^{2}\) eigenproblem
\begin{equation}\label{eq:eigenproblem}
	\vb*{K}\vb*{f}
	= \mu\vb*{f}.
\end{equation}
The matrix \(\vb*{K}\) is real, symmetric and positive definite; hence, the eigenvalues \(\slepian{\mu}\) are always real.
The eigenvalues are a measure of how well concentrated the eigenvectors are, \cf{} \cref{eq:spatial_concentration_ratio_pixel}.
Hence, the eigenvalues satisfy \(1 > \mu_{1} \geq \mu_{2} \geq \ldots \geq \mu_{\lmax^{2}} > 0\), with corresponding eigenvectors \(\vb*{f}_{1},\ \vb*{f}_{2},\ \ldots,\ \vb*{f}_{\lmax^{2}}\). 
The largest eigenvalue \(\mu_{1}\) is strictly less than one as no bandlimited function may be restricted exactly within a region \(R\), and the smallest eigenvalue \(\mu_{\lmax^{2}}\) is strictly greater than zero due to the positive definiteness of the matrix \(\vb*{K}\).
Consider the sum of the eigenvalues
\begin{align}\label{eq:shannon}
	N
	= \harmonicSum K_{\ell m,\ell m}
	 & = \harmonicSum \integrateRegion{\sphereVolume} \pixel{\harmonic{Y}} \pixel{\conj{\harmonic{Y}}} \nonumber{} \\
	 & = \sum\limits_{\ell} \factor P_{\ell}(1) \integrateRegion{\sphereVolume} \nonumber{}                        \\
	& = \frac{A}{4\pi} \lmax^{2},
\end{align}
where the second line follows by the addition theorem \cref{eq:addition_theorem}, and \(A\) is the area of the region \(R\).
This \(N\) is a spherical analogue interpretation of the \emph{Shannon number}~\cite{Simons2006}, which is a good estimate of the number of significant eigenvalues~\cite{Percival1993}.

\subsubsection{Slepian Decomposition}

The Slepian functions offer an alternative orthogonal basis on the sphere, decomposing a function \(f \in \hilbert{\twoSphere}\) into this basis is
\begin{equation}
	\pixel{f}
	= \sum\limits_{p=1}^{\lmax^{2}} \slepian{f}\pixel{\slepian{S}},
\end{equation}
where \(\pixel{\slepian{S}}\) are the Slepian functions.
If the function \(\pixel{f}\) is well-localised in the region \(R\) (\ie{} \(f \in \hilbert{R}\)), then \(N\) well-localised Slepian functions can describe it
\begin{equation}\label{eq:slepian_inverse}
	\pixel{f}
	\approx \sum\limits_{p=1}^{N} \slepian{f}\pixel{\slepian{S}} \\
	= \slepianSum \slepian{f}\pixel{\slepian{S}},
\end{equation}
where a shorthand notation has been introduced.
The usual projection onto the basis functions permits one to calculate the Slepian coefficients \(\slepian{f} = \braket*{f}{\slepian{S}}\).

For a well-localised function one instead may compute the Slepian coefficients with an integral over the region \(R\) rather than \(\twoSphere{}\)
\begin{equation}
	\slepian{f}
	\approx \frac{1}{\slepian{\mu}} \integrateRegion{\sphereVolume} \pixel{f} \pixel{\conj{\slepian{S}}},
\end{equation}
as
\begin{align}
	\integrateRegion{\sphereVolume} \pixel{f} \pixel{\conj{\slepian{S}}}
	 & \approx \slepianSum['] \slepian[']{f} \integrateRegion{\sphereVolume} \pixel{\slepian[']{S}} \pixel{\conj{\slepian{S}}} \nonumber{} \\
	 & = \slepianSum['] \slepian[']{f} \hermitian{\slepian{\vb*{S}}} \vb*{K} \slepian[']{\vb*{S}}
	= \slepian{f} \slepian{\mu},
\end{align}
where \(\slepian{\vb*{S}}\) are the harmonic coefficients of \(\pixel{\slepian{S}}\).
Although, there will always be a small amount of signal leakage out of the region.
Note the use of the orthogonality results
\begin{equation}\label{eq:orthogonality_sphere}
	\integrateSphere{\omega} \pixel{\slepian{S}} \pixel{\conj{\slepian[']{S}}}
	= \hermitian{\slepian[']{\vb*{S}}} \slepian{\vb*{S}}
	= \delta_{pp'},
\end{equation}
and
\begin{equation}
	\integrateRegion{\sphereVolume} \pixel{\slepian{S}} \pixel{\conj{\slepian[']{S}}}
	= \hermitian{\slepian[']{\vb*{S}}} \vb*{K} \slepian{\vb*{S}}
	= \slepian{\mu} \hermitian{\slepian[']{\vb*{S}}} \slepian{\vb*{S}}
	= \slepian{\mu} \delta_{pp'}.
\end{equation}

One may transform from Slepian coefficients to spherical harmonic coefficients by
\begin{equation}
	\harmonic{f}
	= \integrateSphere{\omega} \pixel{f} \pixel{\conj{\harmonic{Y}}}
	= \slepianSum \slepian{f}\harmonic{(\slepian{S})},
\end{equation}
where
\begin{equation}\label{eq:harmonic_slepian}
	\harmonic{(\slepian{S})}
	= \integrateSphere{\omega} \pixel{\slepian{S}} \pixel{\conj{\harmonic{Y}}}
\end{equation}
are the eigenvectors of \cref{eq:eigenproblem}.
The corresponding transform from spherical harmonic coefficients to Slepian coefficients is
\begin{equation}\label{eq:harmonic_to_slepian}
	\slepian{f}
	= \integrateSphere{\omega} \pixel{f} \pixel{\conj{\slepian{S}}}
	= \harmonicSum \harmonic{f} \conj{\harmonic{(\slepian{S})}}.
\end{equation}

\section{Slepian Wavelets}\label{sec:slepian_wavelets}

This section develops the theory behind Slepian wavelets.
The sifting convolution~\cite{Roddy2021} is first extended to the Slepian basis in \cref{sec:slepian_sifting_convolution}.
This convolution is then used to define the Slepian wavelet transform in \cref{sec:slepian_scale_discretised_wavelets_sphere} with a suitable admissibility condition to ensure exact reconstruction.
The generating functions used to construct Slepian wavelets that satisfy the admissibility condition are presented in \cref{sec:generating_functions}.
Lastly, in \cref{sec:properties} some properties of the wavelets are discussed.

\subsection{Slepian Sifting Convolution}\label{sec:slepian_sifting_convolution}

The central part of a continuous wavelet transform is a convolution, and thus the Slepian functions can be used as a basis to define wavelets in the region.
The sifting convolution on the sphere developed by the authors of the current article~\cite{Roddy2021} and defined in \cref{sec:sifting_convolution_sphere}, can be extended to any arbitrary basis.
Consider the sifting convolution of a region on the sphere with a localised basis given by the Slepian functions.
The translation can be defined by
\begin{equation}
	\pixel{(\translation{\omega'}\slepian{S})}
	\equiv \pixel[']{\slepian{S}} \pixel{\slepian{S}},
\end{equation}
where \(\omega'=(\theta',\phi')\), and the \(\pixel{\slepian{S}}\) are the Slepian functions defined in \cref{sec:slepian_concentration_problem}.
Although the Slepian functions are defined over the whole sphere they are concentrated in a given region, therefore typically \(\omega' \in R\).
Thus, the translation of an arbitrary function \(f \in \hilbert{R}\) is
\begin{equation}
	\pixel{(\translation{\omega'}f)} = \slepianSum \slepian{f} \pixel[']{\slepian{S}} \pixel{\slepian{S}},
\end{equation}
which in Slepian space becomes
\begin{equation}
	\slepian{(\translation{\omega'}f)} = \slepian{f} \pixel[']{\slepian{S}}.
\end{equation}
The sifting convolution between two functions \(f,\ g \in \hilbert{R}\) is as before, \ie{}
\begin{equation}
	\pixel{\convolution{f}{g}}
	\equiv \integrateSphere{\omega'} \pixel[']{(\translation{\omega}f)} \pixel[']{\conj{g}},
\end{equation}
which is a product in Slepian space
\begin{equation}
	\slepian{\convolution{f}{g}}
	= \slepian{f} \conj{\slepian{g}},
\end{equation}
and hence is efficient to compute.
With a convolution defined in Slepian space to hand, one may now define the Slepian wavelets.

\subsection{Slepian Scale-Discretised Wavelets on the Sphere}\label{sec:slepian_scale_discretised_wavelets_sphere}

A tiling of Slepian space can be used to construct the Slepian wavelet transform, where one may restrict the scale parameter \(p\) to \(N=\lmax^{2}A/4\pi{}\) for a region \(R\) (or to \(\lmax^{2}\) for the entire sphere).
One may probe spatially localised, scale-dependent content through a scale-discretised wavelet transform.
These wavelets are defined similarly to~\cite{Wiaux2008,McEwen2018} but expanded in the Slepian basis rather than the spherical harmonics.

Now consider a signal of interest \(f \in \hilbert{R}\) concentrated within a region \(R\).
Wavelet coefficients \(W^{\Psi^{j}} \in \hilbert{R}\) may be defined by a sifting convolution of \(f\) with the wavelet \(\Psi^{j} \in \hilbert{R}\) for wavelet scale \(j\)
\begin{align}\label{eq:slepian_wavelet_omega}
	\pixel{W^{\Psi^{j}}}
	& = \pixel{\convolution{\Psi^{j}}{f}} \nonumber{} \\
	& = \integrateSphere{\omega'} \pixel[']{(\translation{\omega}\Psi^{j})} \pixel[']{\conj{f}},
\end{align}
or in Slepian space
\begin{equation}\label{eq:slepian_wavelet_p}
	\slepian{W}^{\Psi^{j}}
	= \slepian{\Psi}^{j} \conj{\slepian{f}},
\end{equation}
where \(\slepian{\Psi}^{j}\) are the Slepian harmonic coefficients of the wavelet at scale \(j\).

Typically, wavelets are complemented with scaling functions, with each capturing different scales of the underlying function.
Similarly to the wavelet coefficients, scaling coefficients \(W^{\Phi} \in \hilbert{R}\) may be defined by a sifting convolution between \(f\) and the scaling function \(\Phi \in \hilbert{R}\)
\begin{equation}\label{eq:slepian_scaling_omega}
	\pixel{W^{\Phi}}
	= \pixel{\convolution{\Phi}{f}}
	= \integrateSphere{\omega'} \pixel[']{(\translation{\omega}\Phi)} \pixel[']{\conj{f}},
\end{equation}
or in Slepian space
\begin{equation}\label{eq:slepian_scaling_p}
	\slepian{W}^{\Phi}
	= \slepian{\Phi} \conj{\slepian{f}},
\end{equation}
where \(\slepian{\Phi}\) are the Slepian coefficients of the scaling function.

The function \(f\) may be reconstructed from its wavelet and scaling coefficients given that the wavelets and scaling function satisfy an admissibility condition by
\begin{align}\label{eq:synthesis}
	\pixel{f}
	 & = \integrateSphere{\omega'} \pixel[']{(\translation{\omega}\Phi)} \pixel[']{W^{\Phi\ast}} \nonumber{}          \\
	 & + \waveletSum \integrateSphere{\omega'} \pixel[']{(\translation{\omega}\Psi^{j})} \pixel[']{W^{\Psi^{j}\ast}},
\end{align}
or in Slepian space
\begin{equation}\label{eq:synthesis_slepian}
	\slepian{f}
	= \slepian{W}^{\Phi\ast} \slepian{\Phi}
	+ \waveletSum \slepian{W}^{\Psi^{j}\ast} \slepian{\Psi}^{j}.
\end{equation}
The lowest and highest scales \(j\) of the wavelet decomposition are represented by the parameters \(J_{0}\) and \(J\) respectively --- these parameters must be defined consistently to ensure exact reconstruction~\cite{Wiaux2008}.
The admissibility condition on which synthesis of \cref{eq:synthesis} relies, is thus
\begin{equation}\label{eq:admissibility}
	\abs{\slepian{\Phi}}^{2}
	+ \waveletSum \abs{\slepian{\Psi}^{j}}^{2}
	= 1,\ \forall p,
\end{equation}
which is found by substituting \cref{eq:slepian_scaling_omega,eq:slepian_scaling_p} into \cref{eq:synthesis_slepian}.
One may now define wavelets and a scaling function that satisfy this admissibility property.

\subsection{Generating Functions}\label{sec:generating_functions}

To tile the Slepian line, one requires a set of smooth generating functions.
This work utilises a set of such functions defined by~\cite{Wiaux2008}, a brief summary follows.
Consider the \(C^{\infty}\) Schwartz function with compact support on \(\interval{-1}{1}\)
\begin{equation}
	s(t) \equiv
	\begin{cases}
		\exp(1/(t^{2}-1)), & t \in \interval{-1}{1},    \\
		0,                 & t \notin \interval{-1}{1},
	\end{cases}
\end{equation}
for \(t \in \mathbb{R}\).
One may then introduce the positive real parameter \(\lambda \in \realPosParam{}\) to map \(s(t)\) to
\begin{equation}
	s_{\lambda}(t)
	\equiv s\bigg(\frac{2\lambda}{\lambda-1}(t-\lambda^{-1}) - 1\bigg),
\end{equation}
which has compact support in \(\interval{1/\lambda}{1}\).
The smoothly decreasing function \(k_{\lambda}\) is then defined by
\begin{equation}
	k_{\lambda}(t)
	\equiv \int\limits_{t}^{1} \dd{t'} \frac{s^{2}_{\lambda}(t')}{t'}
	\bigg/ \int\limits_{1/\lambda}^{1} \dd{t'} \frac{s^{2}_{\lambda}(t')}{t'},
\end{equation}
which is unity for \(t < 1/\lambda{}\), zero for \(t > 1\), and smoothly decreasing from unity to zero for \(t \in \interval{1/\lambda}{1}\).
The wavelet generating function is defined
\begin{equation}
	\kappa_{\lambda}(t)
	\equiv \sqrt{k_{\lambda}(t/\lambda) - k_{\lambda}(t)},
\end{equation}
and the scaling function generating function
\begin{equation}
	\eta_{\lambda}(t)
	\equiv \sqrt{k_{\lambda}(t)}.
\end{equation}

\begin{figure}
	\centering
	\includegraphics[width=\columnwidth]{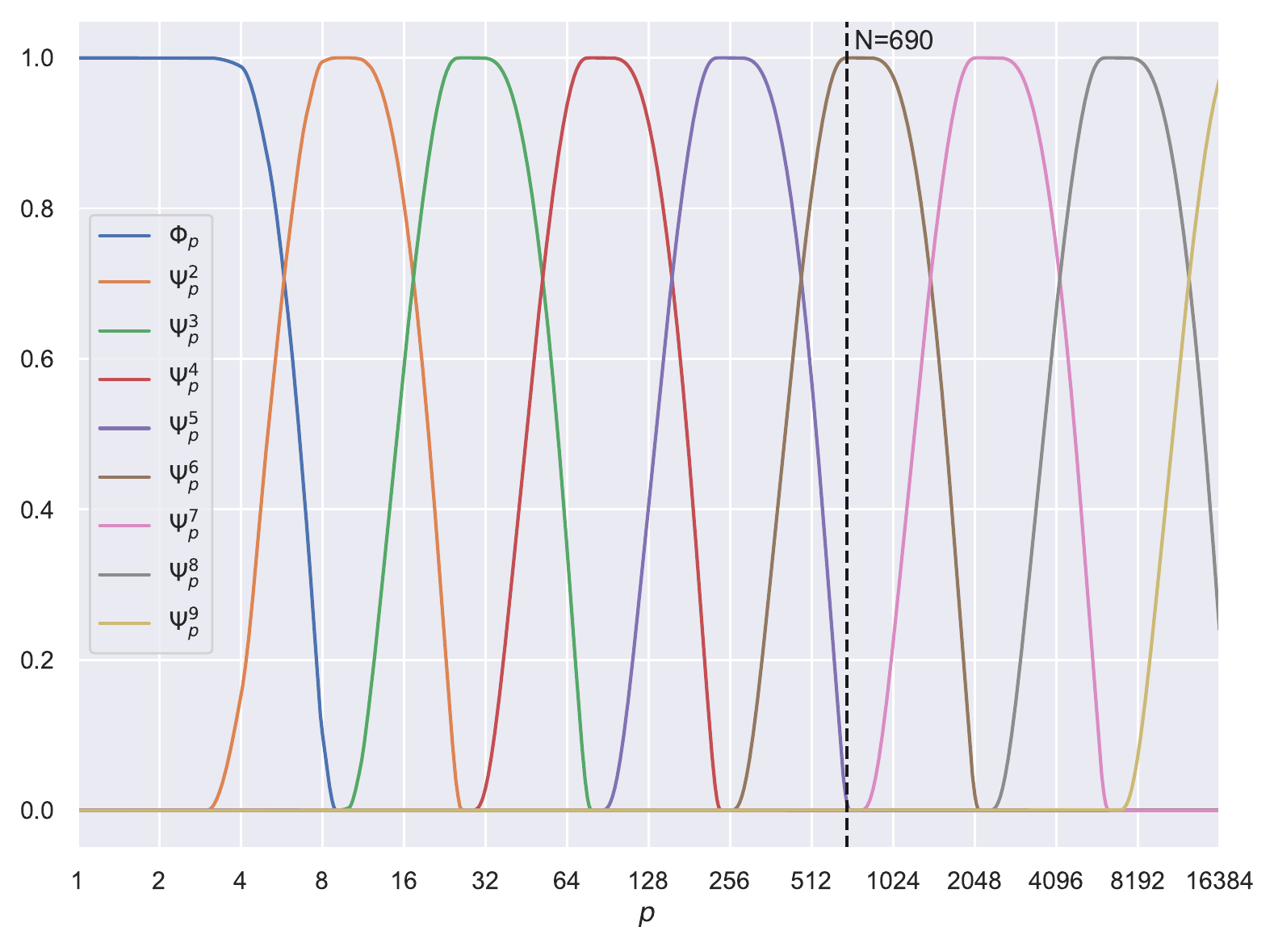}
	\caption{
		The tiling of the Slepian line with parameters \(\lambda=3\), \(J_{0}=2\) and bandlimit \(\lmax=128\), where increasing \(p\) represents worse concentration.
		The black dashed line marks the Shannon number for the South America region \(N=690\).
		The scaling function and the first five wavelets are non-zero, as the coefficients are within the Shannon number.
	}\label{fig:tiling}
\end{figure}

A natural approach is to define the wavelets \(\slepian{\Psi}^{j}\) from the generating functions \(\kappa_{\lambda}\) to have support on \(\interval{\lambda^{j-1}}{\lambda^{j+1}}\), which yields
\begin{equation}
	\slepian{\Psi}^{j}
	\equiv \kappa_{\lambda}\bigg(\frac{p}{\lambda^{j}}\bigg).
\end{equation}
The admissibility condition \cref{eq:admissibility} is satisfied for these wavelets for \(p \geq \lambda^{J_{0}}\), where \(J_{0}\) is the lowest wavelet scale used in the decomposition.
The scaling function \(\Phi{}\) is constructed to extract the modes that cannot be probed by the wavelets (\ie{} modes with \(p < \lambda^{J_{0}}\))
\begin{equation}
	\slepian{\Phi}
	\equiv \eta_{\lambda}\bigg(\frac{p}{\lambda^{J_{0}}}\bigg).
\end{equation}
\(J\) is set to ensure exact reconstruction yielding \(J = \lceil{} \log_{\lambda}(N)\rceil{}\).
The lowest wavelet scale \(J_{0}\) is arbitrary, provided \(0 \leq J_{0} < J\).
The Slepian wavelets are constructed by the tiling of the Slepian line as shown in \cref{fig:tiling}.

\subsection{Properties}\label{sec:properties}

The properties of Slepian wavelets are reviewed here, which are often akin to the standard spherical scale-discretised wavelets constructed on the spherical harmonics, but not always.

\subsubsection{Localisation}\label{sec:localisation}

Typically, scale-discretised wavelets are constructed through a tiling of the harmonic line.
Therefore, the value of \(j\) in the wavelets \(\Psi^{j}\) corresponds to increasingly higher frequencies (smaller scales).
The Slepian setting, however, is less straightforward.
Here, \(p\) is a measure of spatial concentration, where the lower the value of \(p\) the more well-localised the Slepian function is.
The Slepian wavelets are built on a tiling of the Slepian harmonic line, hence the localisation is captured in the wavelets and corresponding wavelet coefficients.

\subsubsection{Wavelet Energy}

The wavelet energy is
\begin{equation}
	\inducedNorm{\Psi^{j}}^{2}
	= \integrateSphere{\omega} \abs{\pixel{\Psi^{j}}}^{2}
	= \slepianSum \abs{\slepian{\Psi}^{j}}^{2}.
\end{equation}
A similar expression exists for the scaling function energy.

\subsubsection{Parseval Frame}

Slepian scale-discretised wavelets on the sphere satisfy the following Parseval frame property
\begin{align}
	A\inducedNorm{f}^{2} \leq
	 & \integrateSphere{\omega} \abs{\braket*{\translation{\omega}\Phi}{f}}^{2} \nonumber{}       \\
	 & + \waveletSum \integrateSphere{\omega} \abs{\braket*{\translation{\omega}\Psi^{j}}{f}}^{2}
	\leq B\inducedNorm{f}^{2},
\end{align}
with \(A,\ B \in \realPosParam{}\).
This can be proved by noting the Slepian representation of the scaling coefficients \cref{eq:slepian_scaling_p} and the wavelet coefficients \cref{eq:slepian_wavelet_p}, and using the orthogonality of the Slepian functions \cref{eq:orthogonality_sphere}
\begin{equation}
	\slepianSum \abs{\slepian{\Phi}}^{2} \abs{\slepian{f}}^{2}
	+ \waveletSum \abs{\slepian{\Psi}^{j}}^{2} \abs{\slepian{f}}^{2}
	= \inducedNorm{f}^{2},
\end{equation}
where the equality follows by the admissibility condition \cref{eq:admissibility}.
Thus, scale-discretised wavelets provide a Parseval frame with \(A = B = 1\), implying that the energy of \(f\) is conserved in wavelet space.

\subsubsection{Wavelet Domain Variance}

For notational brevity, define a quantity
\begin{equation}\label{eq:scaling_and_wavelet}
	\varphi \in \set{\Phi,\Psi^{j}}
\end{equation}
to represent both the scaling function and the wavelets.
The variance of the wavelet/scaling coefficients is given by
\begin{equation}
	\variance{\pixel{W^{\varphi}}}
	= \expval*{\abs{\pixel{W^{\varphi}}}^{2}}
	-\abs{\expval*{\pixel{W^{\varphi}}}}^{2}.
\end{equation}
For the common case of zero-mean Gaussian noise, the expected value of the Slepian coefficients is zero, and hence the wavelet/scaling coefficient expectation is zero.
Thus, the variance of the wavelet/scaling coefficients is
\begin{equation}\label{eq:slepian_isotropic_noise}
	\variance{\pixel{W^{\varphi}}}
	= \slepianSum \slepianSum['] \slepian{\varphi} \conj{\slepian[']{\varphi}} \pixel{\slepian{S}} \pixel{\conj{\slepian[']{S}}} \expval*{\conj{\slepian{f}} \slepian[']{f}}.
\end{equation}

To simplify \cref{eq:slepian_isotropic_noise} further, consider homogenous and isotropic noise defined by its power spectrum
\begin{equation}
	\expval*{\harmonic{f} \conj{\harmonic[']{f}}}
	= \powerSpectrum \delta_{\ell\ell'} \delta_{mm'},
\end{equation}
where \(\powerSpectrum = \sigma^{2}\) for white noise.
The power spectrum in harmonic space may be converted to the Slepian space by
\begin{align}
	\expval*{\slepian{f} \conj{\slepian[']{f}}}
	 & = \harmonicSum \harmonicSum['] \expval*{\harmonic{f} \conj{\harmonic[']{f}}} \conj{\harmonic{(\slepian{S})}} \harmonic[']{(\slepian[']{S})} \nonumber{} \\
	 & = \sigma^{2} \delta_{pp'},
\end{align}
where the first line follows from \cref{eq:harmonic_to_slepian}, and the last line follows from the orthogonality of the Slepian functions \cref{eq:orthogonality_sphere}.
Thus, the final expression for the wavelet domain variance becomes
\begin{equation}\label{eq:wavelet_variance}
	\variance{\pixel{W^{\varphi}}}
	= \sigma^{2} \slepianSum \abs{\slepian{\varphi}}^{2} \abs{\pixel{\slepian{S}}}^{2},
\end{equation}
and hence the variance depends on the position on the sphere.

\section{Numerical Illustration}\label{sec:numerical_illustration}

This section demonstrates the construction and application of Slepian wavelets for an example region on the sphere.
A region is constructed from a topographic map of the Earth in \cref{sec:south_america_region}, and the resulting eigenfunctions and eigenvalues are found.
The Slepian wavelets and wavelet coefficients of this region are shown in \cref{sec:south_america_wavelets}.
A straightforward denoising example is presented in \cref{sec:wavelet_denoising} which demonstrates a possible use case of Slepian wavelets.
All computations are performed with the \texttt{SSHT}\footnote{\url{http://astro-informatics.github.io/ssht/}}~\cite{McEwen2011} and \texttt{S2LET}\footnote{\url{http://astro-informatics.github.io/s2let/}}~\cite{Leistedt2013} codes, on which the Slepian wavelet transforms on the sphere are built.
Further, the \texttt{SLEPLET}~\cite{Roddy2022} code has been developed to perform the work in this article.

\subsection{South America Region}\label{sec:south_america_region}

\begin{figure}
	\centering
	\subfloat[EGM2008]
	{\includegraphics[trim={4 7 3 6},clip,width=.5\columnwidth]{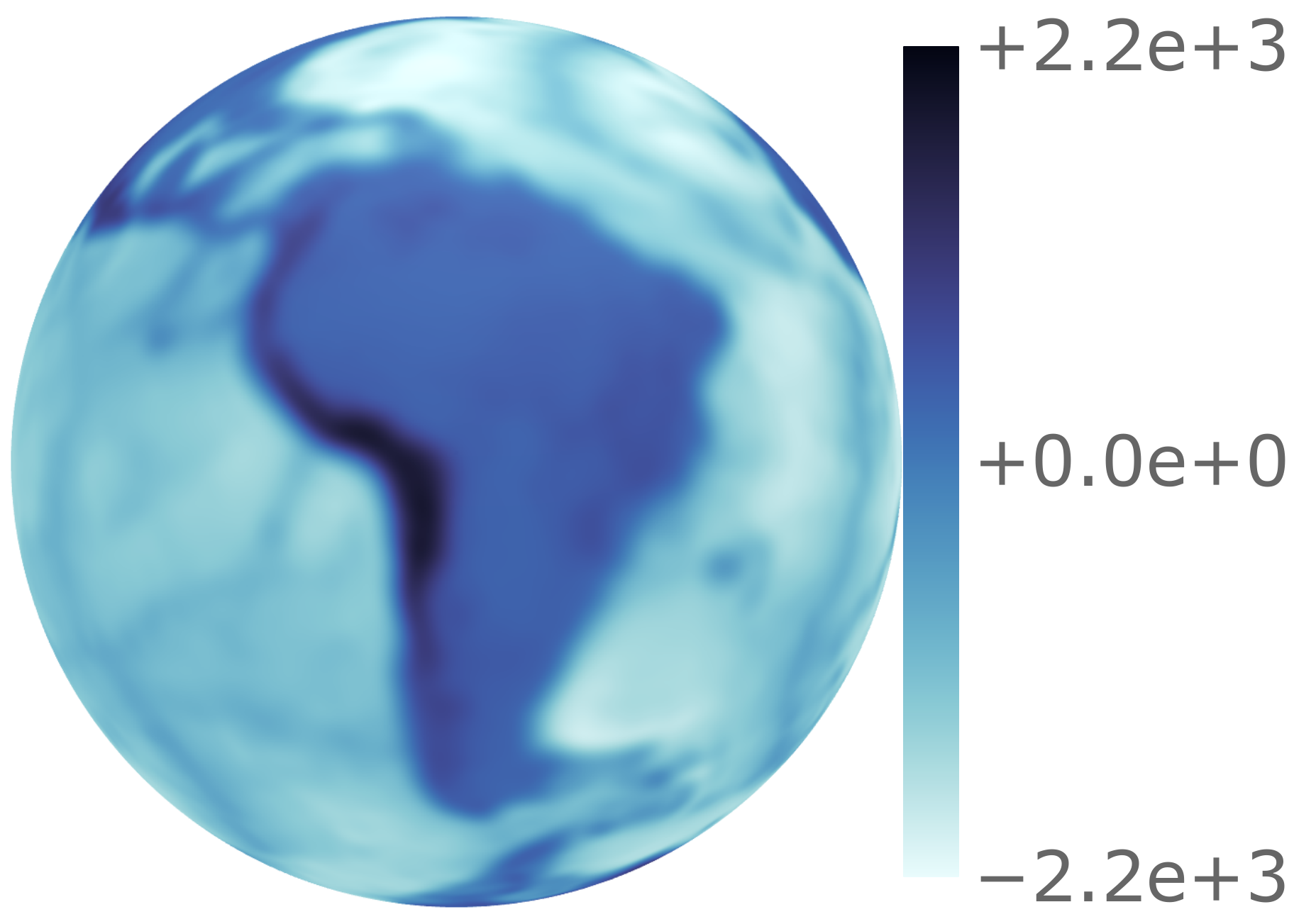}}
	\hfill
	\subfloat[\(R\)]
	{\includegraphics[trim={4 7 3 6},clip,width=.5\columnwidth]{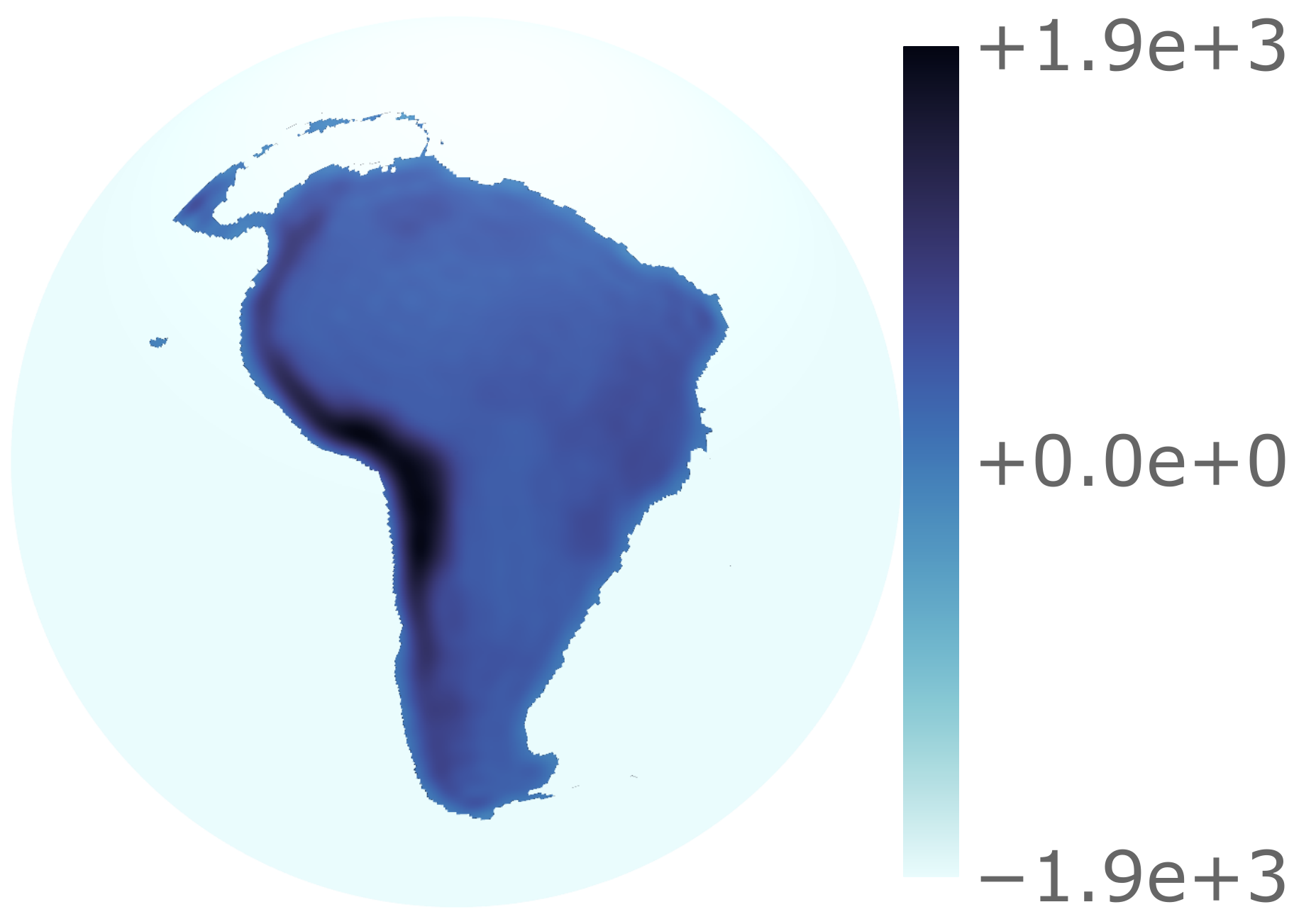}}
	\caption{
		Panel (a) corresponds to a topographic map of the Earth (from the EGM2008 dataset) centred on a view of South America.
		The dataset is bandlimited at \(\lmax=128\), and smoothed with \(\fwhm{1.17}\).
		Panel (b) presents the region \(R\), the shape of which is constructed from the Slepian coefficients of the South America mask.
		The field value outside the region in panel (b) is set to negative infinity for illustrative purposes.
		The amplitude of the right panel is set by the height of the Andes, rather than the lowest depths of the sea.
	}\label{fig:south_america_region}
\end{figure}

A region on the sphere is constructed from the Earth Gravitational Model EGM2008 dataset~\cite{Pavlis2013}, which is a topographic map of the Earth.
The left panel of \cref{fig:south_america_region} presents the dataset up to an order of \(\lmax=128\), smoothed with a \(\fwhm{1.17}\) to avoid aliasing, and centred on a view of South America.
The smoothing is performed purely to improve the visuals from bandlimiting the initial dataset.
A masked region \(R\) is constructed by centring a polar cap of angular opening \(\SI{40}{\degree}\) over South America, and then:
\begin{inparaenum}[(i)]
	\item setting the field value to zero outside the cap; and
	\item setting the negative field value inside the cap to zero.
\end{inparaenum}
The resulting region is shown in the right panel of \cref{fig:south_america_region}.
Another region is considered in \cref{sec:appendix}.

\begin{figure}
	\centering
	\subfloat[\(\Re\big\{\pixel{S_{1}}\big\},\ \mu_{1}=1.00\)] 
	{\includegraphics[trim={4 7 3 6},clip,width=.5\columnwidth]{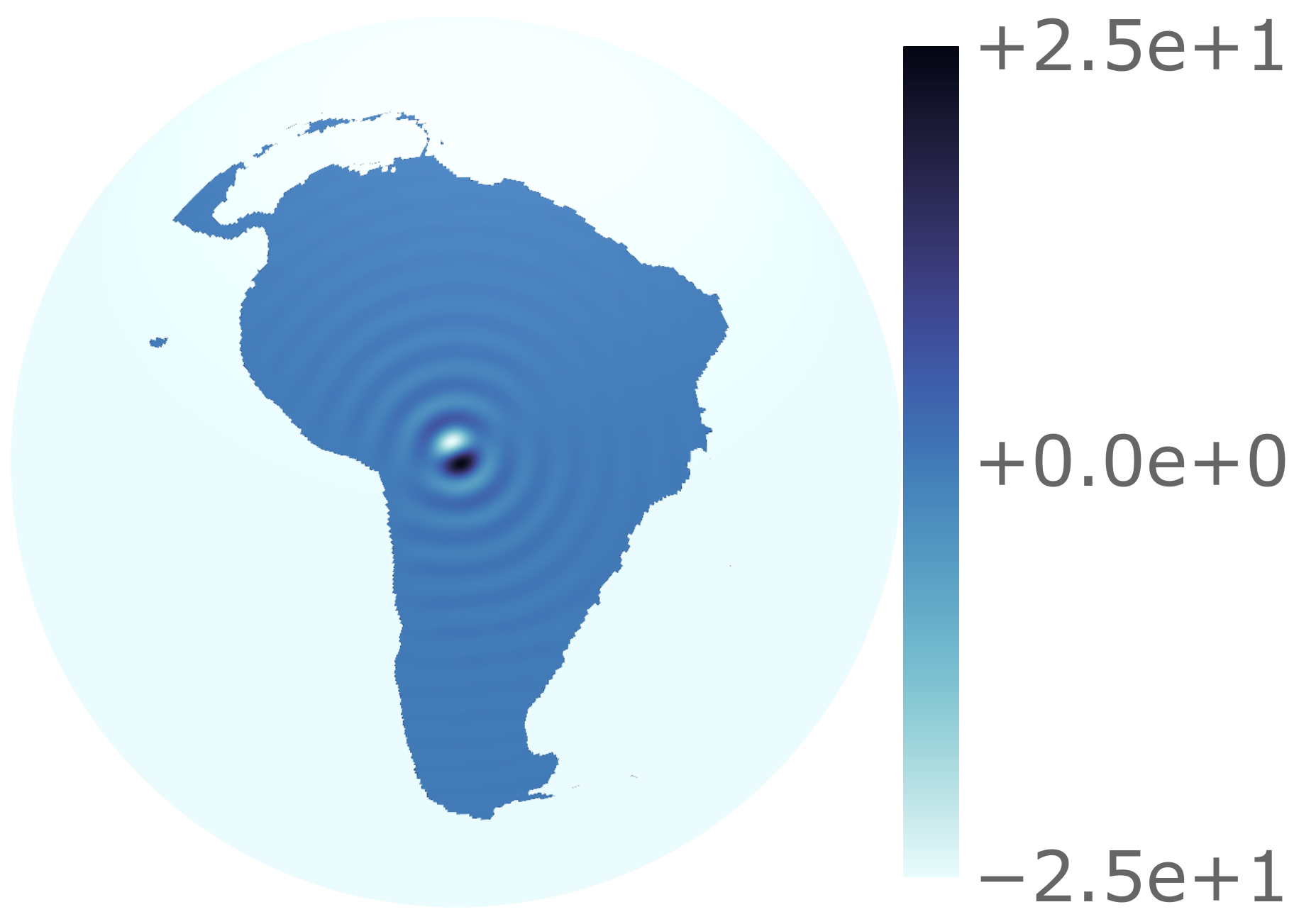}} 
	\hfill
	\subfloat[\(\Re\big\{\pixel{S_{10}}\big\},\ \mu_{10}=1.00\)] 
	{\includegraphics[trim={4 7 3 6},clip,width=.5\columnwidth]{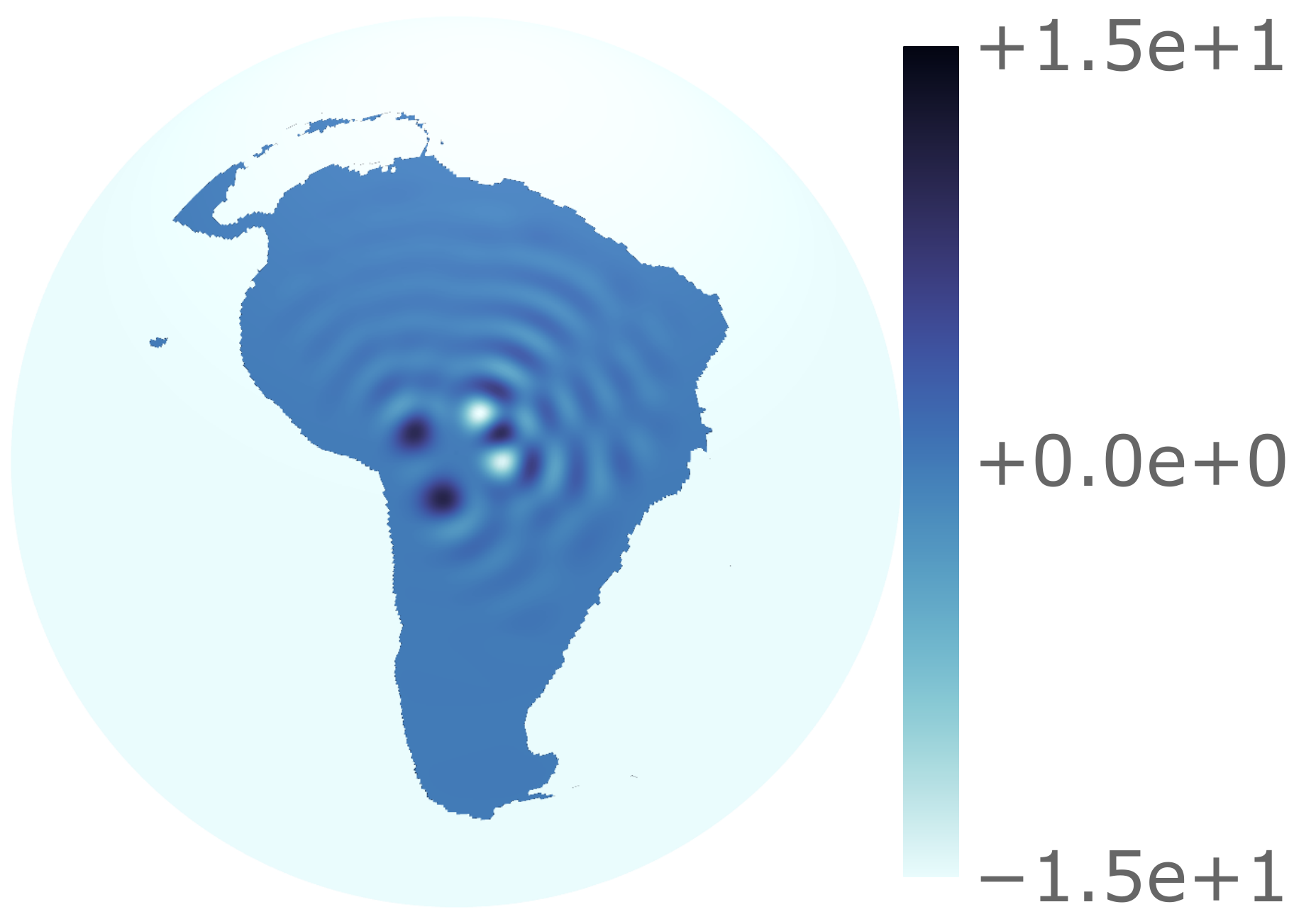}} 
	\newline
	\subfloat[\(\Re\big\{\pixel{S_{25}}\big\},\ \mu_{25}=1.00\)] 
	{\includegraphics[trim={4 7 3 6},clip,width=.5\columnwidth]{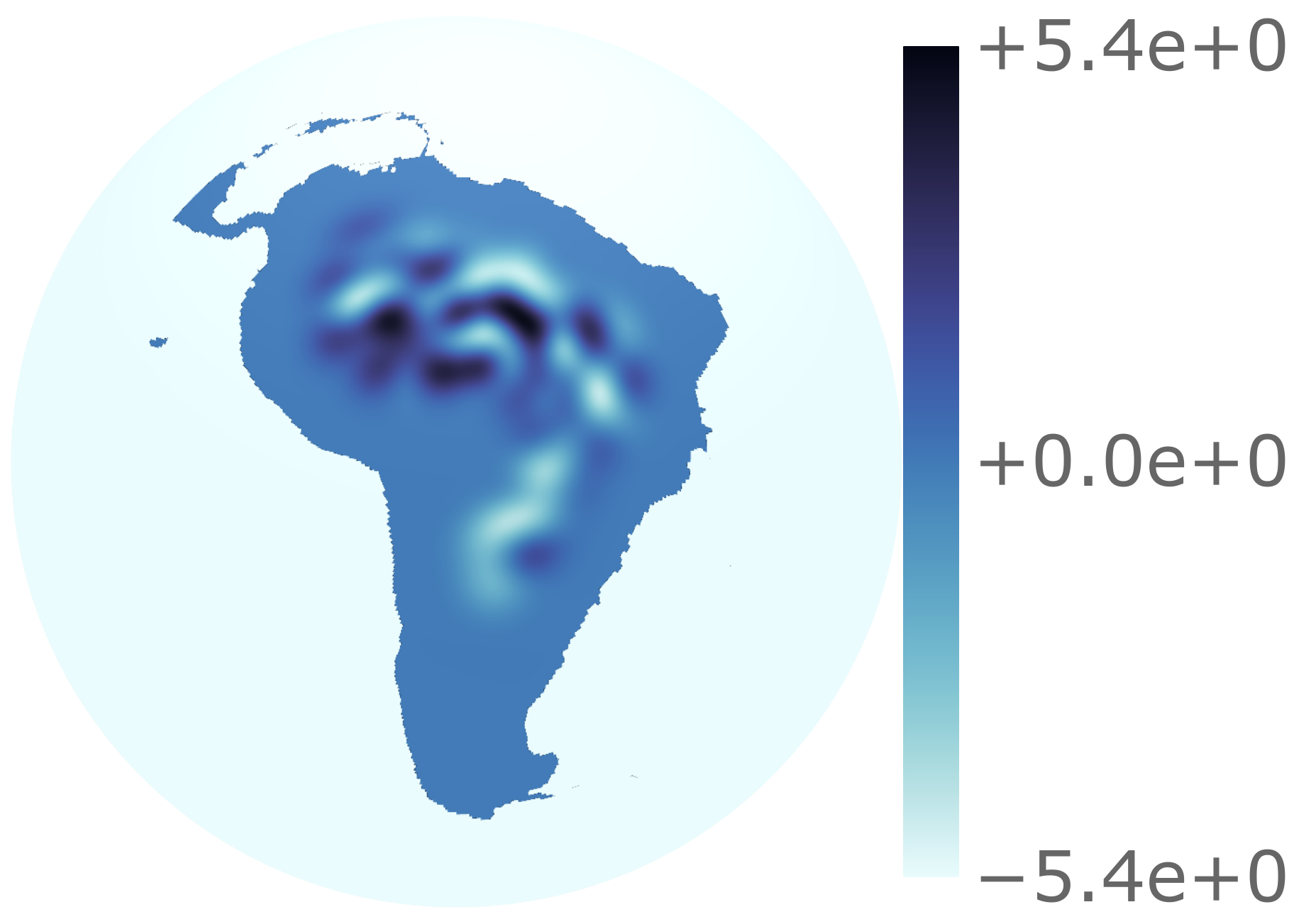}} 
	\hfill
	\subfloat[\(\Re\big\{\pixel{S_{50}}\big\},\ \mu_{50}=1.00\)] 
	{\includegraphics[trim={4 7 3 6},clip,width=.5\columnwidth]{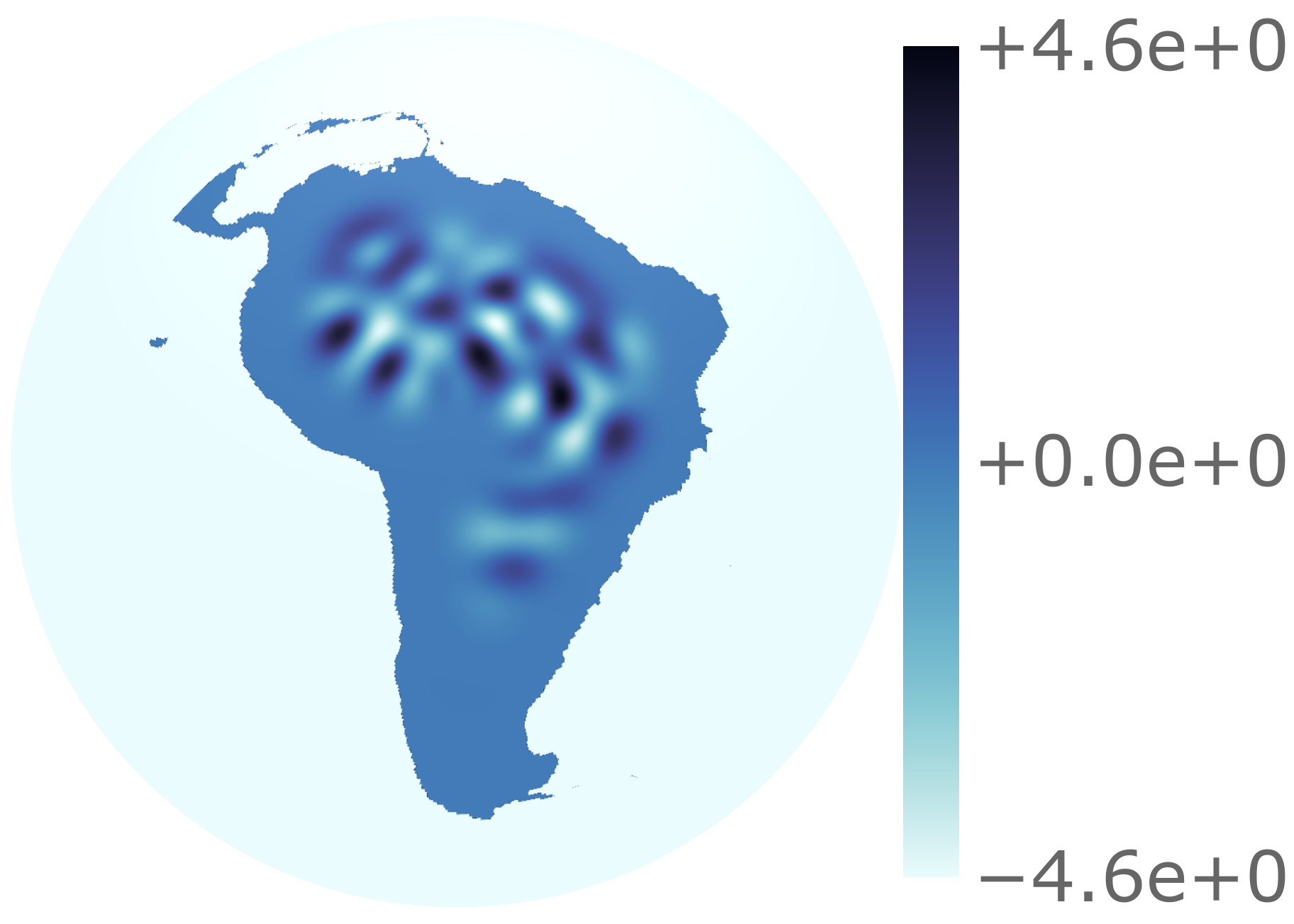}} 
	\newline
	\subfloat[\(\Re\big\{\pixel{S_{100}}\big\},\ \mu_{100}=1.00\)] 
	{\includegraphics[trim={4 7 3 6},clip,width=.5\columnwidth]{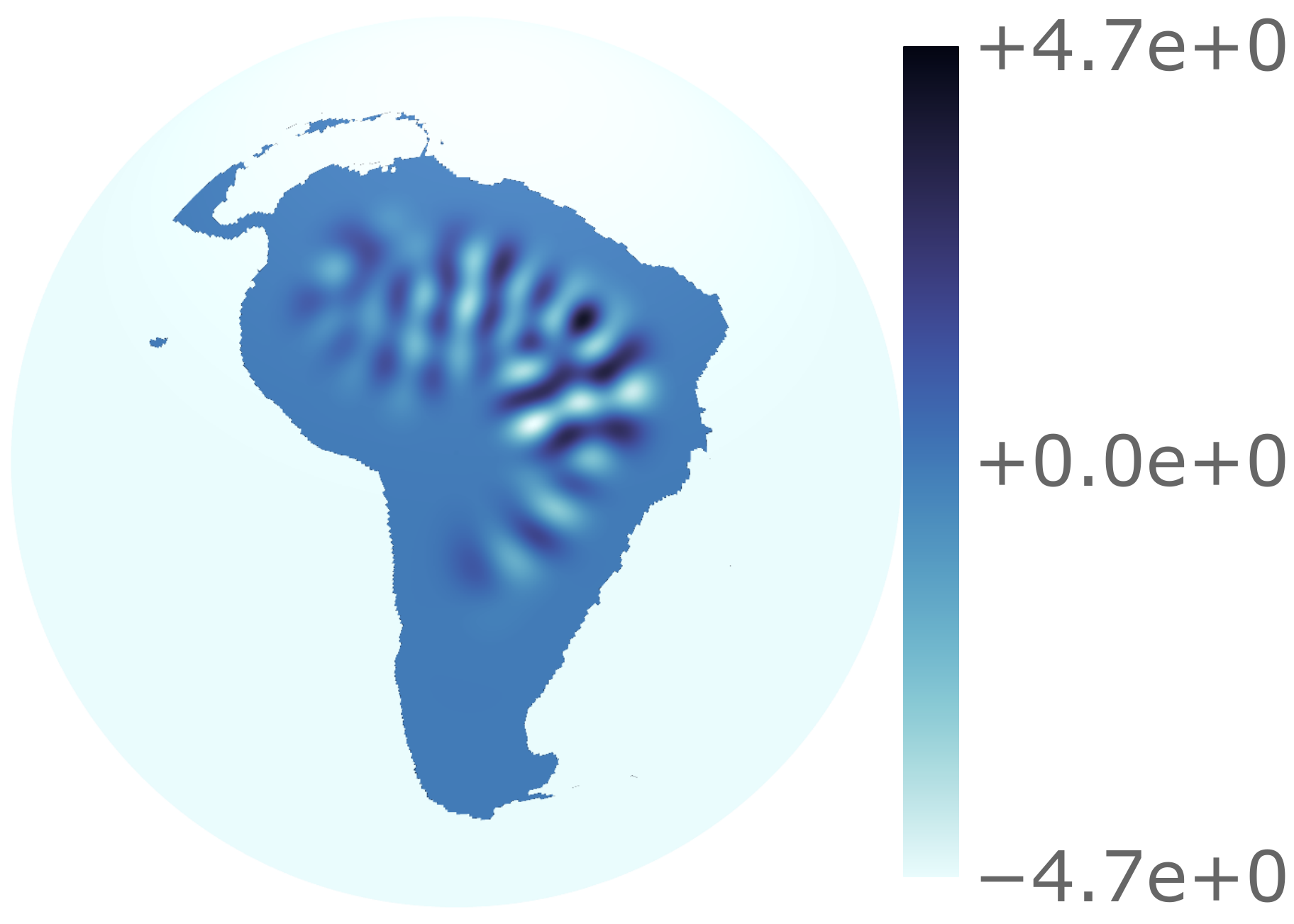}} 
	\hfill
	\subfloat[\(\Re\big\{\pixel{S_{200}}\big\},\ \mu_{200}=1.00\)] 
	{\includegraphics[trim={4 7 3 6},clip,width=.5\columnwidth]{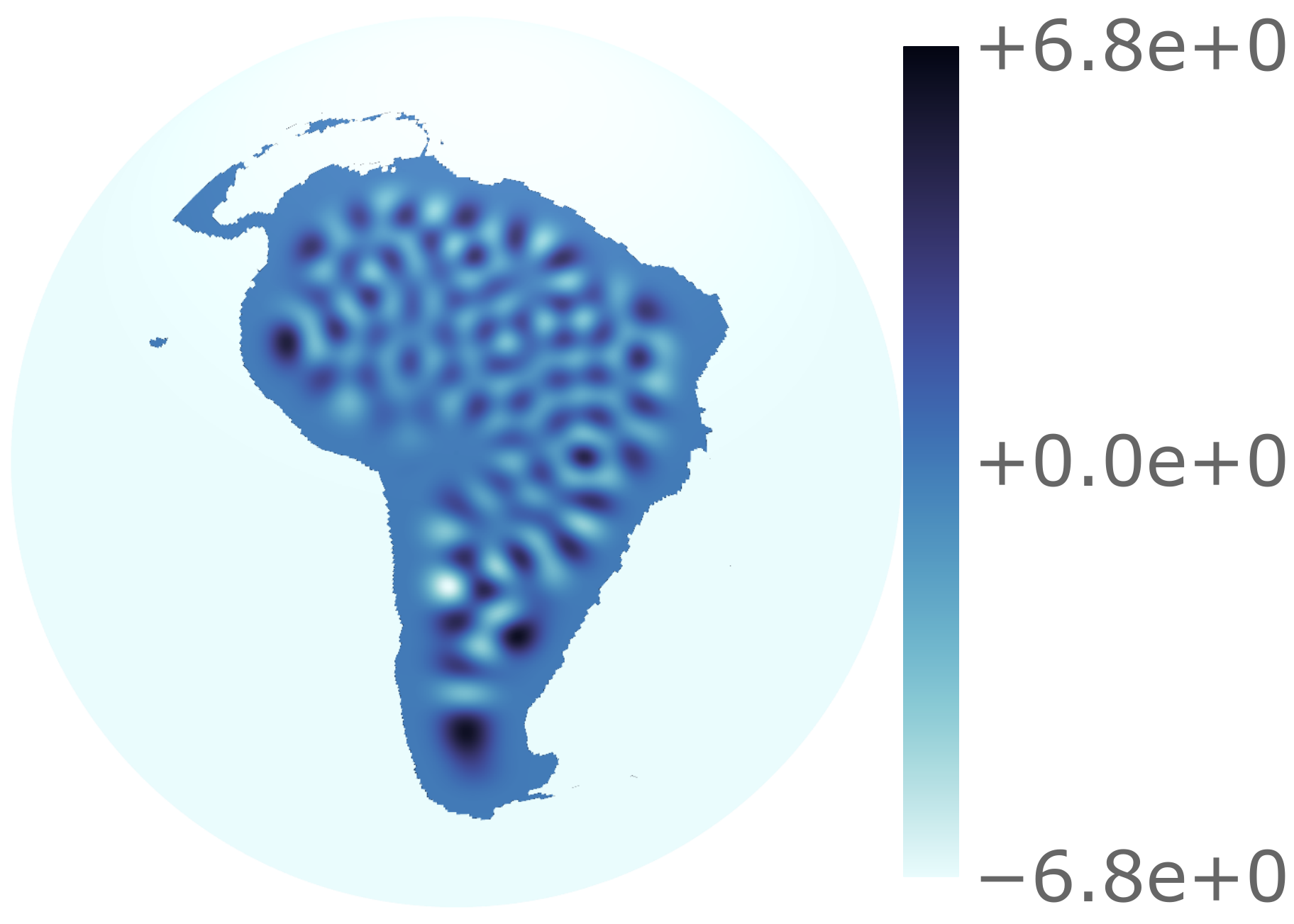}} 
	\caption{
		The Slepian functions of the South America region \(S_p(\omega)\) for \(p \in \set{1, 10, 25, 50, 100, 200}\) shown left-to-right, top-to-bottom.
		The corresponding eigenvalue \(\slepian{\mu}\) is a measure of the concentration within the given region \(R\), which remain \(\almost{1}\) for many \(p\) values before decreasing towards zero.
		Note the radial structure of the Slepian functions for a region on \(\twoSphere{}\) like that of the spherical harmonics.
	}\label{fig:south_america_eigenfunctions}
\end{figure}

\begin{figure}
	\centering
	\includegraphics[width=\columnwidth]{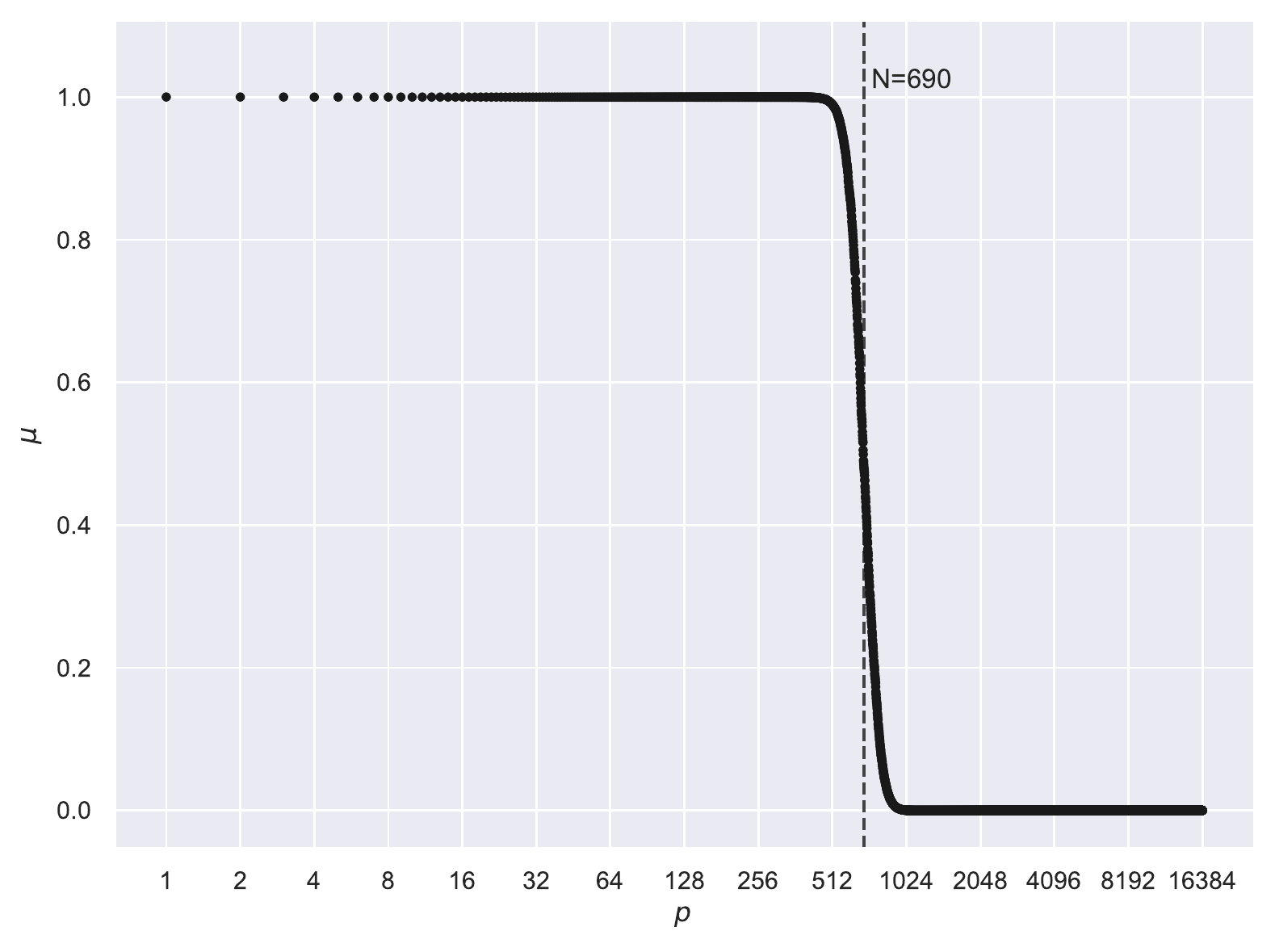}
	\caption{
		The eigenvalues of the South America region concentrated within the Shannon number \(N=690\).
		The majority of the eigenvalues are \(\almost{1}\) before decreasing rapidly towards zero around the Shannon number.
	}\label{fig:south_america_eigenvalues}
\end{figure}

The Slepian functions of this region are found by solving the eigenproblem \cref{eq:eigenproblem}, and then performing an inverse spherical harmonic transform.
The Shannon number \cref{eq:shannon} of \(R\) is \(N=690\).
An example set of Slepian functions for this region is given in \cref{fig:south_america_eigenfunctions} for \(p \in \set{1, 10, 25, 50, 100, 200}\), with the corresponding eigenvalue \(\slepian{\mu}\), which is a measure of the concentration within the region.
The Slepian functions initially peak in the middle of the region representing good concentration, and gradually spread out as \(p\) increases.
The boundaries of the region are captured by those Slepian functions where \(p \lesssim N\) corresponding to the wavelets in panels (e--f) of \cref{fig:south_america_slepian_wavelets}. 
The corresponding \(N\) eigenvalues are shown in \cref{fig:south_america_eigenvalues}, where the eigenvalues remain \(\almost{1}\) for many \(p\) values until decreasing towards zero.

\subsection{Wavelets and Wavelet Coefficients}\label{sec:south_america_wavelets}

Recall the Slepian scaling function and wavelets described in \cref{sec:generating_functions}.
These functions are constructed from a tiling of the Slepian line with parameters \(\lambda=3\), \(J_{0}=2\) and bandlimit \(\lmax=128\) as shown in \cref{fig:tiling}, where the Shannon number of the South America region has been highlighted.
In practice, this means that the scaling function and the wavelets for scales \(j \in \set{2, 3, 4, 5, 6}\) are the only non-zero basis functions for this region.

\begin{figure}
	\centering
	\subfloat[\(\Re\big\{\pixel{\Phi}\big\}\)] 
	{\includegraphics[trim={4 7 3 6},clip,width=.5\columnwidth]{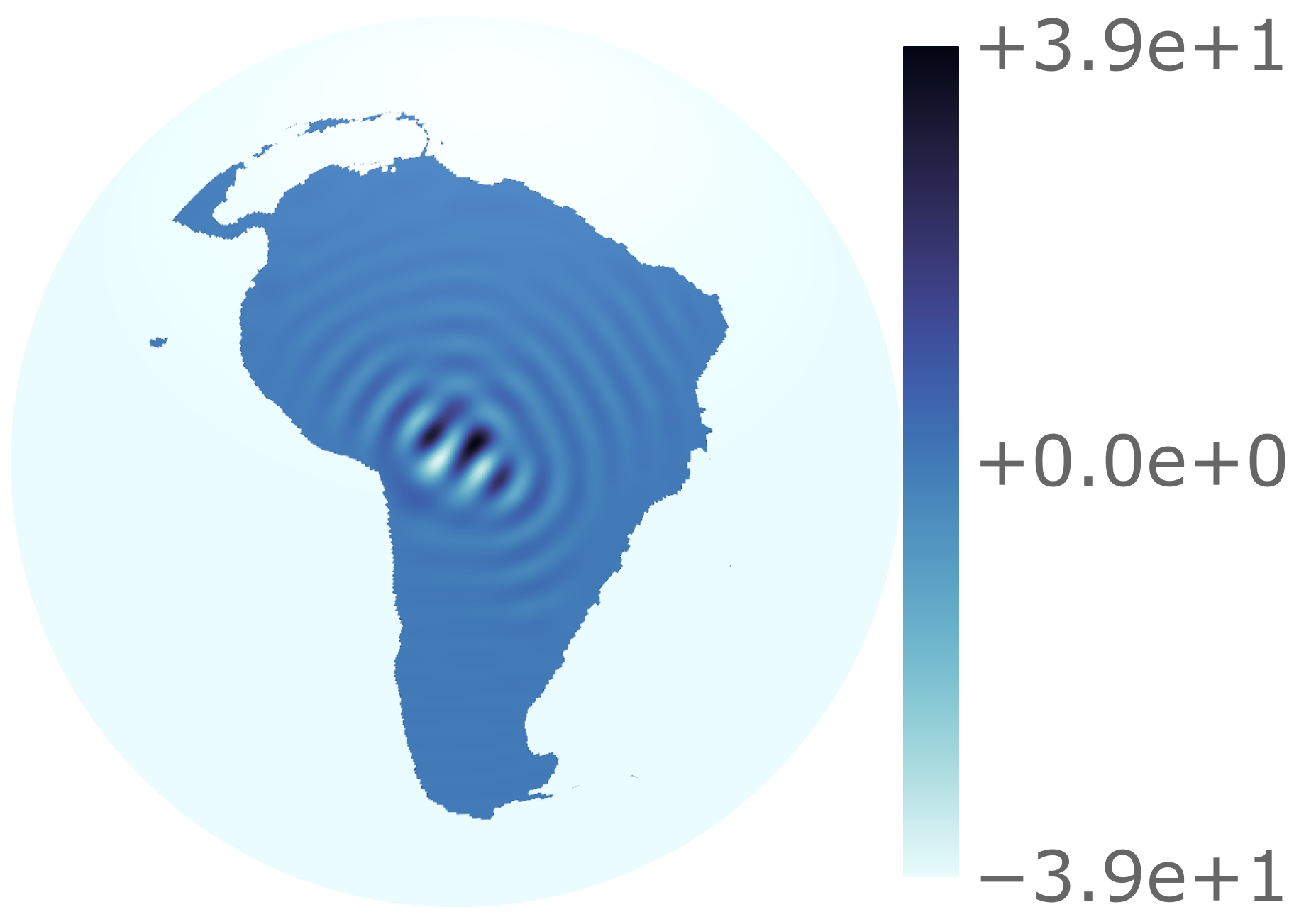}}
	\hfill
	\subfloat[\(\Re\big\{\pixel{\Psi^{2j}}\big\}\)] 
	{\includegraphics[trim={4 7 3 6},clip,width=.5\columnwidth]{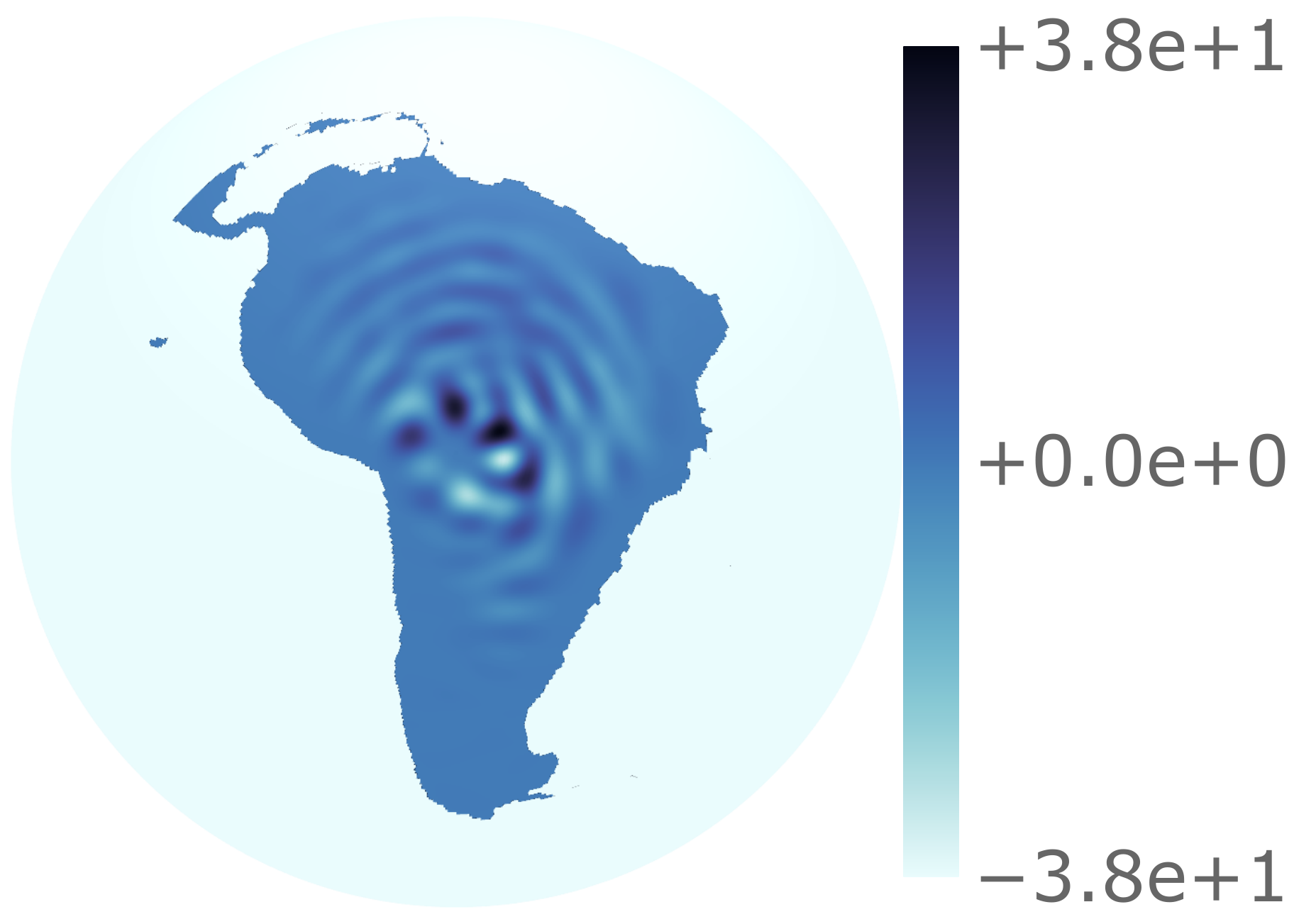}}
	\newline
	\subfloat[\(\Re\big\{\pixel{\Psi^{3j}}\big\}\)] 
	{\includegraphics[trim={4 7 3 6},clip,width=.5\columnwidth]{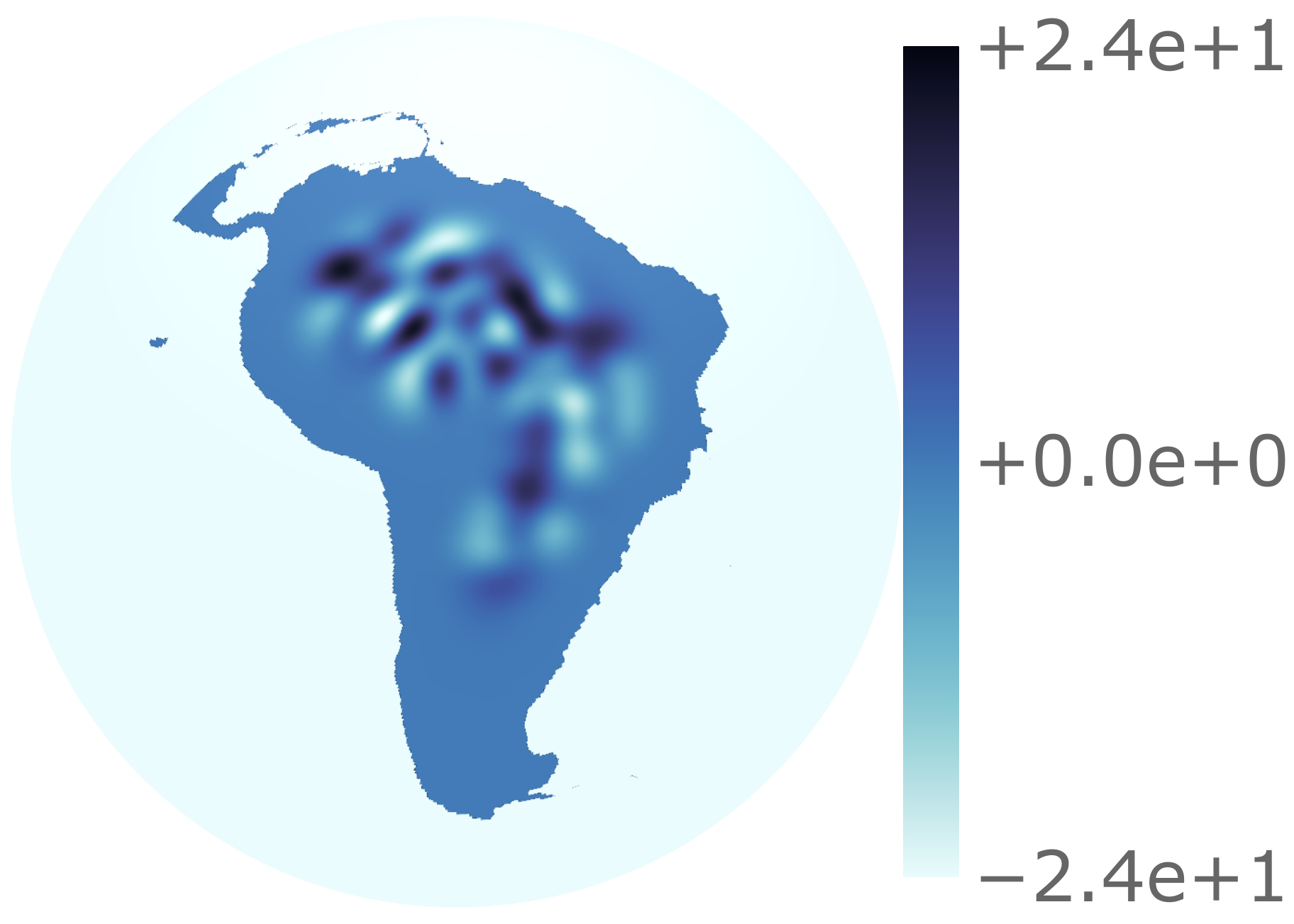}}
	\hfill
	\subfloat[\(\Re\big\{\pixel{\Psi^{4j}}\big\}\)] 
	{\includegraphics[trim={4 7 3 6},clip,width=.5\columnwidth]{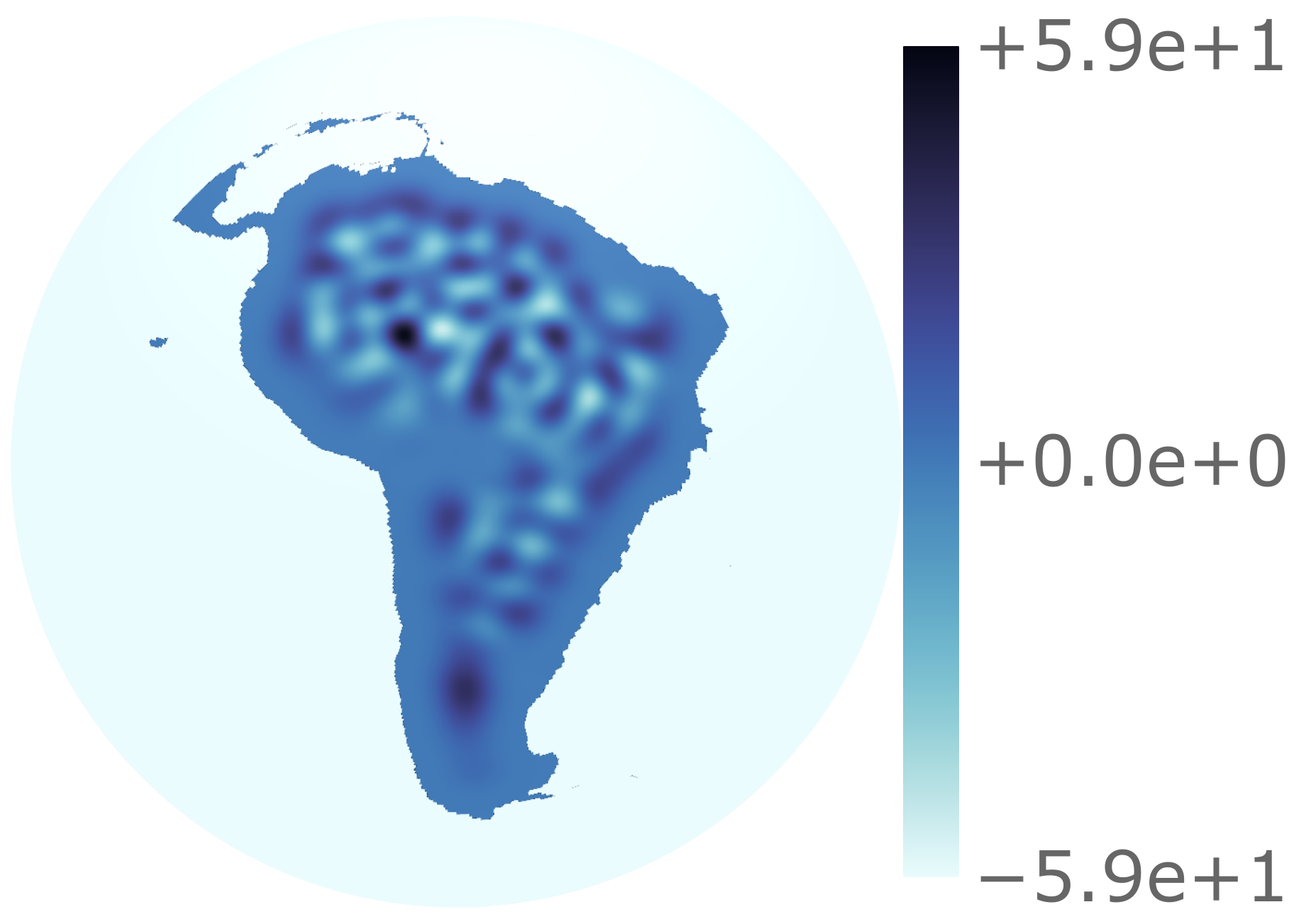}}
	\newline
	\subfloat[\(\Re\big\{\pixel{\Psi^{5j}}\big\}\)] 
	{\includegraphics[trim={4 7 3 6},clip,width=.5\columnwidth]{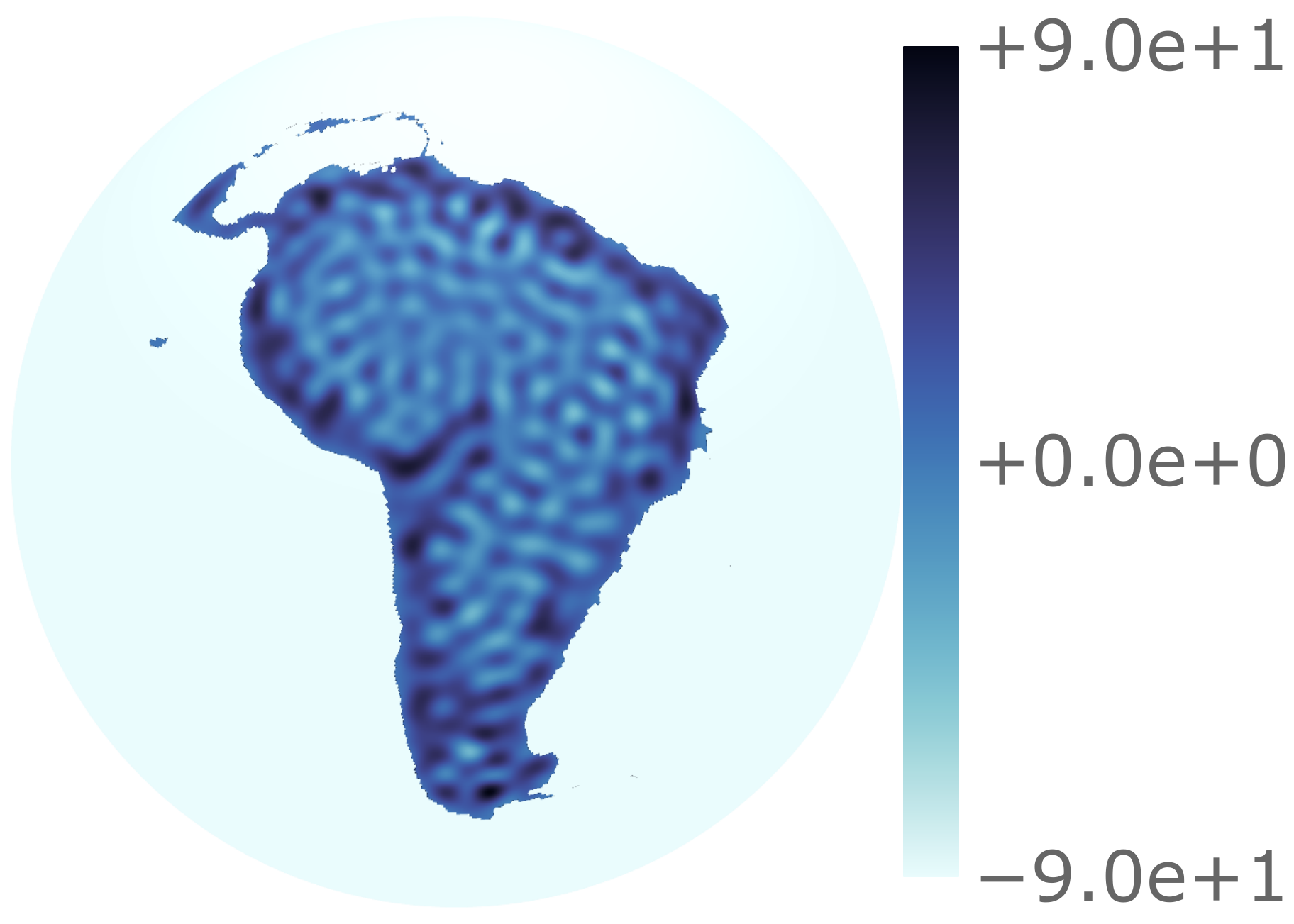}}
	\hfill
	\subfloat[\(\Re\big\{\pixel{\Psi^{6j}}\big\}\)] 
	{\includegraphics[trim={4 7 3 6},clip,width=.5\columnwidth]{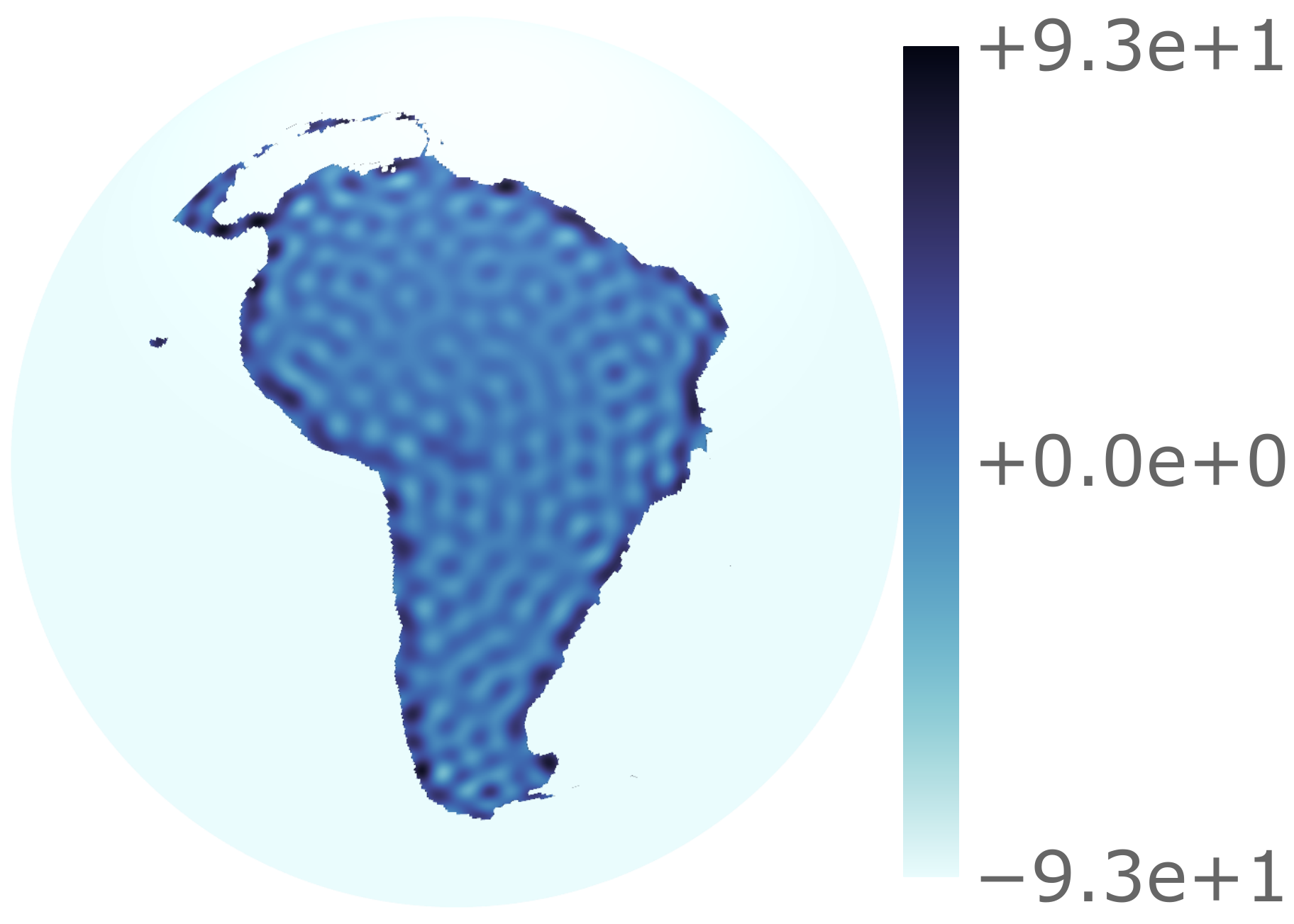}}
	\caption{
		The scaling function and the wavelets for scales \(j \in \set{2, 3, 4, 5, 6}\) for the South America region shown left-to-right, top-to-bottom.
		The wavelets are constructed through a tiling of the Slepian line using scale-discretised functions, with parameters \(\lambda=3\), \(J_{0}=2\), and bandlimit \(\lmax=128\).
		In contrast to the standard axisymmetric scale-discretised wavelets, Slepian wavelets have a radial structure due to radial nature of the Slepian functions in \cref{fig:south_america_eigenfunctions}.
	}\label{fig:south_america_slepian_wavelets}
\end{figure}

\begin{figure}
	\centering
	\subfloat[\(\Re\big\{\pixel{W^{\Phi}}\big\}\)] 
	{\includegraphics[trim={4 7 3 6},clip,width=.5\columnwidth]{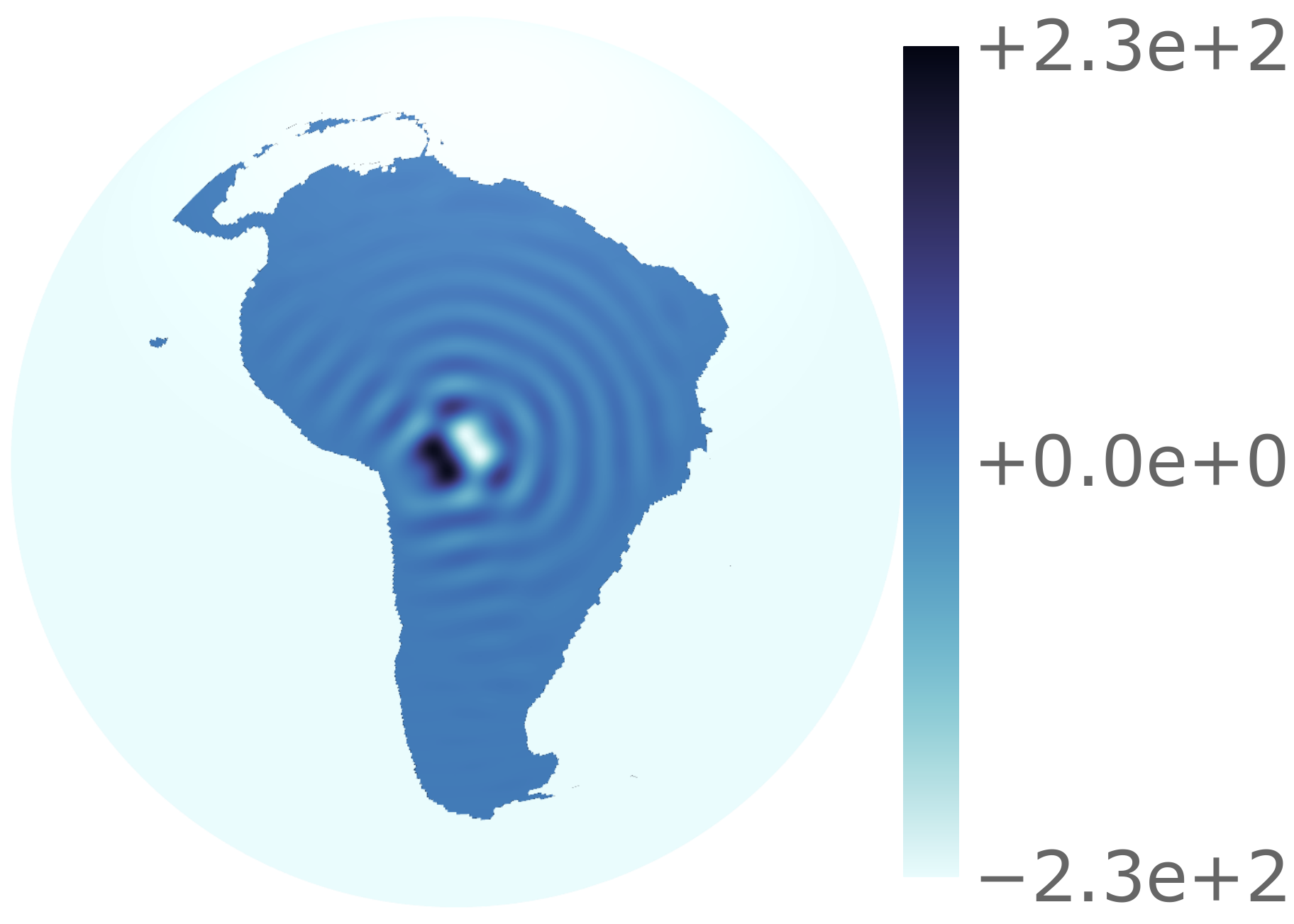}}
	\hfill
	\subfloat[\(\Re\big\{\pixel{W^{\Psi^{2j}}}\big\}\)] 
	{\includegraphics[trim={4 7 3 6},clip,width=.5\columnwidth]{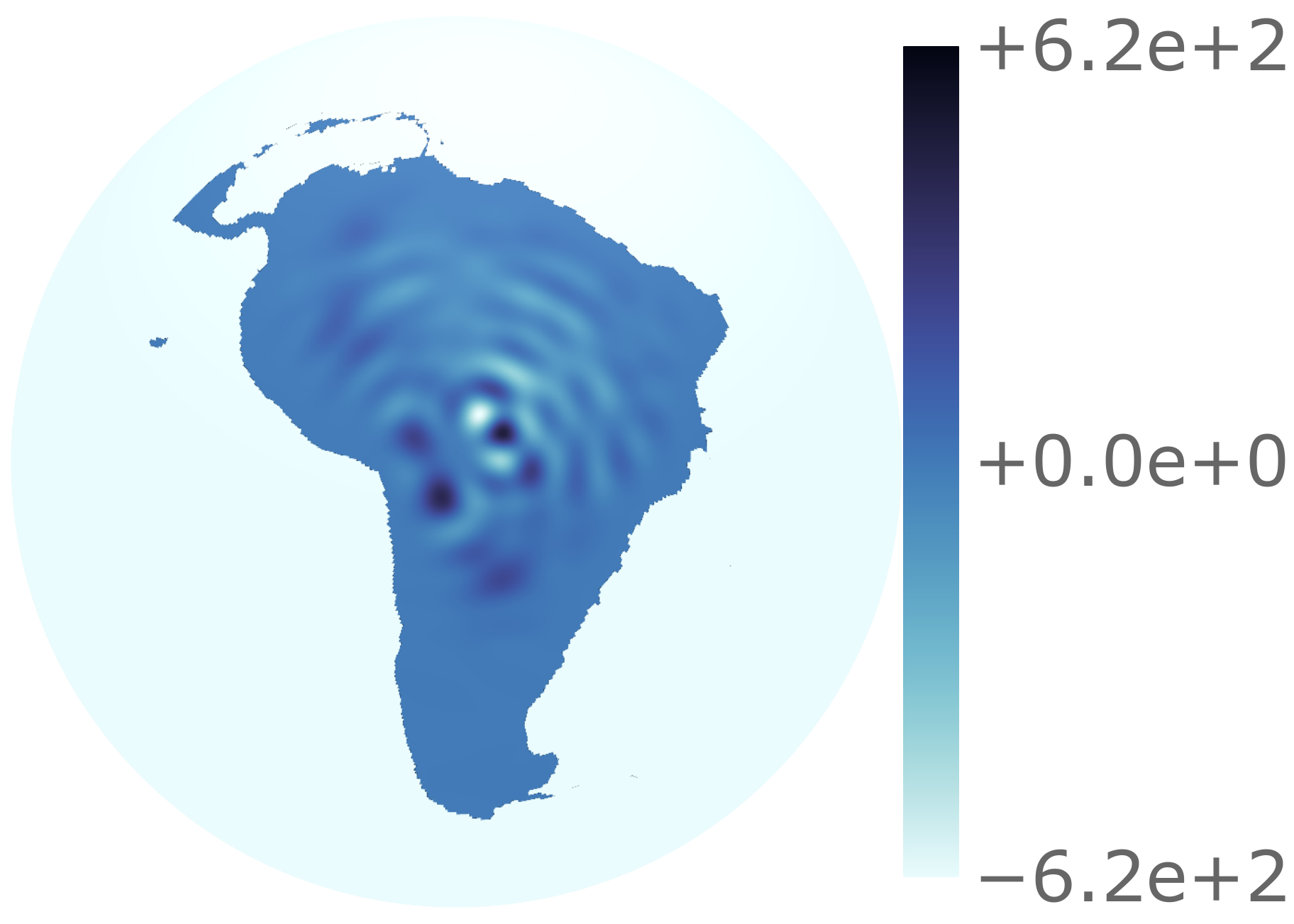}}
	\newline
	\subfloat[\(\Re\big\{\pixel{W^{\Psi^{3j}}}\big\}\)] 
	{\includegraphics[trim={4 7 3 6},clip,width=.5\columnwidth]{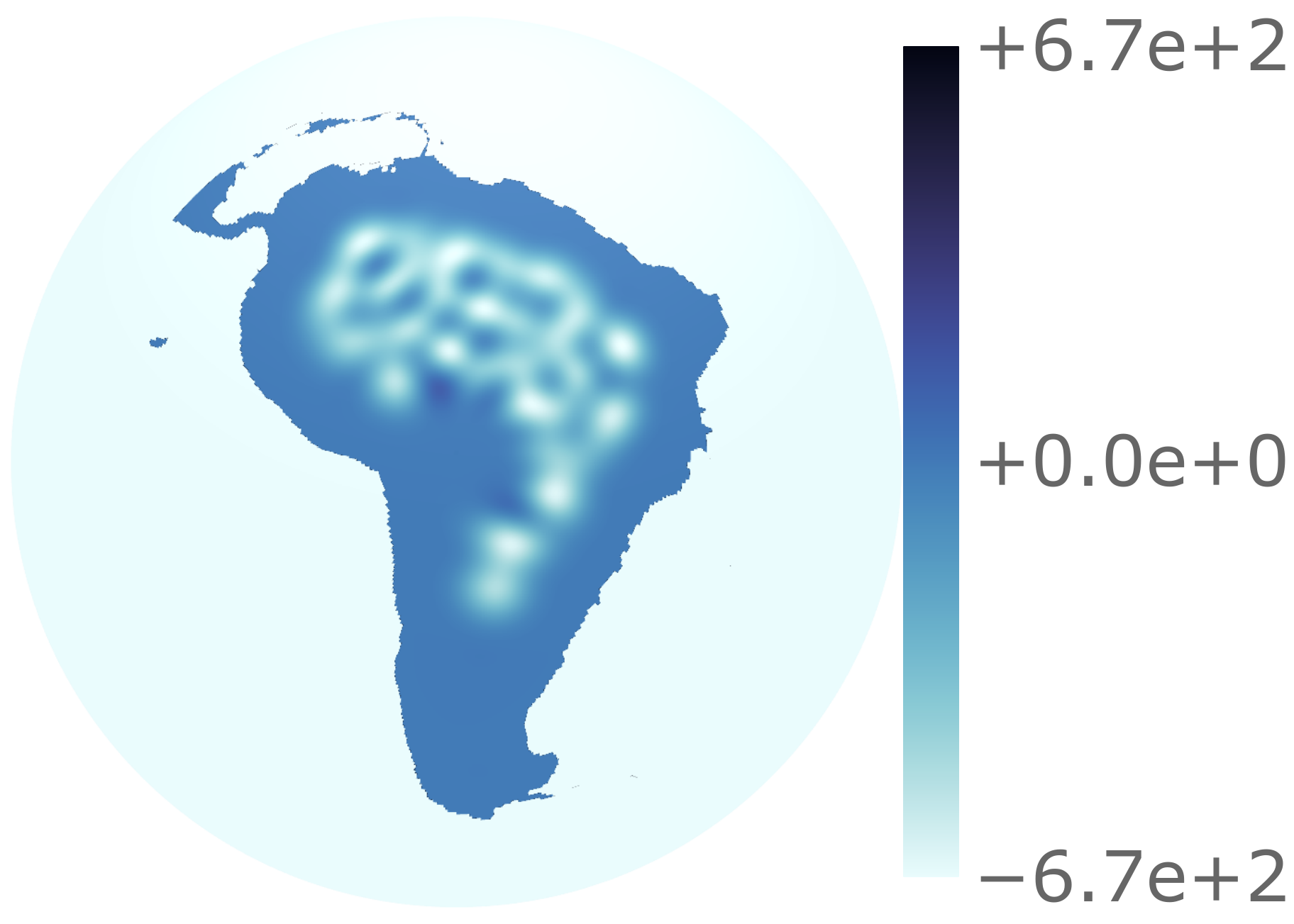}}
	\hfill
	\subfloat[\(\Re\big\{\pixel{W^{\Psi^{4j}}}\big\}\)] 
	{\includegraphics[trim={4 7 3 6},clip,width=.5\columnwidth]{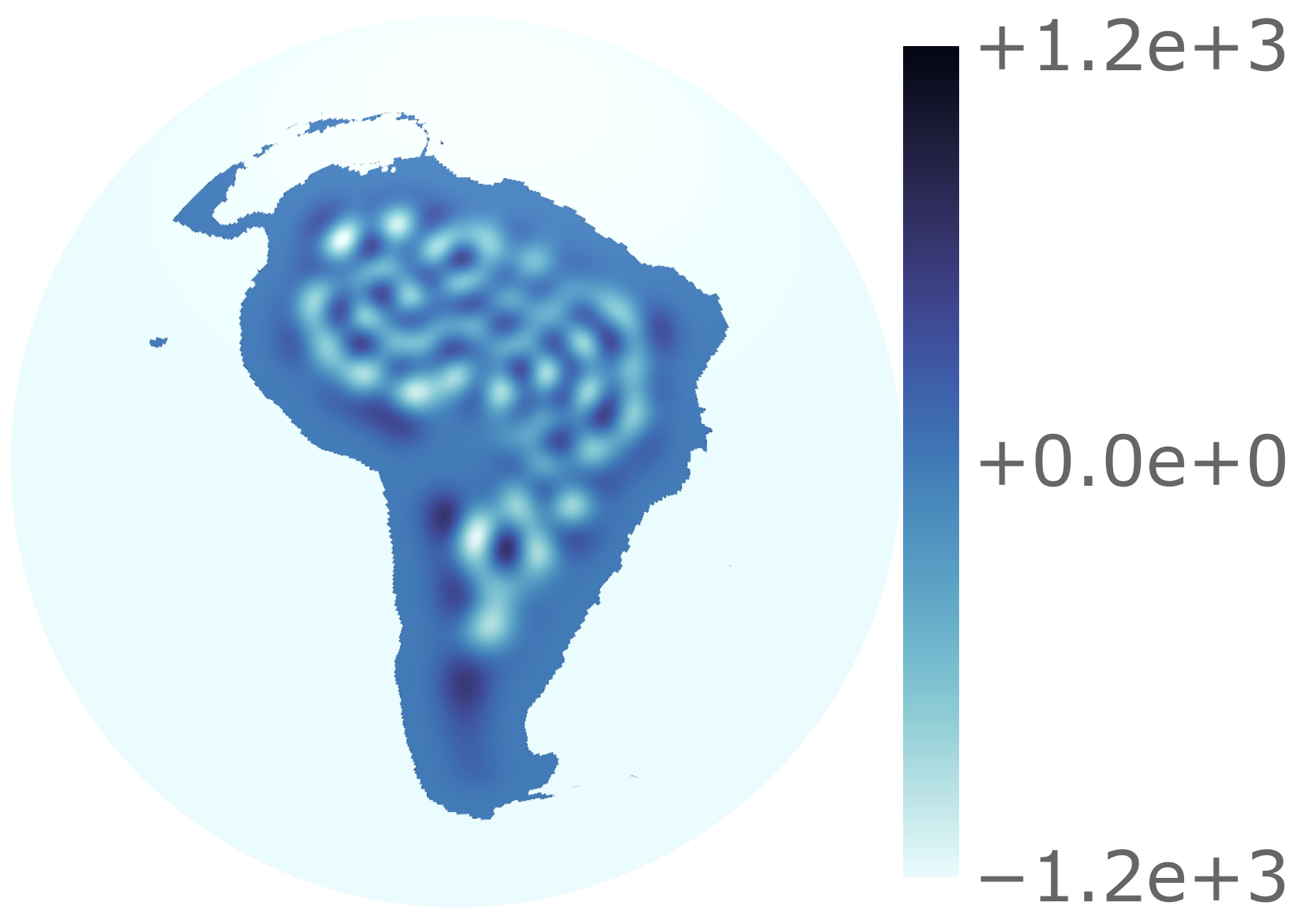}}
	\newline
	\subfloat[\(\Re\big\{\pixel{W^{\Psi^{5j}}}\big\}\)] 
	{\includegraphics[trim={4 7 3 6},clip,width=.5\columnwidth]{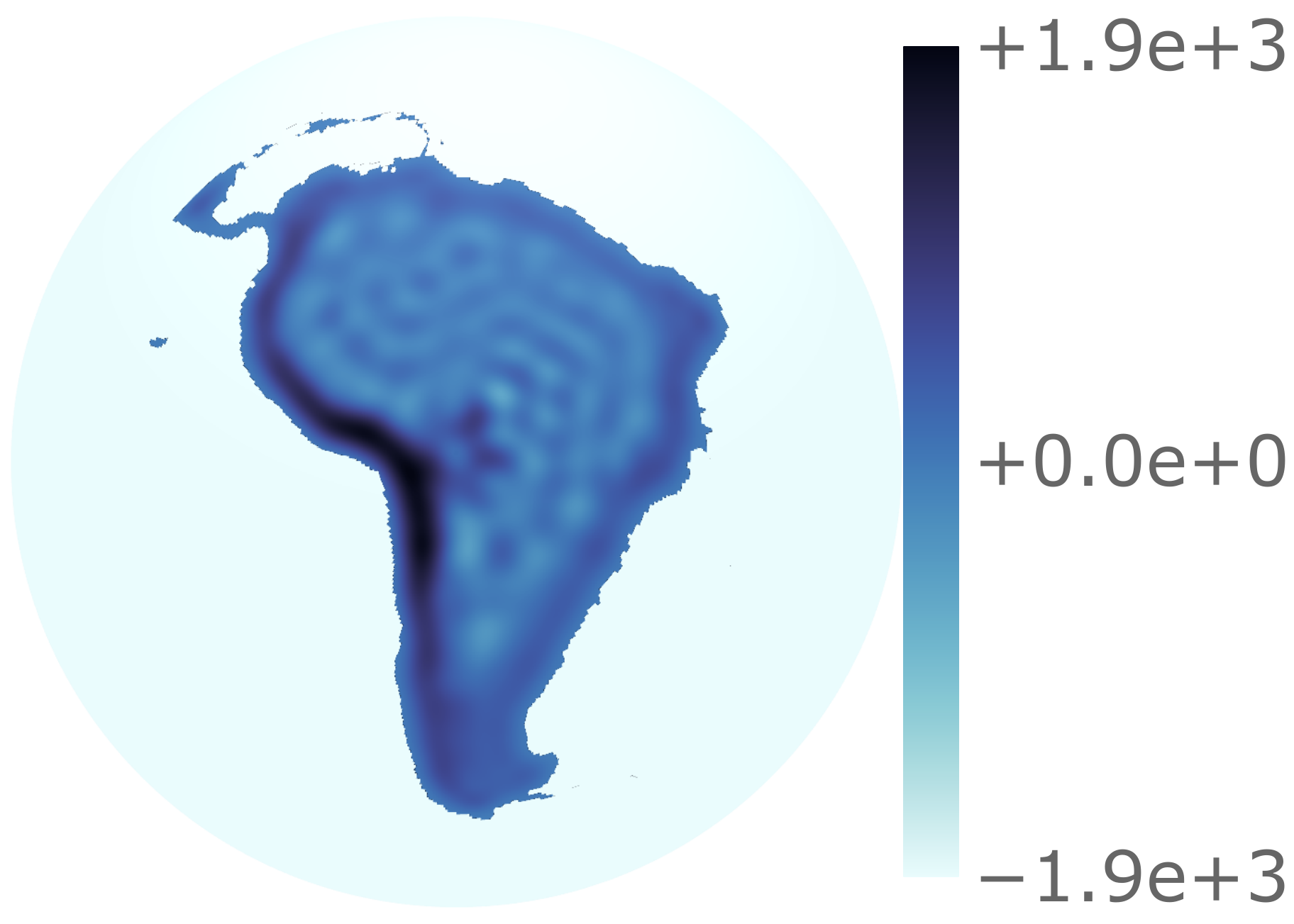}}
	\hfill
	\subfloat[\(\Re\big\{\pixel{W^{\Psi^{6j}}}\big\}\)] 
	{\includegraphics[trim={4 7 3 6},clip,width=.5\columnwidth]{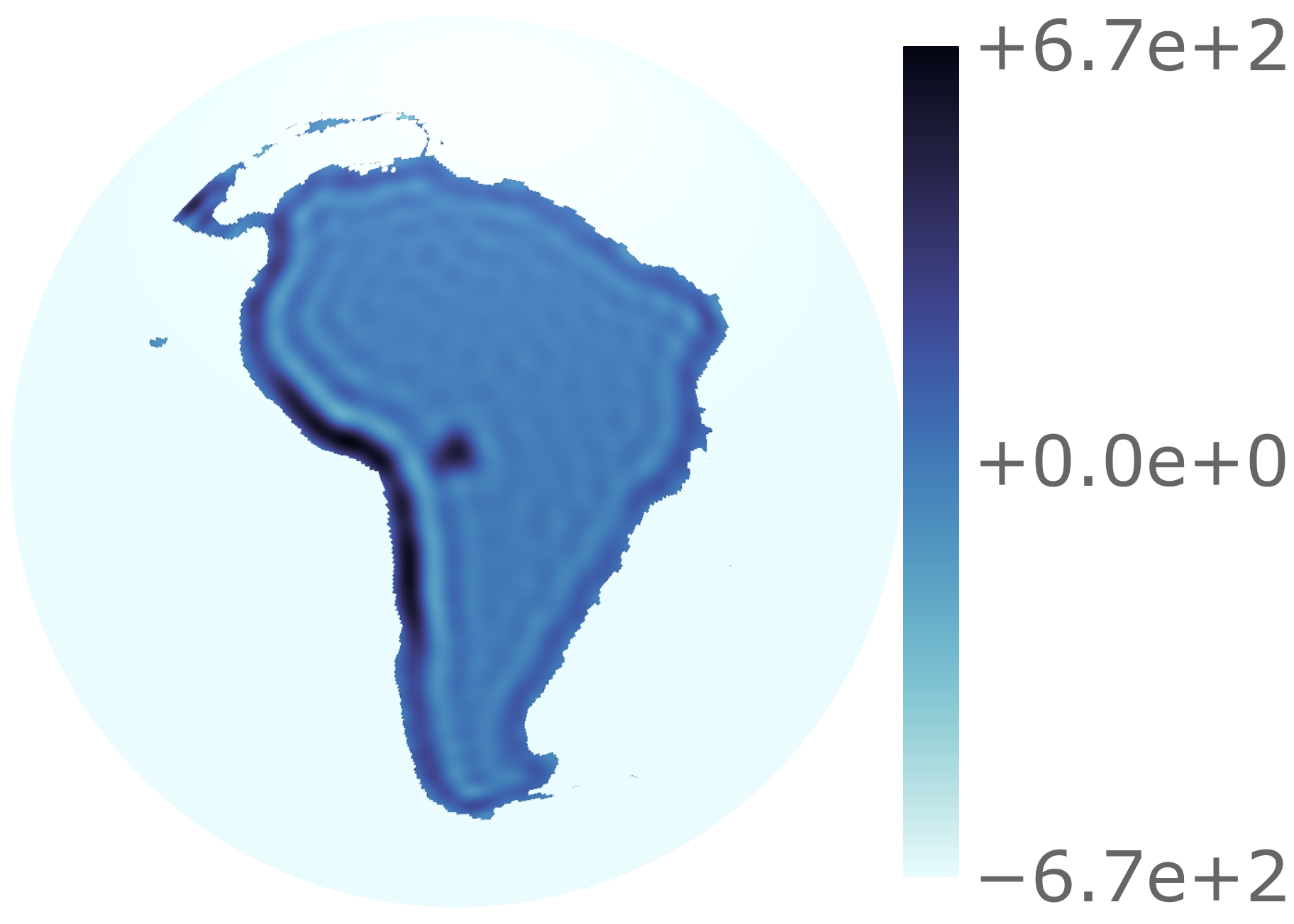}}
	\caption{
		The scale-discretised wavelet transform of the topographic map of South America for parameters \(\lambda=3\), \(J_{0}=2\), and bandlimit \(\lmax=128\); \ie{} with the wavelets shown in \cref{fig:south_america_slepian_wavelets}.
		Spatially localised, scale-dependent features of the bandlimited signal may be extracted by the wavelet coefficients given by the wavelet transform.
		The scaling coefficients are given in the top left plot, while the wavelet coefficients at scales \(j \in \set{2, 3, 4, 5, 6}\) are shown left-to-right, top-to-bottom.
	}\label{fig:south_america_slepian_wavelet_coefficients}
\end{figure}

The scaling function and corresponding wavelets are shown in \cref{fig:south_america_slepian_wavelets}.
The scaling function and the lower wavelet scales are more concentrated within the region as they are based on the Slepian functions corresponding to the eigenvalues on the left of \cref{fig:south_america_eigenvalues}.
In contrast, the higher wavelet scales are focused on the boundaries of the region due to the less well-concentrated Slepian functions.
One may calculate the corresponding scaling and wavelet coefficients using the wavelet transforms in \cref{eq:slepian_scaling_omega,eq:slepian_wavelet_omega} respectively.
The wavelet and scaling coefficients for the Earth topographic data of the South America region are shown in \cref{fig:south_america_slepian_wavelet_coefficients}.

\subsection{Wavelet Denoising}\label{sec:wavelet_denoising}

A typical use case of wavelets is to denoise signals.
Wavelets localise features in the data to different scales, hence the important parts of the signal can be preserved whilst removing the noise.
To test the efficacy of Slepian wavelets, white noise is added to the constructed region \(R\) in the right panel of \cref{fig:south_america_region}.
A straightforward hard-thresholding scheme is developed to perform the denoising.

To calculate the covariance of a filtered field (wavelet coefficients), consider a signal localised in \(R\) in the presence of noise
\begin{equation}\label{eq:noised_signal}
	\pixel{x} = \pixel{s} + \pixel{n},
\end{equation}
where \(\pixel{s}\) and \(\pixel{n}\) are the signal and noise, respectively.
The power spectrum of the noise in Slepian space is as before
\begin{equation}
	\expval*{\slepian{n} \conj{\slepian[']{n}}}
	= \sigma^{2} \delta_{pp'}.
\end{equation}
To assess the fidelity of the observed signal, a signal-to-noise-ratio in the region is given by
\begin{equation}
	\snr{x}
	\equiv 10 \log_{10} \frac{\inducedNorm{s}^{2}}{\inducedNorm{x - s}^{2}}.
\end{equation}
One seeks a denoised version of \(x\), denoted \(d \in \hilbert{R}\), with large \(\snr{d}\) such that \(d\) isolates the informative signal \(s\).
In contrast to standard spherical wavelets --- where the scaling function is often \emph{not} used in denoising --- with Slepian wavelets, one treats both the scaling function and the wavelets similarly.
The Slepian functions, on which the Slepian wavelet scaling function is constructed, can be relatively well-localised and hence the Slepian wavelet scaling coefficients are not necessarily a low-frequency representation of the signal (see \cref{sec:localisation}).

\begin{figure*}[!b]
	\centering
	\subfloat[Initial Noisy Data \newline
		\(\snr{x} = \SI{4.11}{\dB}\)]
	{\includegraphics[trim={4 7 3 6},clip,width=.5\columnwidth]{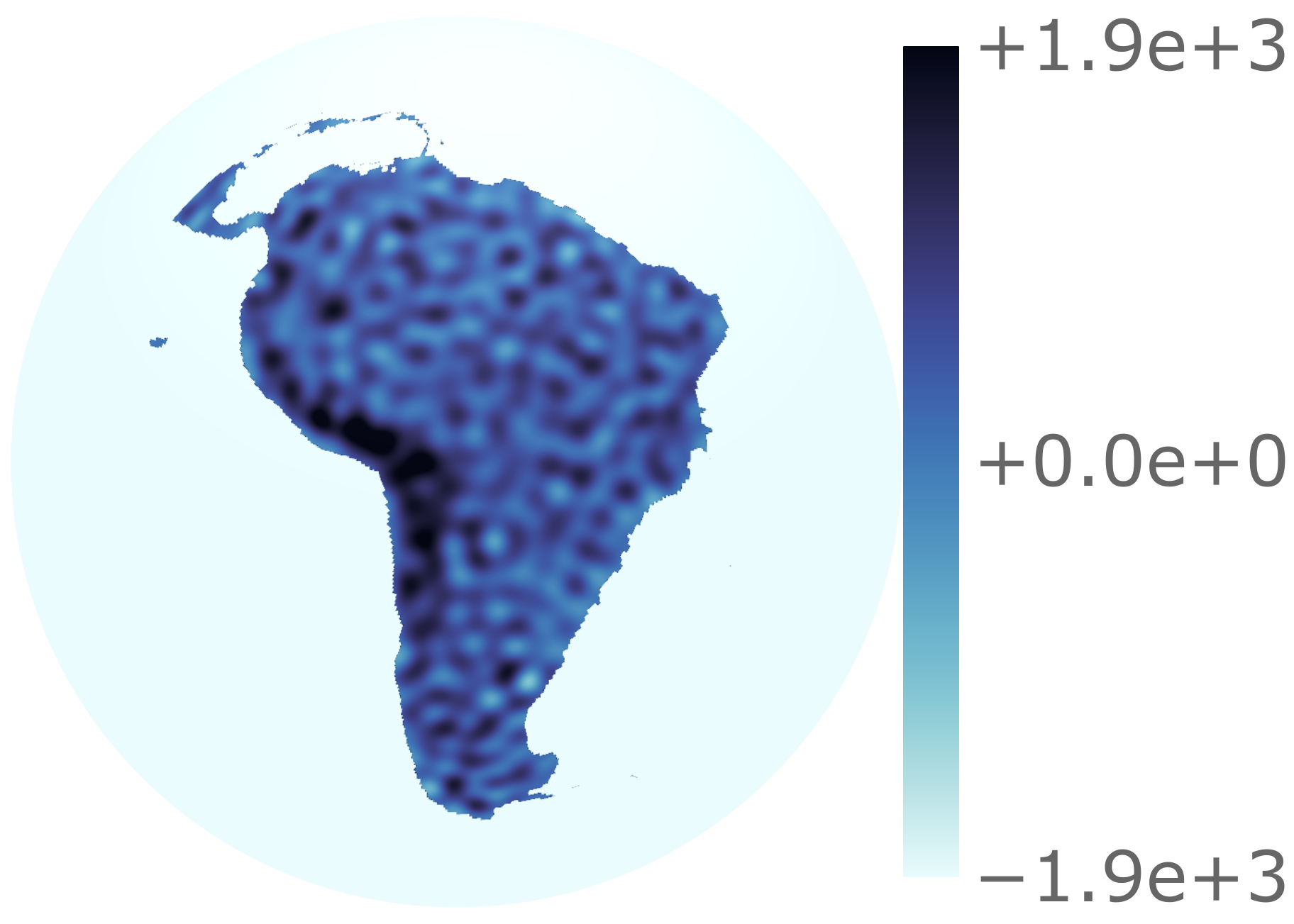}}
	\hfill
	\subfloat[Denoised \(N_{\sigma}=2\) \newline
		\(\snr{d} = \SI{5.67}{\dB}\)]
	{\includegraphics[trim={4 7 3 6},clip,width=.5\columnwidth]{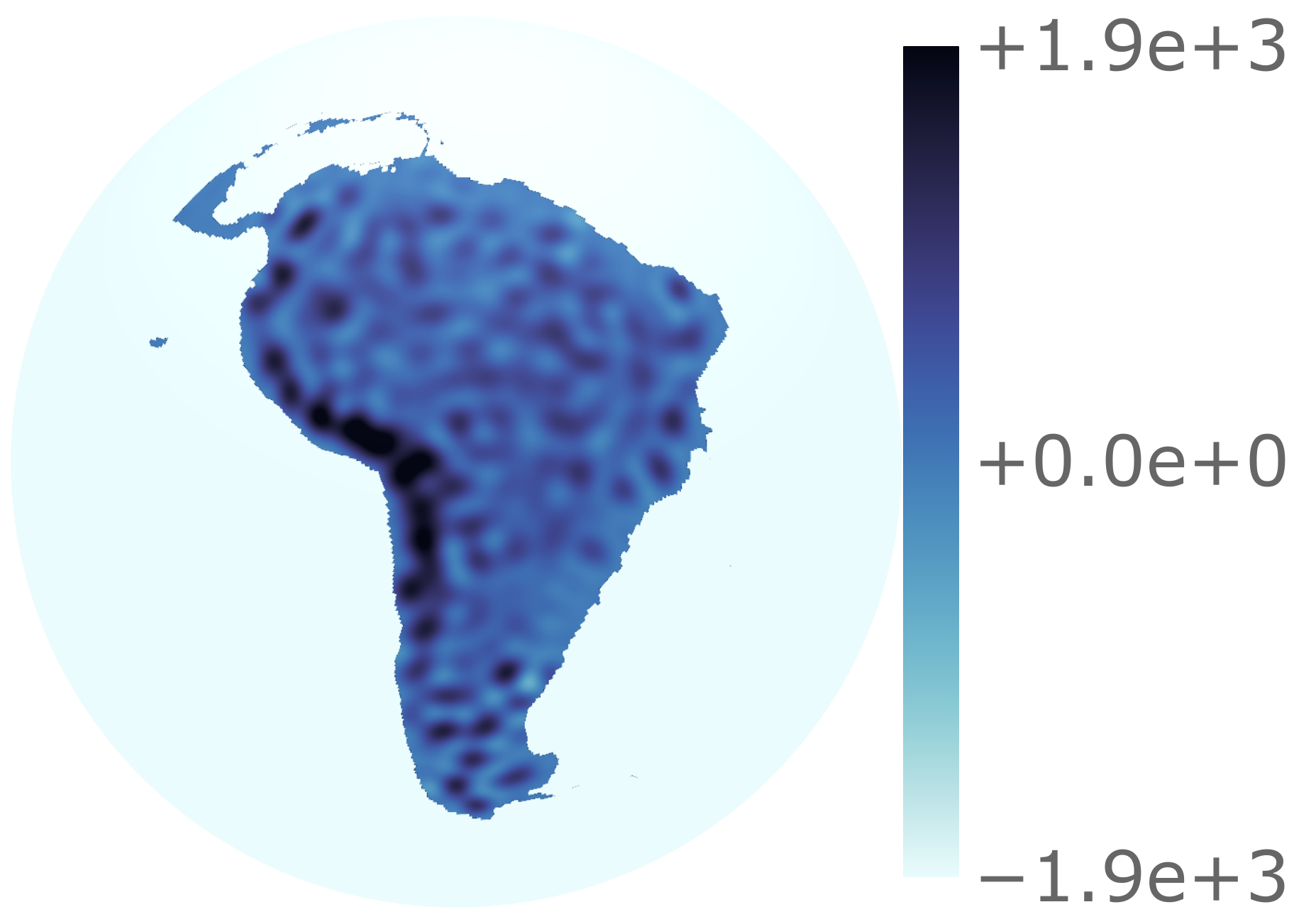}}
	\hfill
	\subfloat[Denoised \(N_{\sigma}=3\) \newline
		\(\snr{d} = \SI{4.60}{\dB}\)]
	{\includegraphics[trim={4 7 3 6},clip,width=.5\columnwidth]{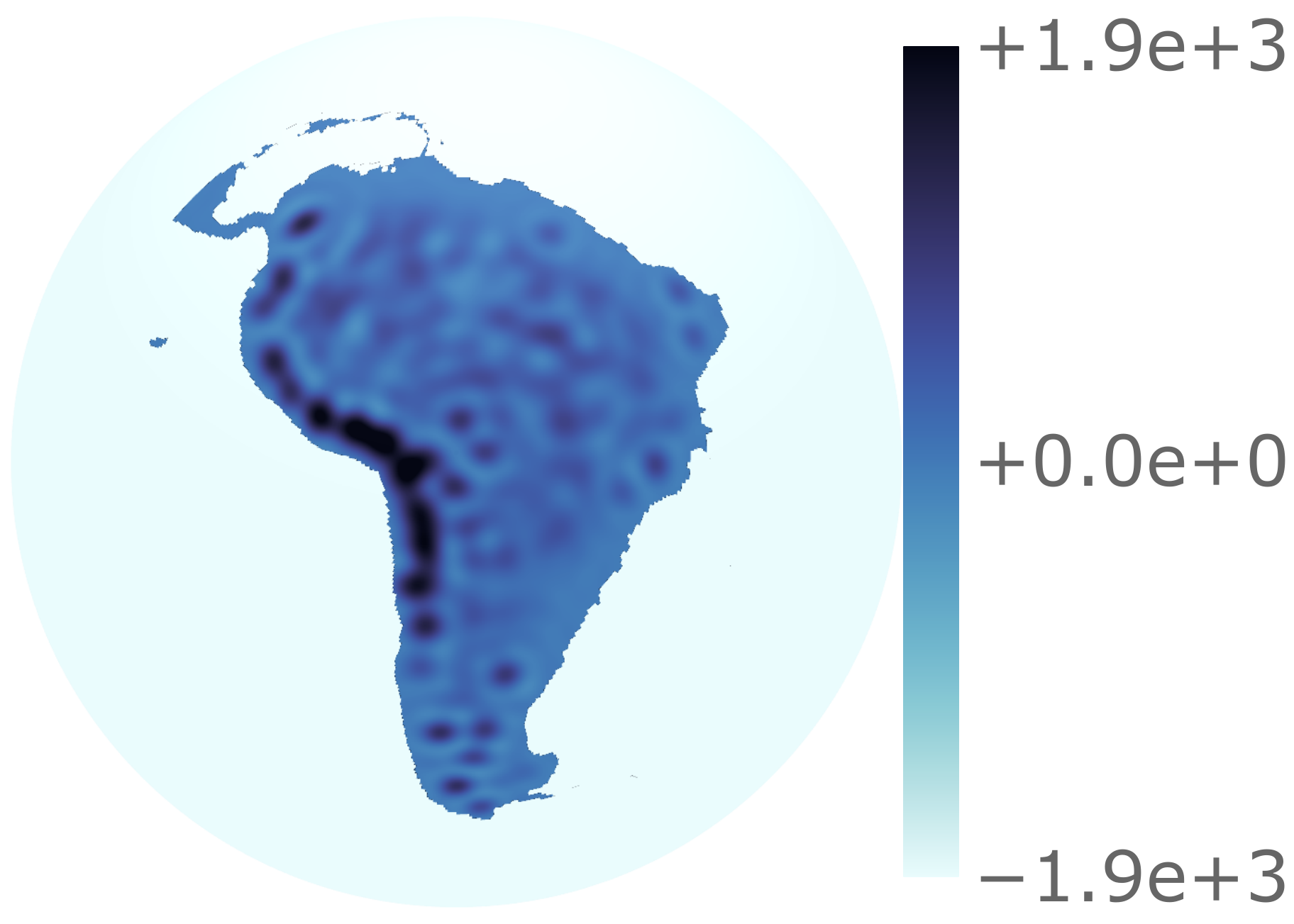}}
	\hfill
	\subfloat[Denoised \(N_{\sigma}=5\) \newline
		\(\snr{d} = \SI{1.27}{\dB}\)]
	{\includegraphics[trim={4 7 3 6},clip,width=.5\columnwidth]{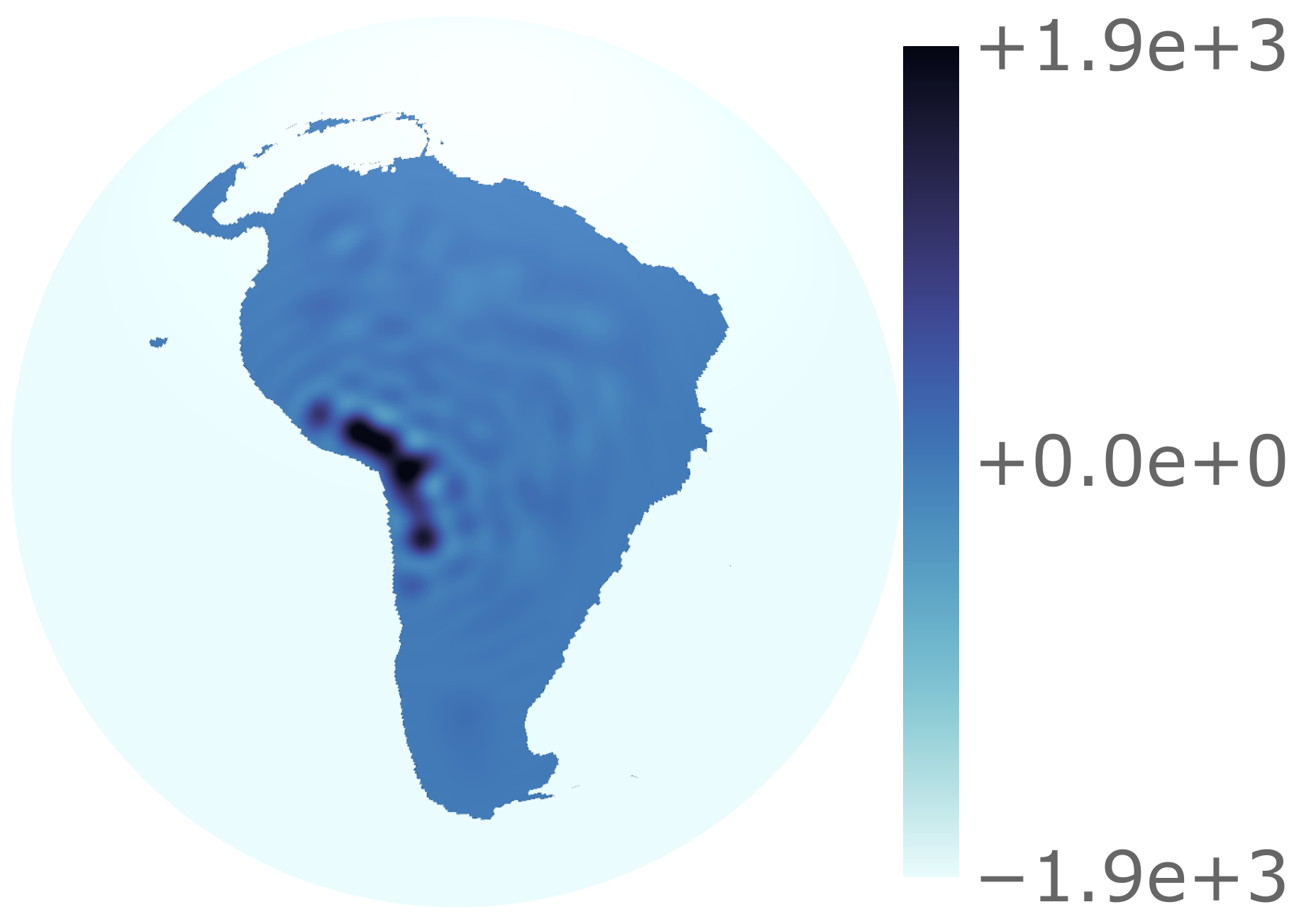}}
	\caption{
		Panel (a) shows the initial noisy signal of the South America region with a signal-to-noise ratio of \(\SI{4.11}{\dB}\).
		The scaling and wavelet coefficients of the noisy signal are calculated and are then hard-thresholded for a few \(N_{\sigma}\) values.
		The corresponding denoised plots for \(N_{\sigma} \in \set{2, 3, 5}\) are shown in panels (b--d). 
		At \(N_{\sigma}=2\) the signal-to-noise ratio is boosted by \(\SI{1.56}{\dB}\) to \(\SI{5.67}{\dB}\).
		As more signal is removed the signal-to-noise ratio decreases to \(\SI{4.60}{\dB}\) at \(N_{\sigma}=3\), which is still higher than the initial noisy signal.
		At \(N_{\sigma}=5\) the signal-to-noise ratio is \(\SI{1.27}{\dB}\), where only the Andes remains.
	}\label{fig:south_america_denoising}
\end{figure*}

The wavelet/scaling coefficients of \cref{eq:noised_signal} are given by the sum of the individual elements since the wavelet transform is linear
\begin{equation}
	\pixel{X^{\varphi}}
	= \pixel{S^{\varphi}} + \pixel{N^{\varphi}}.
\end{equation}
Here, capital letters denote the wavelet coefficients, \ie{} \(\pixel{X^{\varphi}} = \pixel{\convolution{\varphi}{x}}\).
The noise in wavelet space is as before
\begin{equation}
	\variance{\pixel{N^{\varphi}}}
	= \sigma^{2} \slepianSum \abs{\slepian{\varphi}}^{2} \abs{\pixel{\slepian{S}}}^{2}
	\equiv {\pixel{\sigma^{\varphi}}}^{2},
\end{equation}
where the quantity \(\sigma^{\varphi}\) defines the standard deviation of the noise in wavelet space.
One may perform denoising by hard-thresholding the wavelet/scaling coefficients with a threshold \(T\) proportional to the standard deviation of the noise \(\sigma^{\varphi}\).
The denoised wavelet coefficients \(\pixel{D^{\varphi}} = \pixel{\convolution{\varphi}{d}}\) become
\begin{equation}
	\pixel{D^{\varphi}} =
	\begin{cases}
		0,
		 & \pixel{X^{\varphi}} < \pixel{T^{\varphi}},    \\
		\pixel{X^{\varphi}},
		 & \pixel{X^{\varphi}} \geq \pixel{T^{\varphi}},
	\end{cases}
\end{equation}
where \(\pixel{T^{\varphi}} = N_{\sigma}\pixel{\sigma^{\varphi}}\), with \(N_{\sigma} \in \realPosParam{}\).
The denoised signal \(d\) may then be reconstructed from its wavelet/scaling coefficients.
The denoising formalism described above is straightforward, and more sophisticated denoising strategies can be developed; this procedure is merely provided to show a practical use case of Slepian wavelets.

A noisy signal is constructed by adding Gaussian white noise to the South America signal (set to zero outside the region) described in \cref{sec:south_america_region}.
The initial signal-to-noise ratio of the noised signal is \(\SI{4.11}{\dB}\) as shown in panel (a) of \cref{fig:south_america_denoising}.
The denoising procedure described above is performed with \(N_{\sigma} \in \set{2, 3, 5}\), shown in panels (b--d). 
This leads to a signal-to-noise ratio of \(\SI{5.67}{\dB}\), \(\SI{4.60}{\dB}\) and \(\SI{1.27}{\dB}\) respectively.
Thus, this hard-thresholding scheme increases the power of the signal unless a significant amount of signal is removed at which point the signal-to-noise begins to decrease.

\section{Conclusions}\label{sec:conclusion}

This work presents the construction of Slepian scale-discretised wavelets on the sphere, which are wavelets restricted to a region of the sphere.
These wavelets can be used in many fields of science and engineering where data are measured on the sphere but are missing on a particular region.
A common approach to analysing data of this form is by using spherical wavelets, which allow one to probe spatially localised, scale-dependent features of the signal.
However, wavelet methods on the whole sphere suffer problems when data are only defined on a particular region, as the wavelet coefficients are contaminated near the boundaries of the region.
These distorted coefficients may then be detected and removed for accurate analysis.
However, by removing the wavelet coefficients near the boundary, the power of the data is not fully utilised.
Slepian wavelets offer a solution to this problem.

Slepian wavelets are constructed on the eigenfunctions of the Slepian spatial-spectral concentration problem on the sphere, which are the basis functions of the region.
Through a tiling of the Slepian line, one may define Slepian scale-discretised wavelets that are localised within the region of interest.
The current work generalises the sifting convolution~\cite{Roddy2021} on the sphere beyond the spherical harmonic setting to any basis, where here the Slepian functions define the basis.
The Slepian wavelet transform then follows by performing convolutions over the incomplete sphere with the Slepian wavelets.
The original function can then be reconstructed from its Slepian wavelet coefficients.

An example South America region is constructed from a topographic map of the Earth, and the Slepian functions and corresponding eigenvalues of this region are presented.
Through a wavelet transform, the wavelets and wavelet coefficients of this region are found.
The South America signal is corrupted with Gaussian white noise in the region, and a straightforward hard-thresholding denoising formalism is described.
The denoised signal plots are shown for different values of \(N_{\sigma}\), and an improvement to the signal-to-noise ratio is observed.
Through a direct connection to the harmonic space, one can trace a variance directly to the Slepian domain.
Slepian wavelets can be used in many standard wavelet applications, such as sparse regularisation approaches to solve inverse problems~\cite{McEwen2013a,Wallis2017,Price2021}.
This work defines the Slepian functions in the spherical domain; however, the eigenfunctions of the Slepian concentration problem can be found for other manifolds.
The sifting convolution can be defined for a given manifold through a translation of the eigenfunctions and, as such, Slepian wavelets can be generalised to arbitrary manifolds, which is the focus of future work.

\appendix[Africa Example]\label{sec:appendix}

\begin{figure}
	\centering
	\subfloat[EGM2008]
	{\includegraphics[trim={4 7 3 6},clip,width=.5\columnwidth]{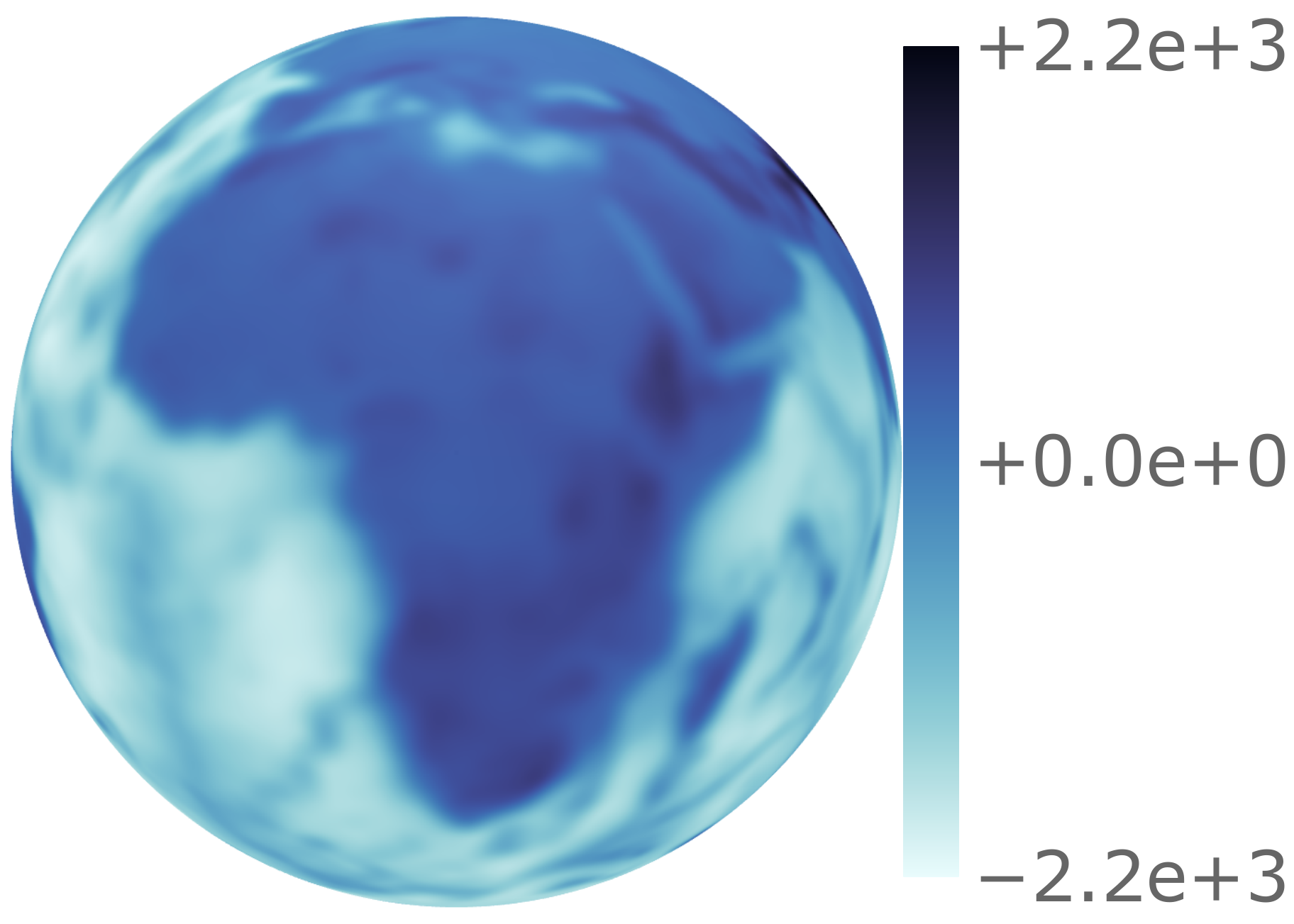}}
	\hfill
	\subfloat[\(R'\)]
	{\includegraphics[trim={4 7 3 6},clip,width=.5\columnwidth]{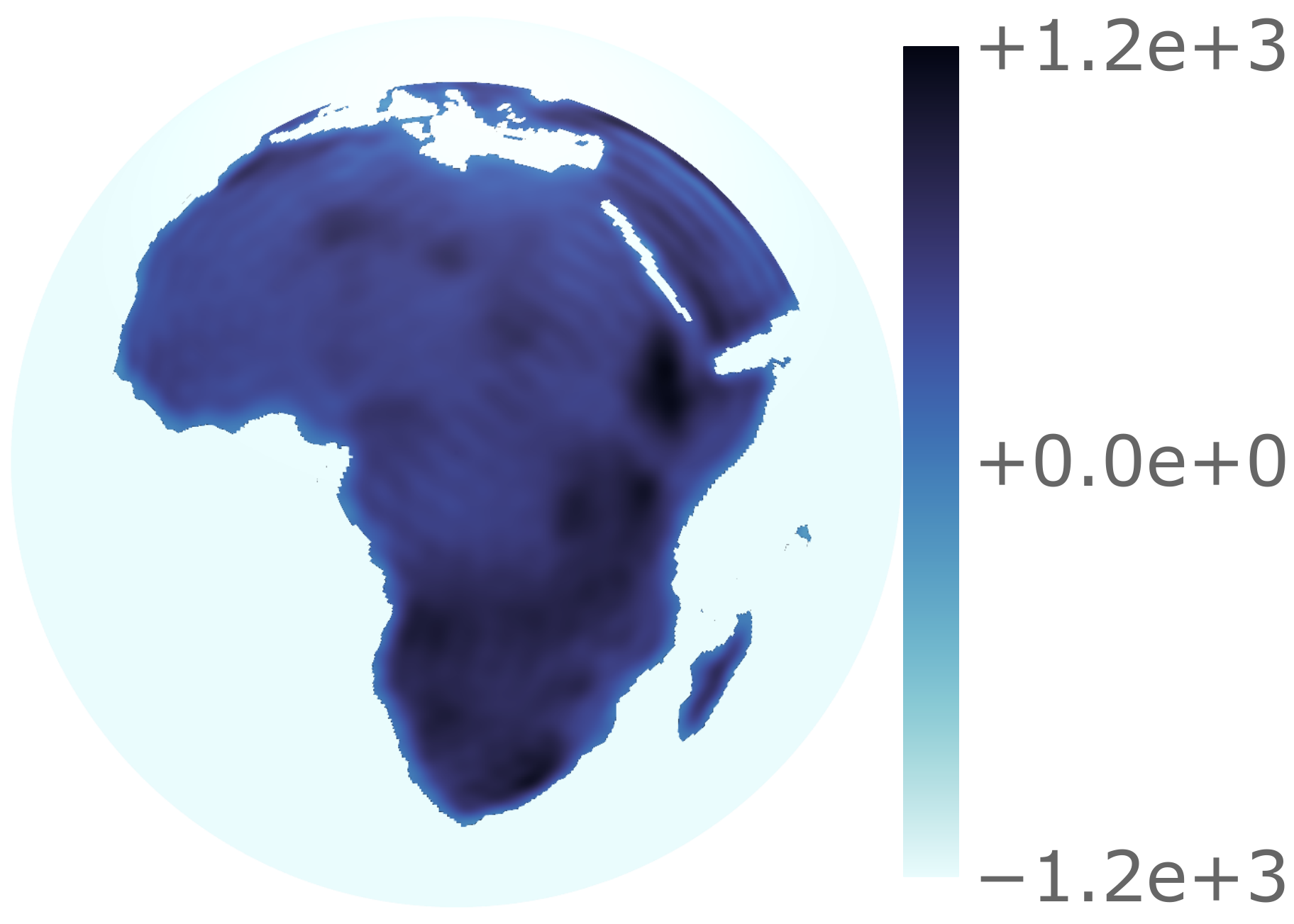}}
	\caption{
		Panel (a) corresponds to the EGM2008 dataset centred on a view of Africa.
		As before, the dataset is bandlimited at \(\lmax=128\), and smoothed with \(\fwhm{1.17}\).
		Panel (b) presents the region \(R'\), the shape of which is constructed from the Slepian coefficients of the Africa mask.
		The field value outside the region in panel (b) is set to negative infinity for illustrative purposes.
		The amplitude of the right panel is set by the height of Mount Kilimanjaro, rather than the lowest depths of the sea.
	}\label{fig:africa_region}
\end{figure}

\begin{figure}
	\centering
	\includegraphics[width=\columnwidth]{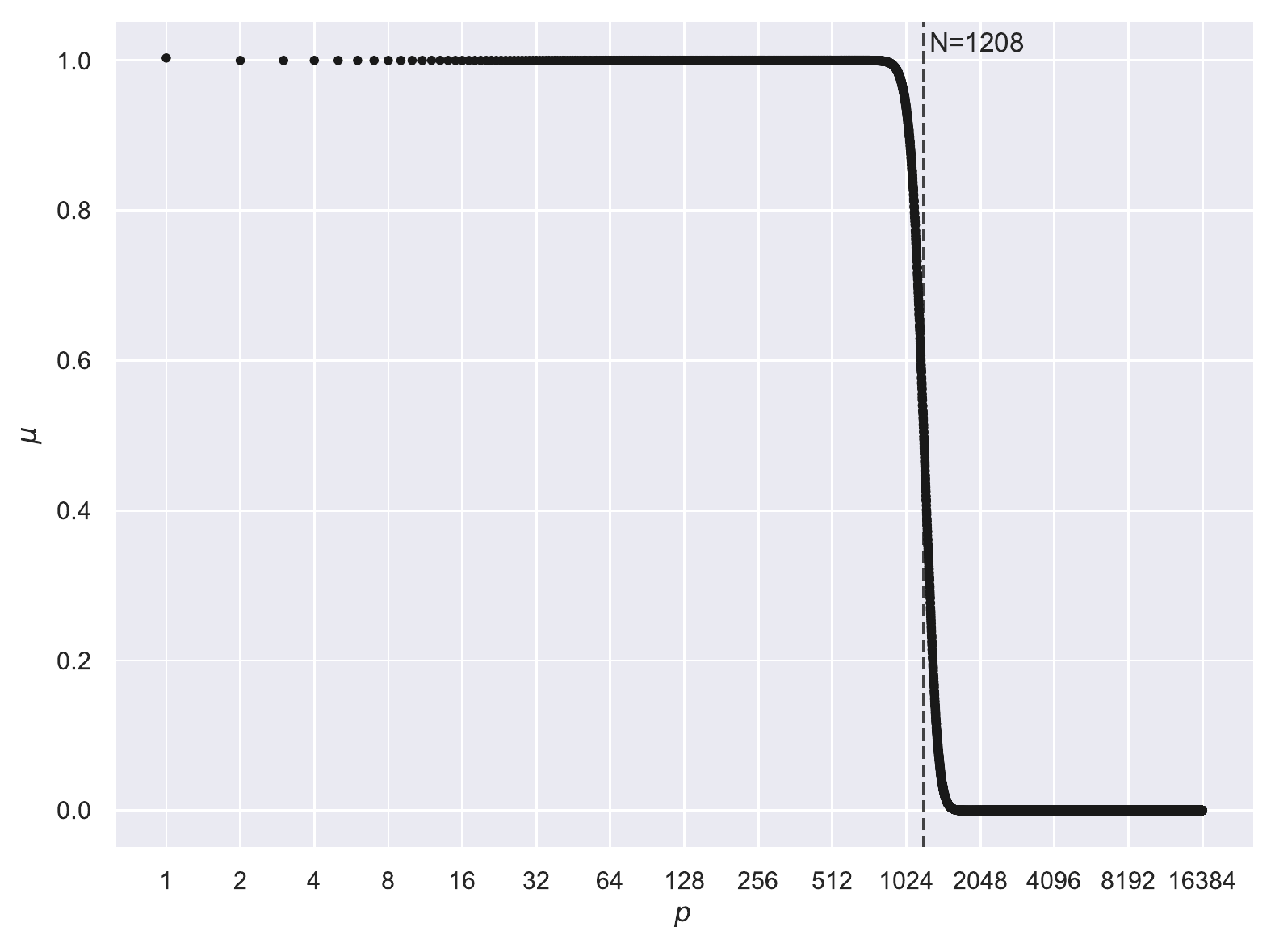}
	\caption{
		The eigenvalues of the Africa region concentrated within the Shannon number \(N=1208\).
		The majority of the eigenvalues are \(\almost{1}\) before decreasing rapidly towards zero around the Shannon number.
	}\label{fig:africa_eigenvalues}
\end{figure}

\begin{figure}
	\centering
	\subfloat[\(\Re\big\{\pixel{S_{1}}\big\},\ \mu_{1}=1.00\)] 
	{\includegraphics[trim={4 7 3 6},clip,width=.5\columnwidth]{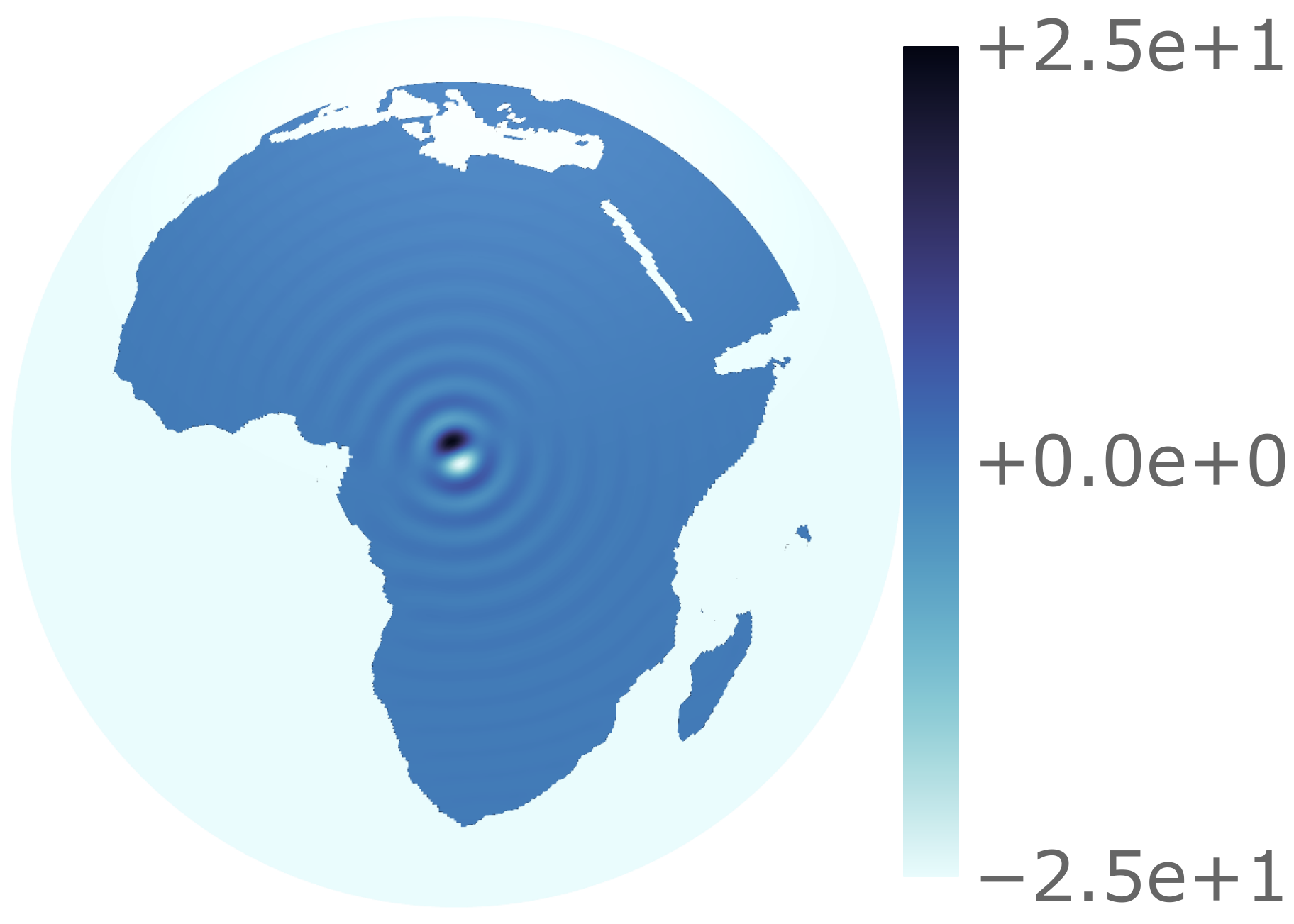}} 
	\hfill
	\subfloat[\(\Re\big\{\pixel{S_{10}}\big\},\ \mu_{10}=1.00\)] 
	{\includegraphics[trim={4 7 3 6},clip,width=.5\columnwidth]{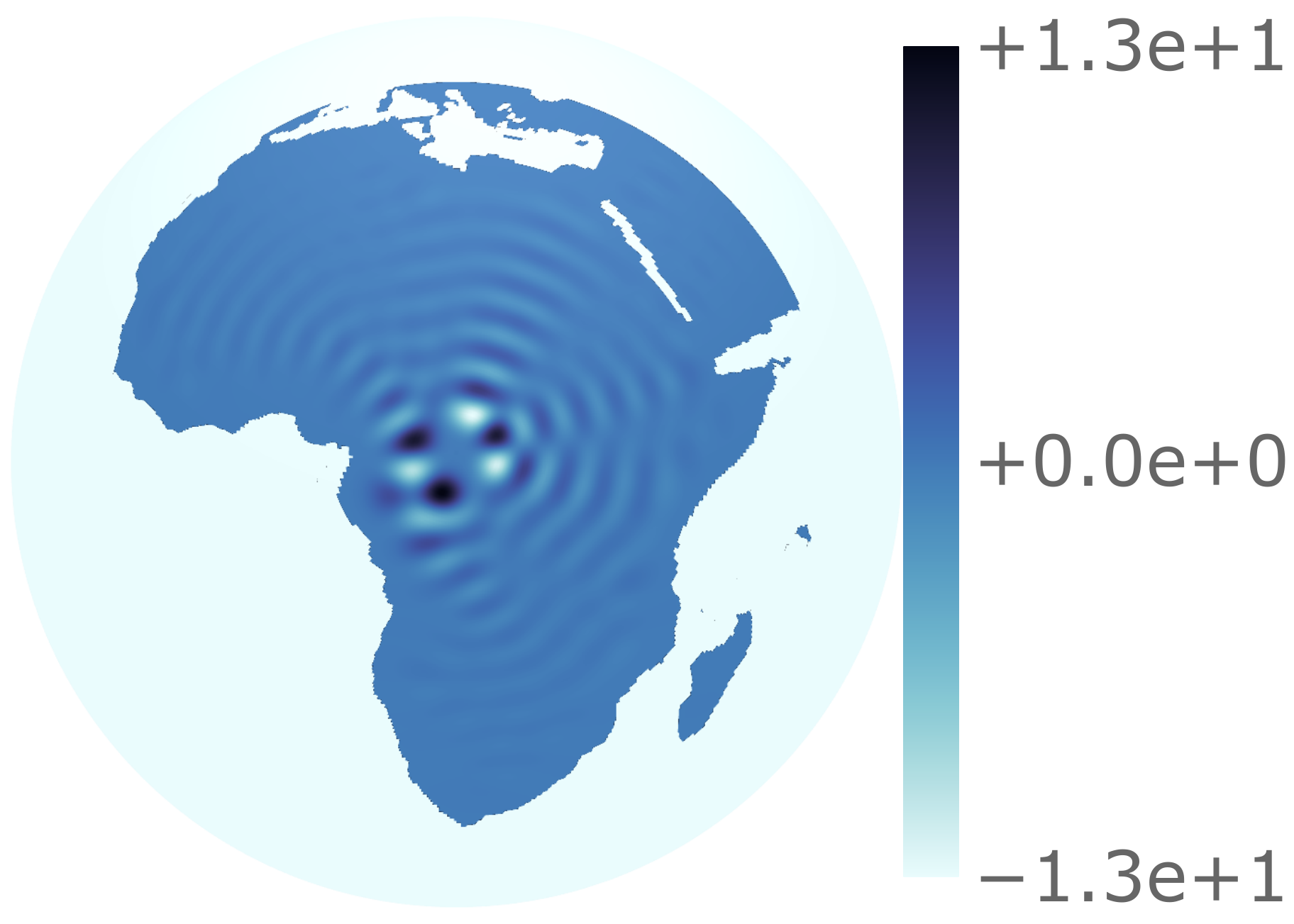}} 
	\newline
	\subfloat[\(\Re\big\{\pixel{S_{25}}\big\},\ \mu_{25}=1.00\)] 
	{\includegraphics[trim={4 7 3 6},clip,width=.5\columnwidth]{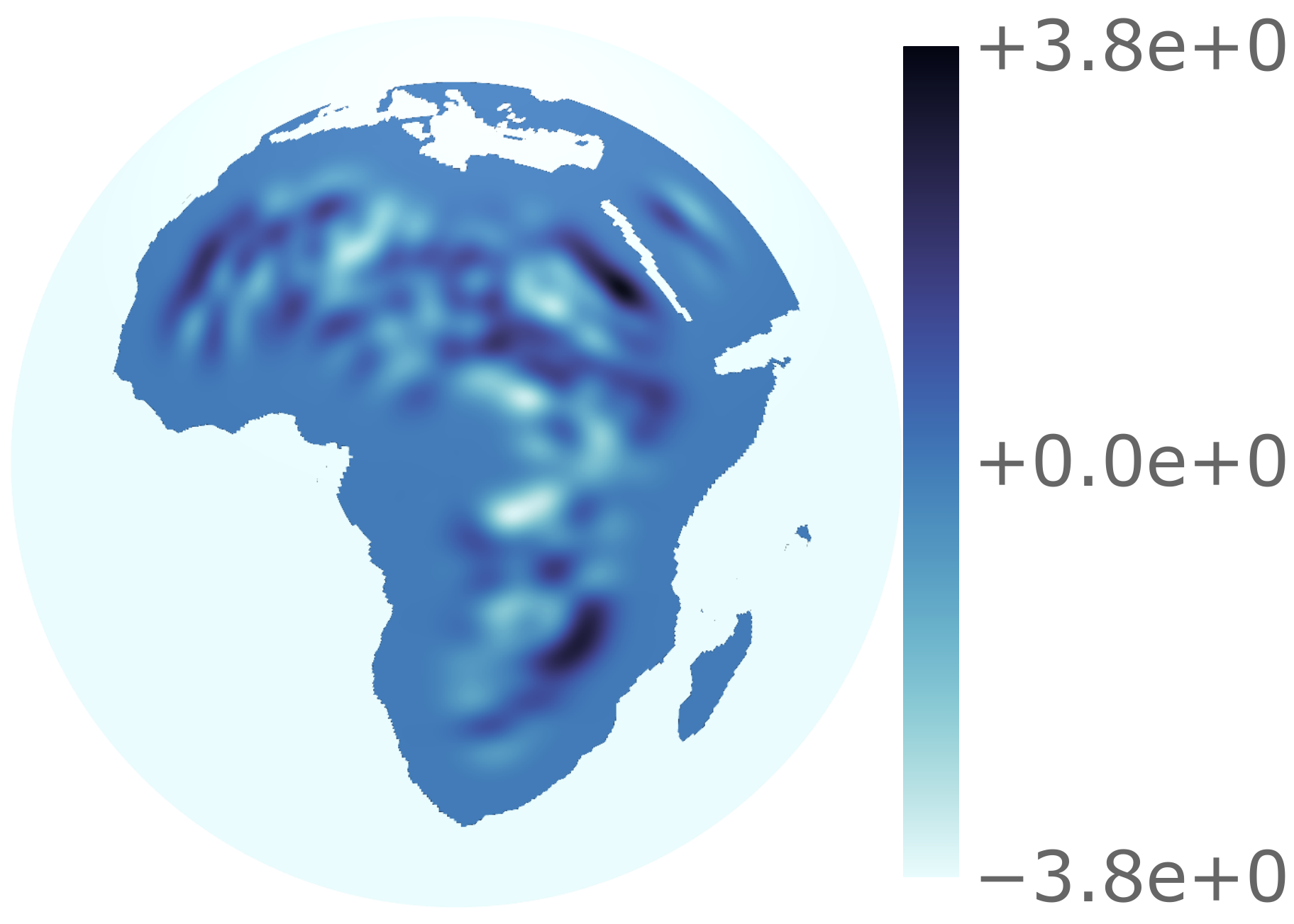}} 
	\hfill
	\subfloat[\(\Re\big\{\pixel{S_{50}}\big\},\ \mu_{50}=1.00\)] 
	{\includegraphics[trim={4 7 3 6},clip,width=.5\columnwidth]{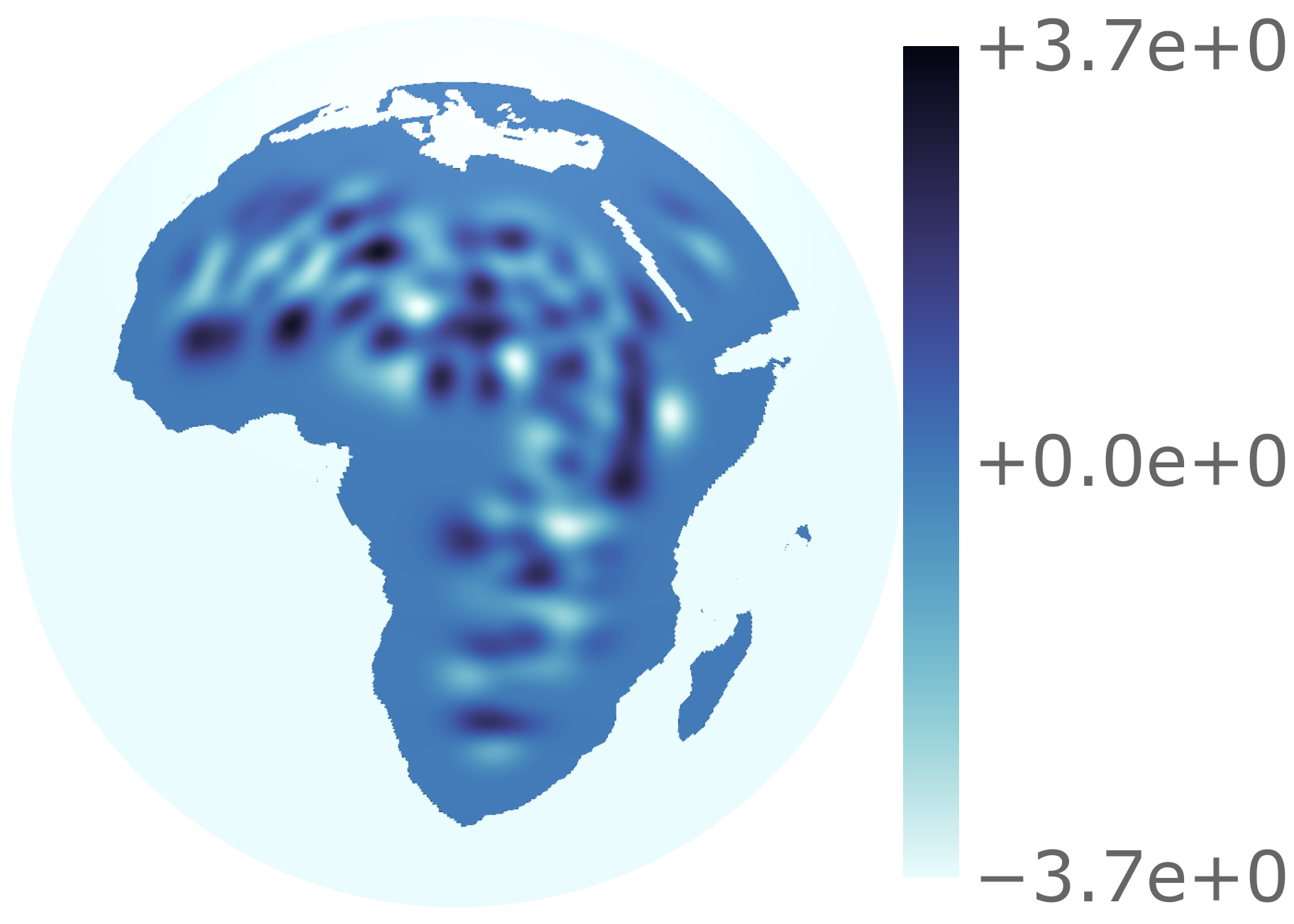}} 
	\newline
	\subfloat[\(\Re\big\{\pixel{S_{100}}\big\},\ \mu_{100}=1.00\)] 
	{\includegraphics[trim={4 7 3 6},clip,width=.5\columnwidth]{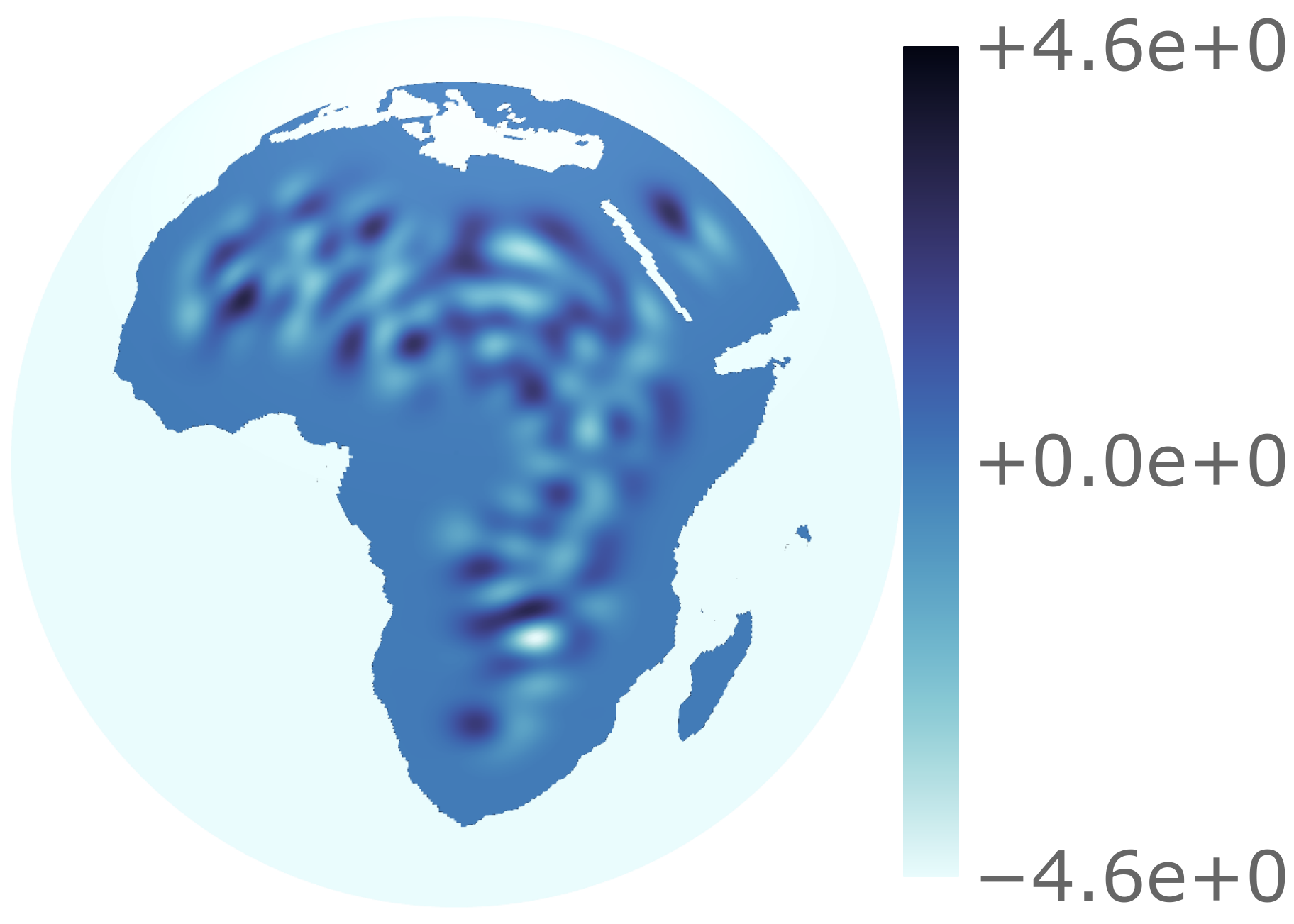}} 
	\hfill
	\subfloat[\(\Re\big\{\pixel{S_{200}}\big\},\ \mu_{200}=1.00\)] 
	{\includegraphics[trim={4 7 3 6},clip,width=.5\columnwidth]{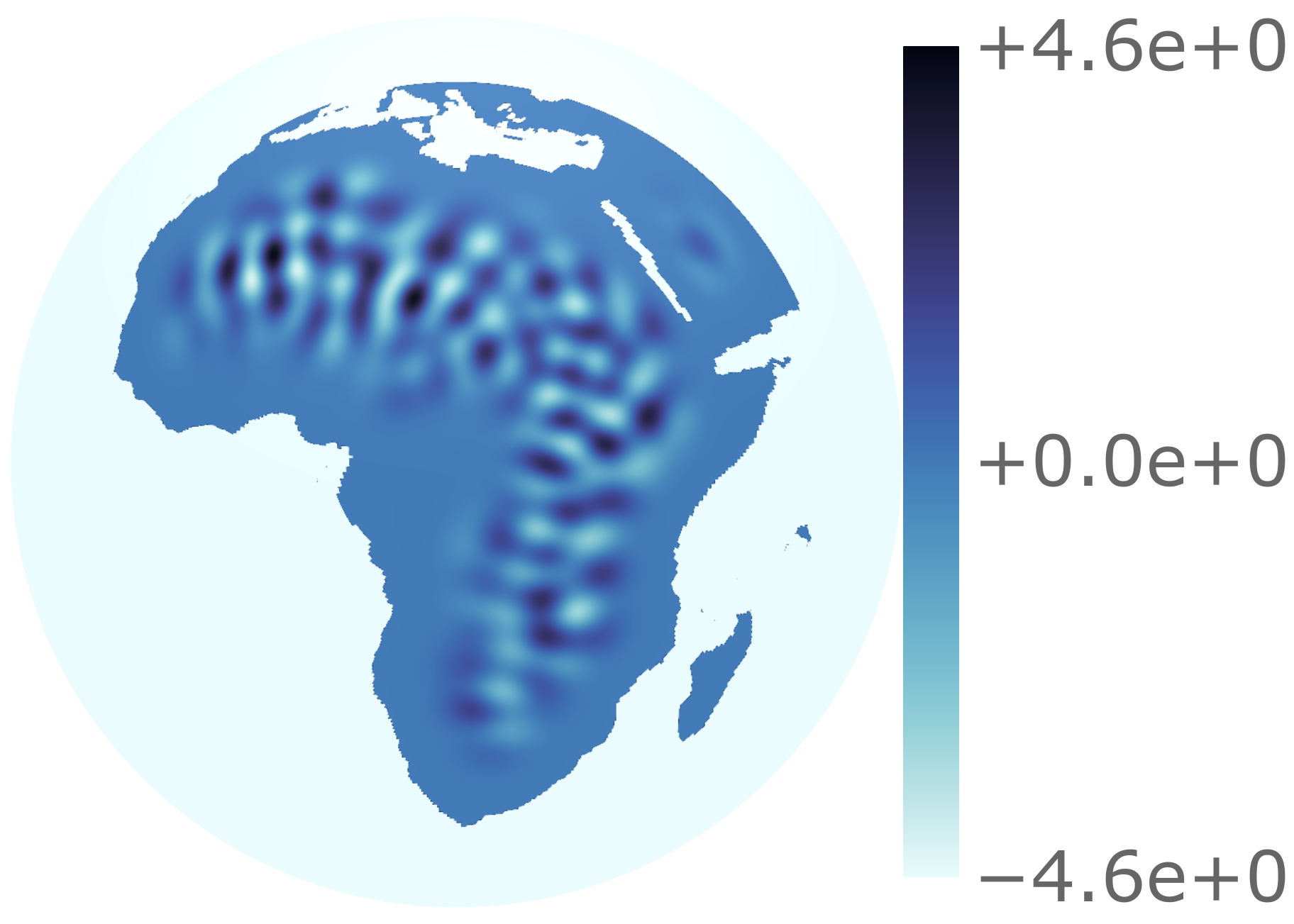}} 
	\caption{
		The Slepian functions of the Africa region \(S_p(\omega)\) for \(p \in \set{1, 10, 25, 50, 100, 200}\) shown left-to-right, top-to-bottom.
		The corresponding eigenvalue \(\slepian{\mu}\) is a measure of the concentration within the given region \(R'\), which remain \(\almost{1}\) for many \(p\) values before decreasing towards zero.
	}\label{fig:africa_eigenfunctions}
\end{figure}

\begin{figure}
	\centering
	\subfloat[\(\Re\big\{\pixel{\Phi}\big\}\)] 
	{\includegraphics[trim={4 7 3 6},clip,width=.5\columnwidth]{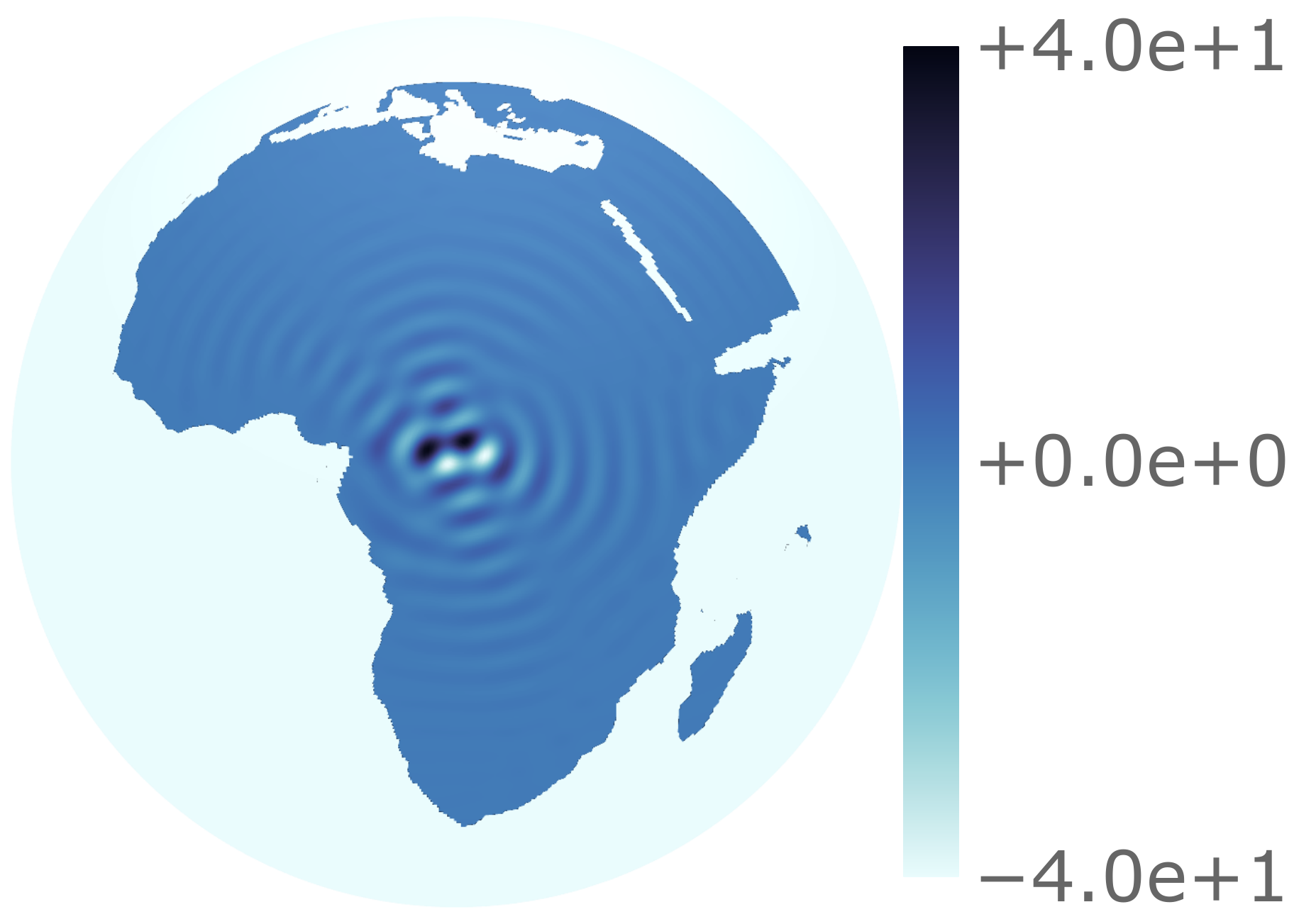}}
	\hfill
	\subfloat[\(\Re\big\{\pixel{\Psi^{2j}}\big\}\)] 
	{\includegraphics[trim={4 7 3 6},clip,width=.5\columnwidth]{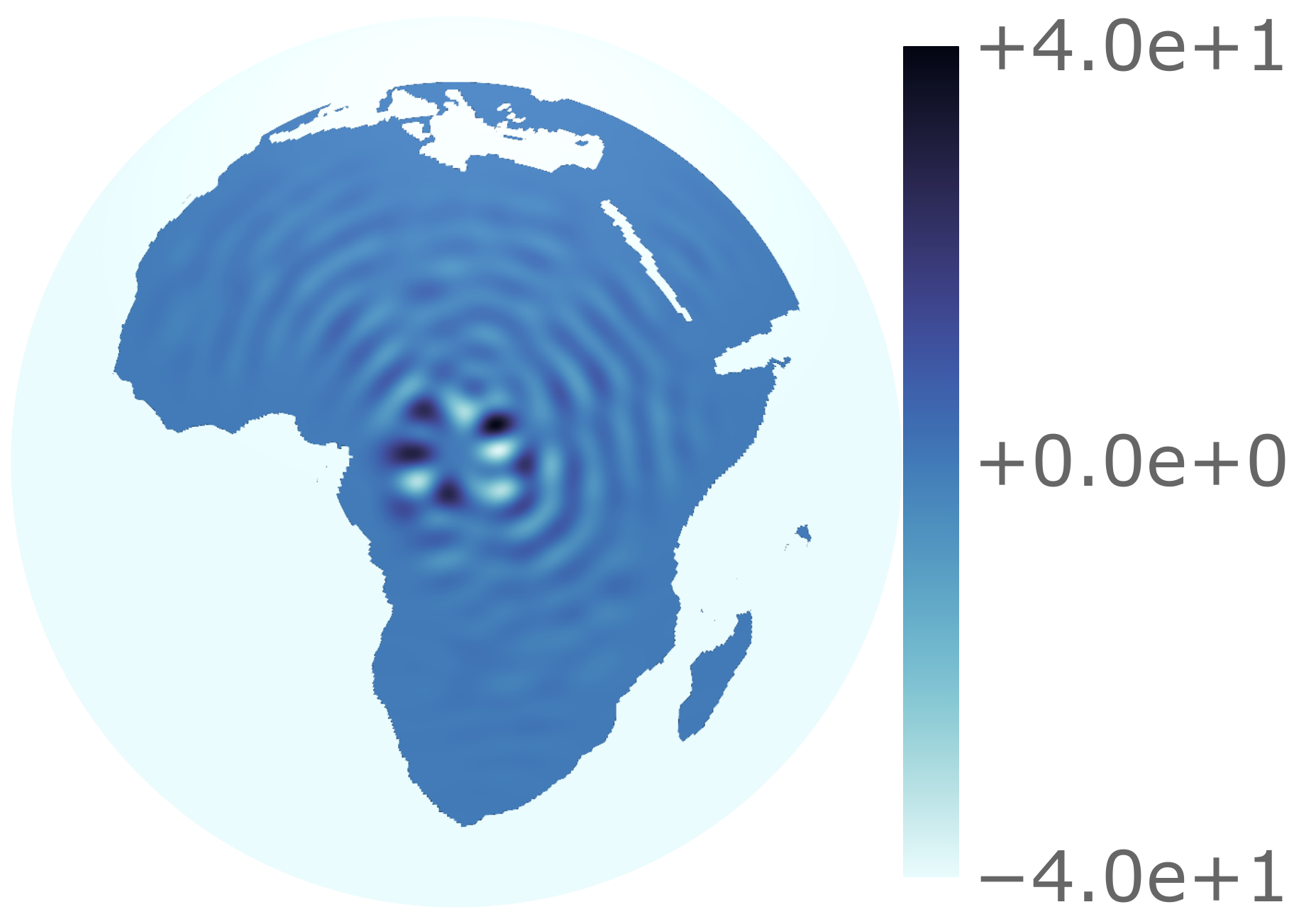}}
	\newline
	\subfloat[\(\Re\big\{\pixel{\Psi^{3j}}\big\}\)] 
	{\includegraphics[trim={4 7 3 6},clip,width=.5\columnwidth]{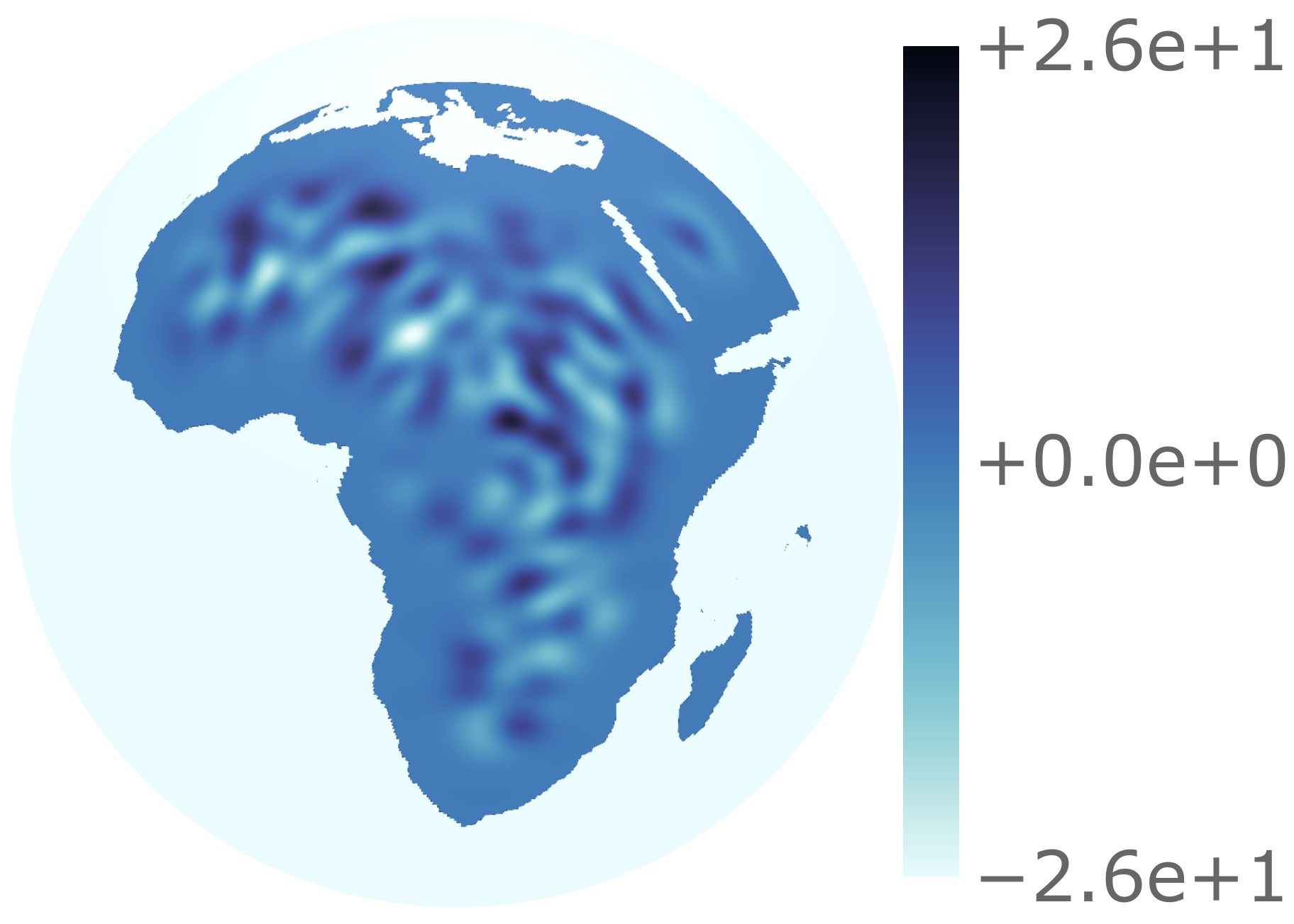}}
	\hfill
	\subfloat[\(\Re\big\{\pixel{\Psi^{4j}}\big\}\)] 
	{\includegraphics[trim={4 7 3 6},clip,width=.5\columnwidth]{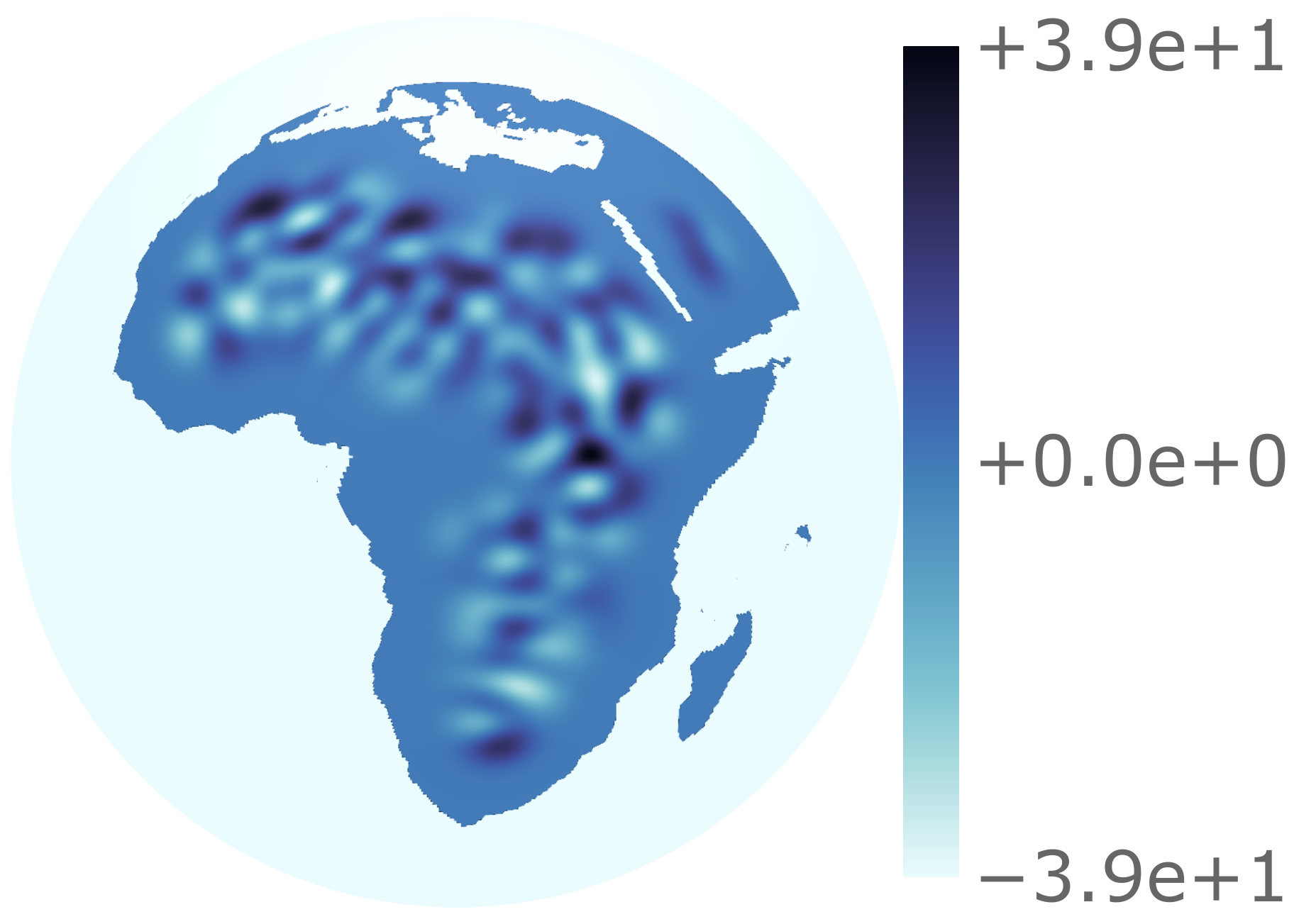}}
	\newline
	\subfloat[\(\Re\big\{\pixel{\Psi^{5j}}\big\}\)] 
	{\includegraphics[trim={4 7 3 6},clip,width=.5\columnwidth]{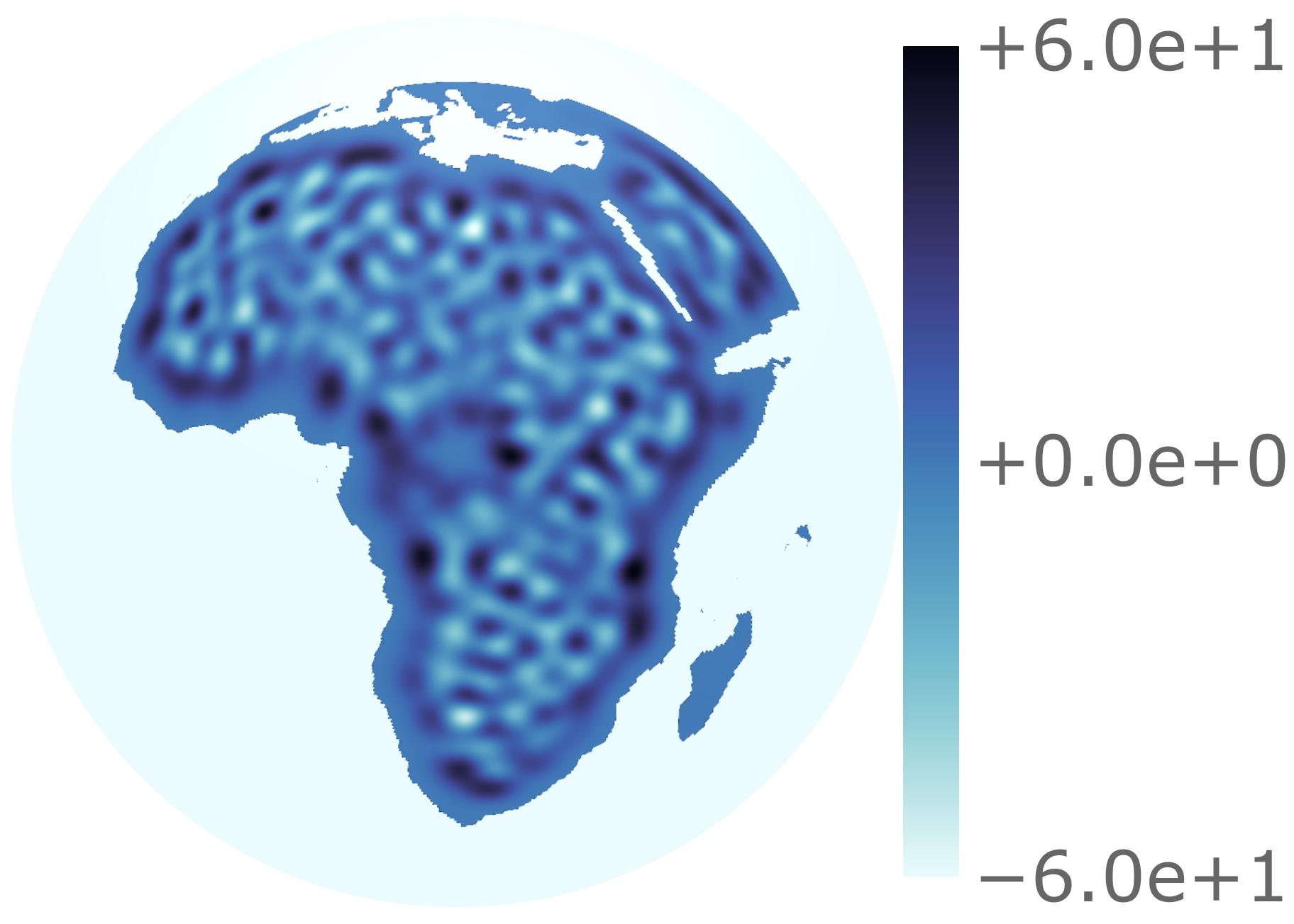}}
	\hfill
	\subfloat[\(\Re\big\{\pixel{\Psi^{6j}}\big\}\)] 
	{\includegraphics[trim={4 7 3 6},clip,width=.5\columnwidth]{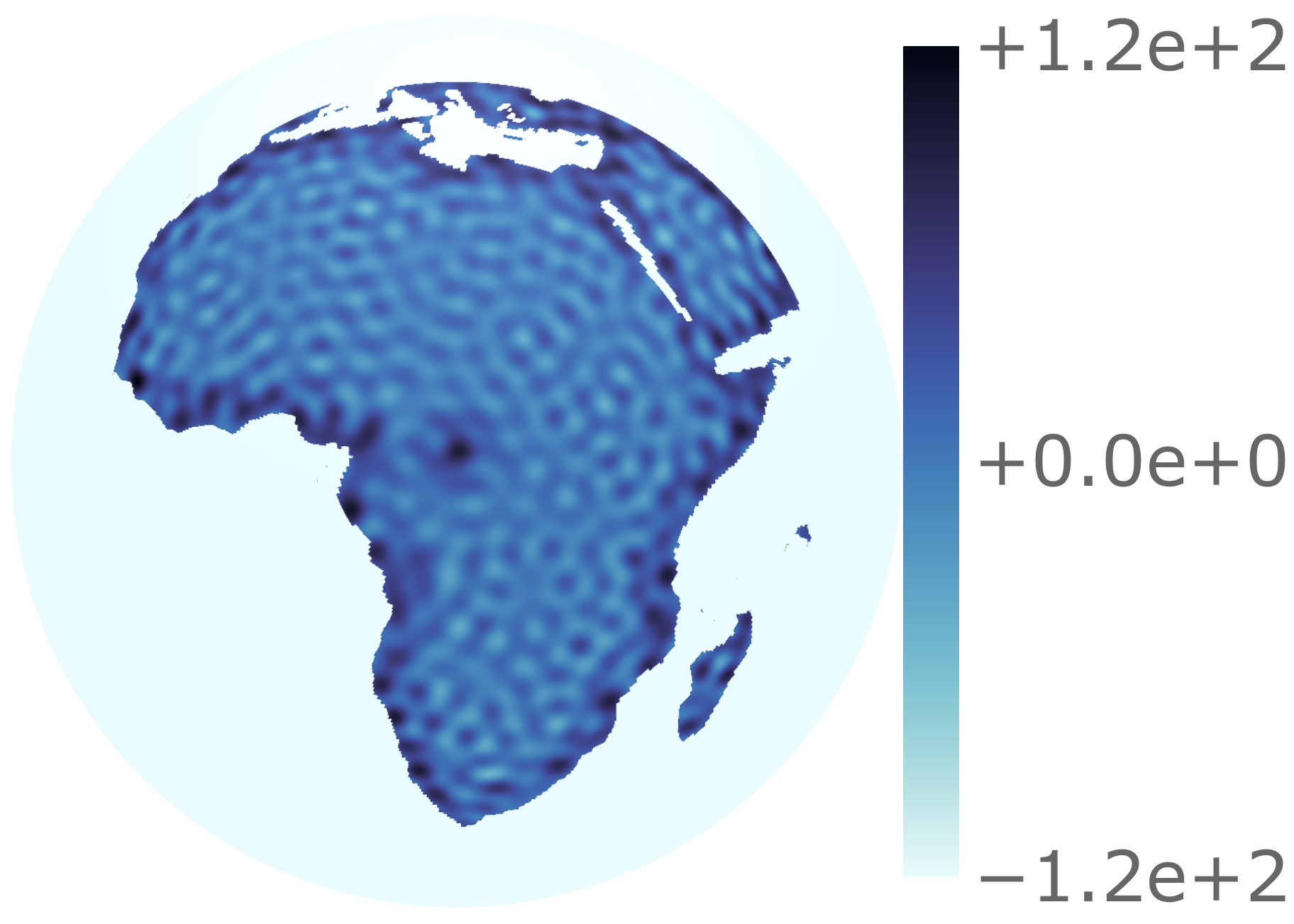}}
	\newline
	\subfloat[\(\Re\big\{\pixel{\Psi^{7j}}\big\}\)] 
	{\includegraphics[trim={4 7 3 6},clip,width=.5\columnwidth]{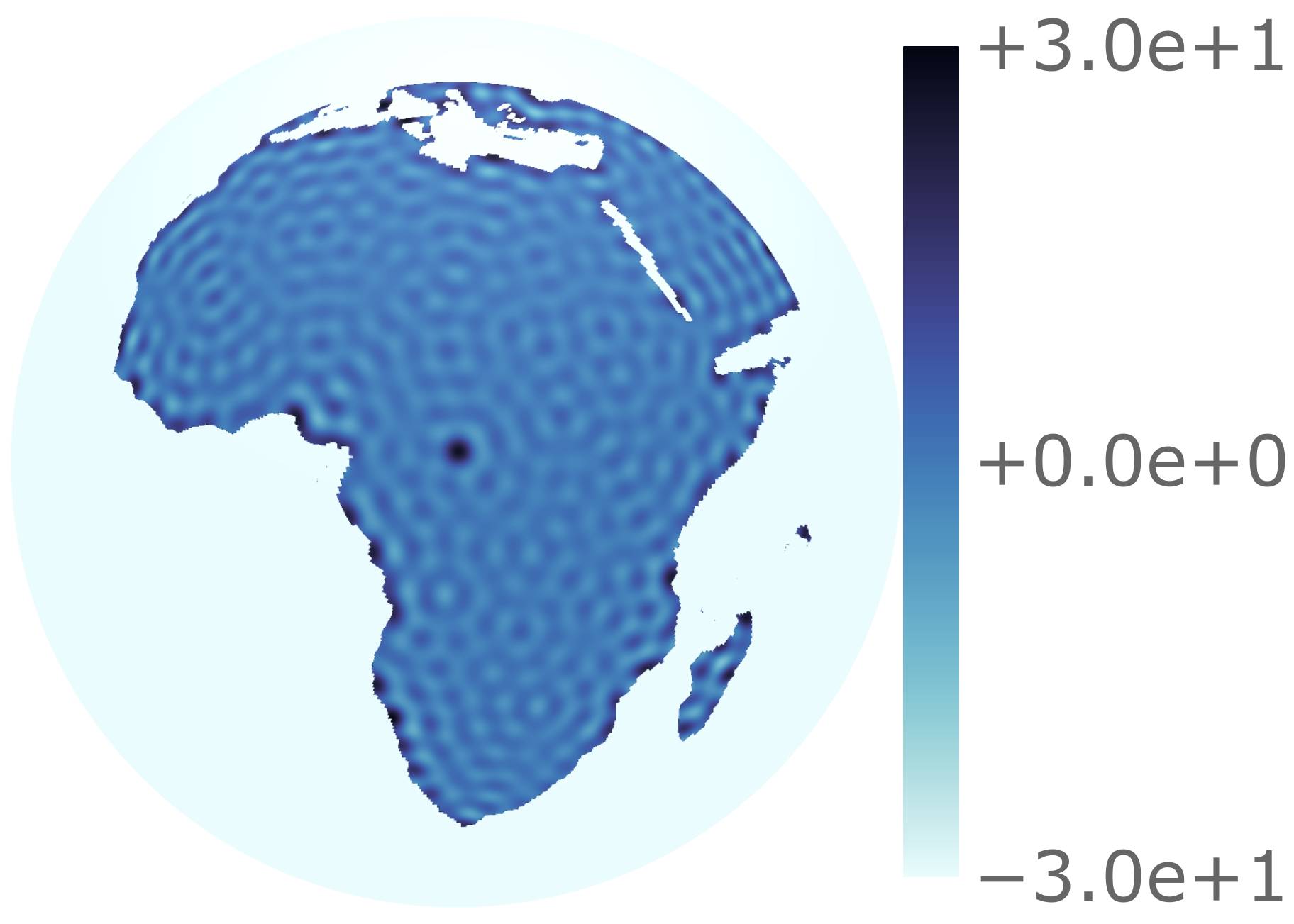}}
	\caption{
		The scaling function and the wavelets for scales \(j \in \set{2, 3, 4, 5, 6, 7}\) for the Africa region shown left-to-right, top-to-bottom.
		The wavelets are constructed through a tiling of the Slepian line using scale-discretised functions, with parameters \(\lambda=3\), \(J_{0}=2\), and bandlimit \(\lmax=128\).
	}\label{fig:africa_slepian_wavelets}
\end{figure}

\begin{figure}
	\centering
	\subfloat[\(\Re\big\{\pixel{W^{\Phi}}\big\}\)] 
	{\includegraphics[trim={4 7 3 6},clip,width=.5\columnwidth]{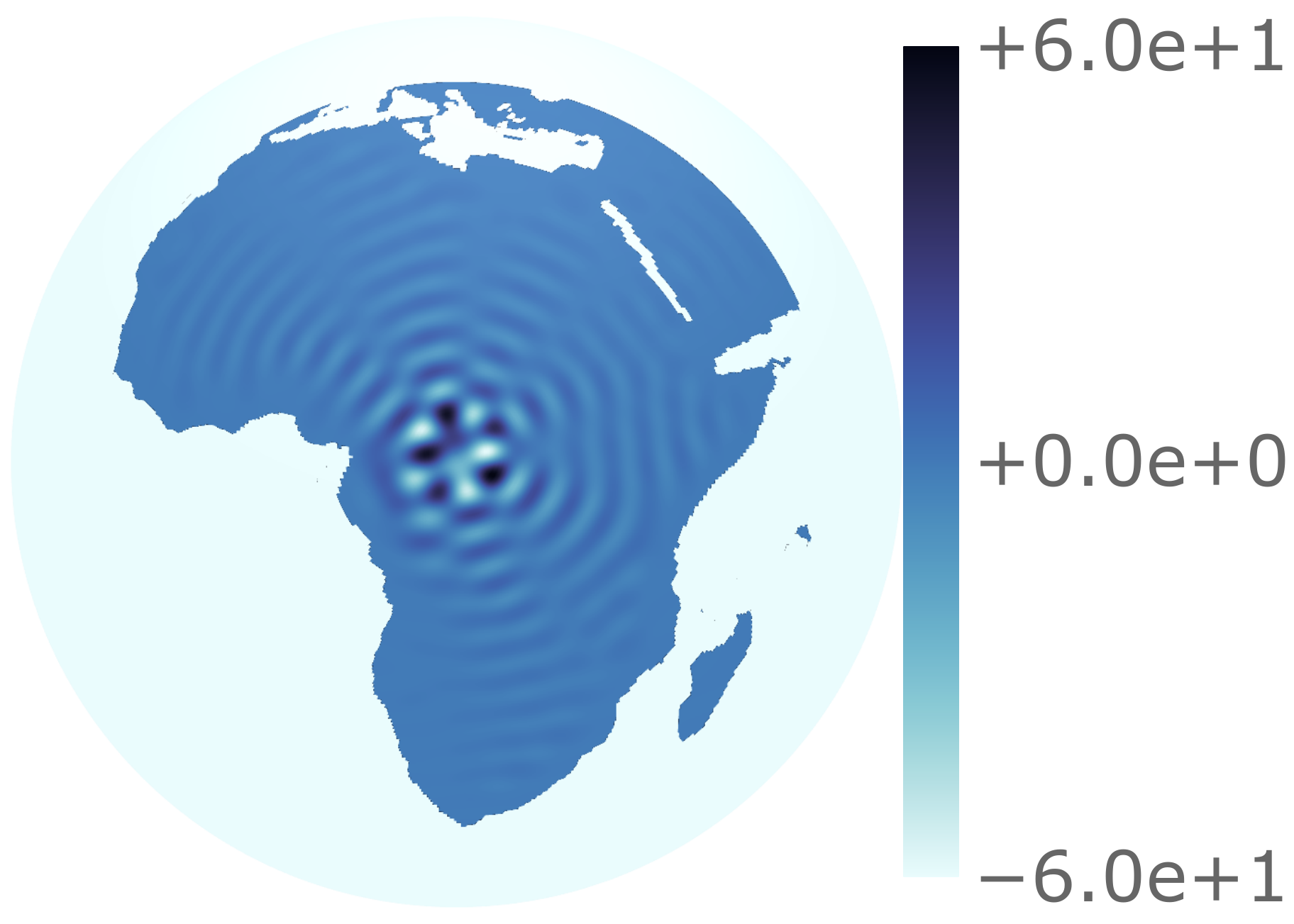}}
	\hfill
	\subfloat[\(\Re\big\{\pixel{W^{\Psi^{2j}}}\big\}\)] 
	{\includegraphics[trim={4 7 3 6},clip,width=.5\columnwidth]{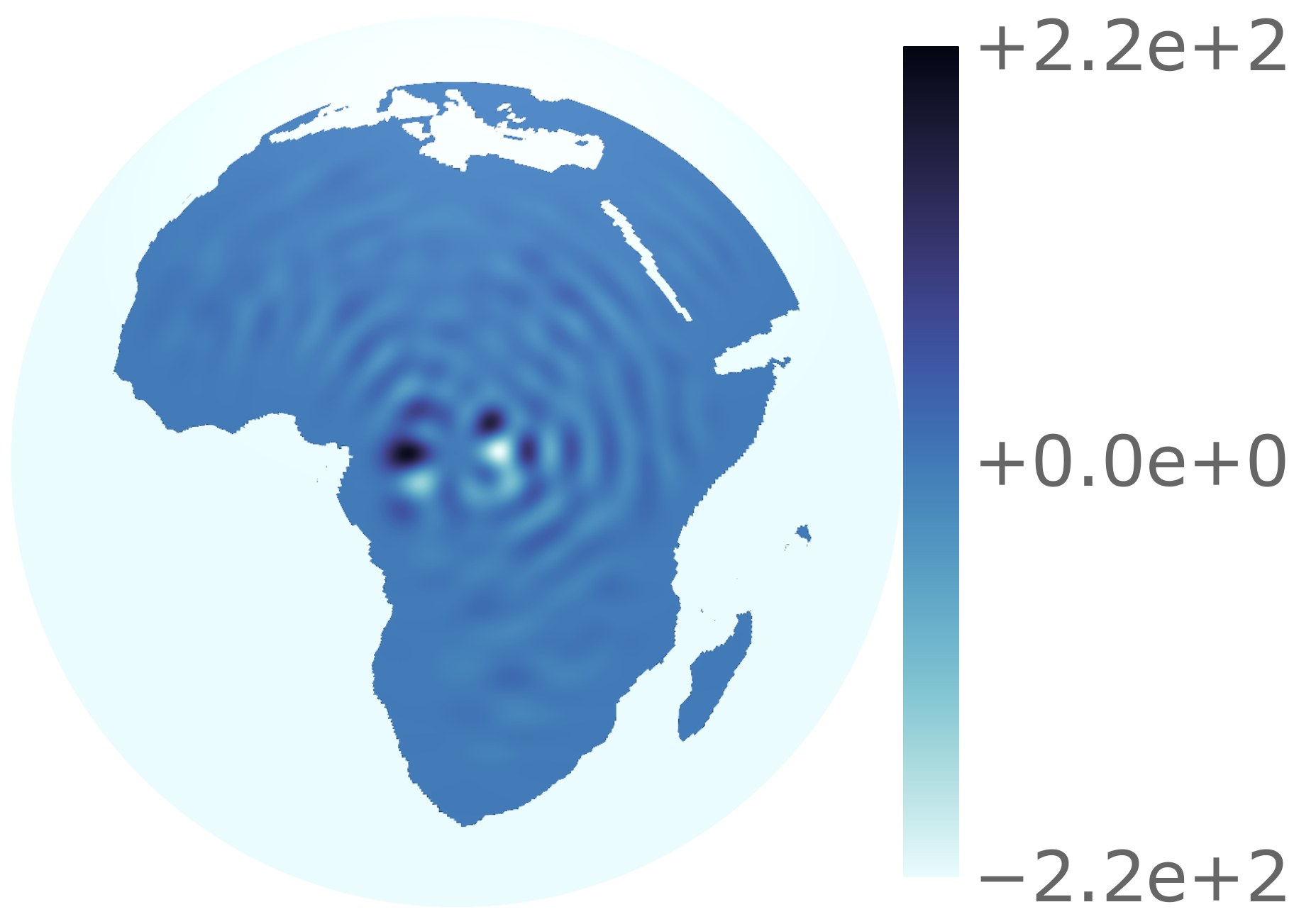}}
	\newline
	\subfloat[\(\Re\big\{\pixel{W^{\Psi^{3j}}}\big\}\)] 
	{\includegraphics[trim={4 7 3 6},clip,width=.5\columnwidth]{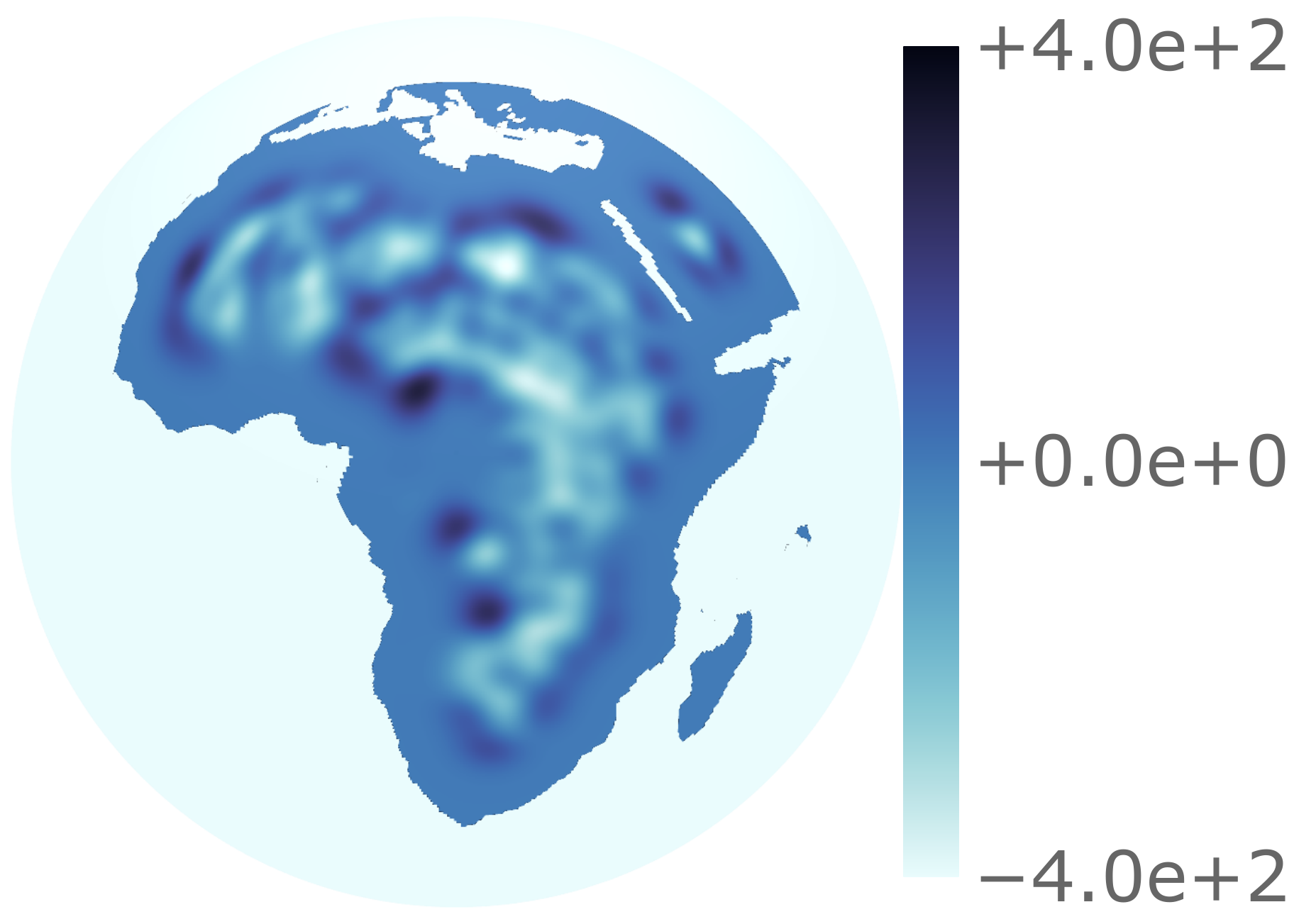}}
	\hfill
	\subfloat[\(\Re\big\{\pixel{W^{\Psi^{4j}}}\big\}\)] 
	{\includegraphics[trim={4 7 3 6},clip,width=.5\columnwidth]{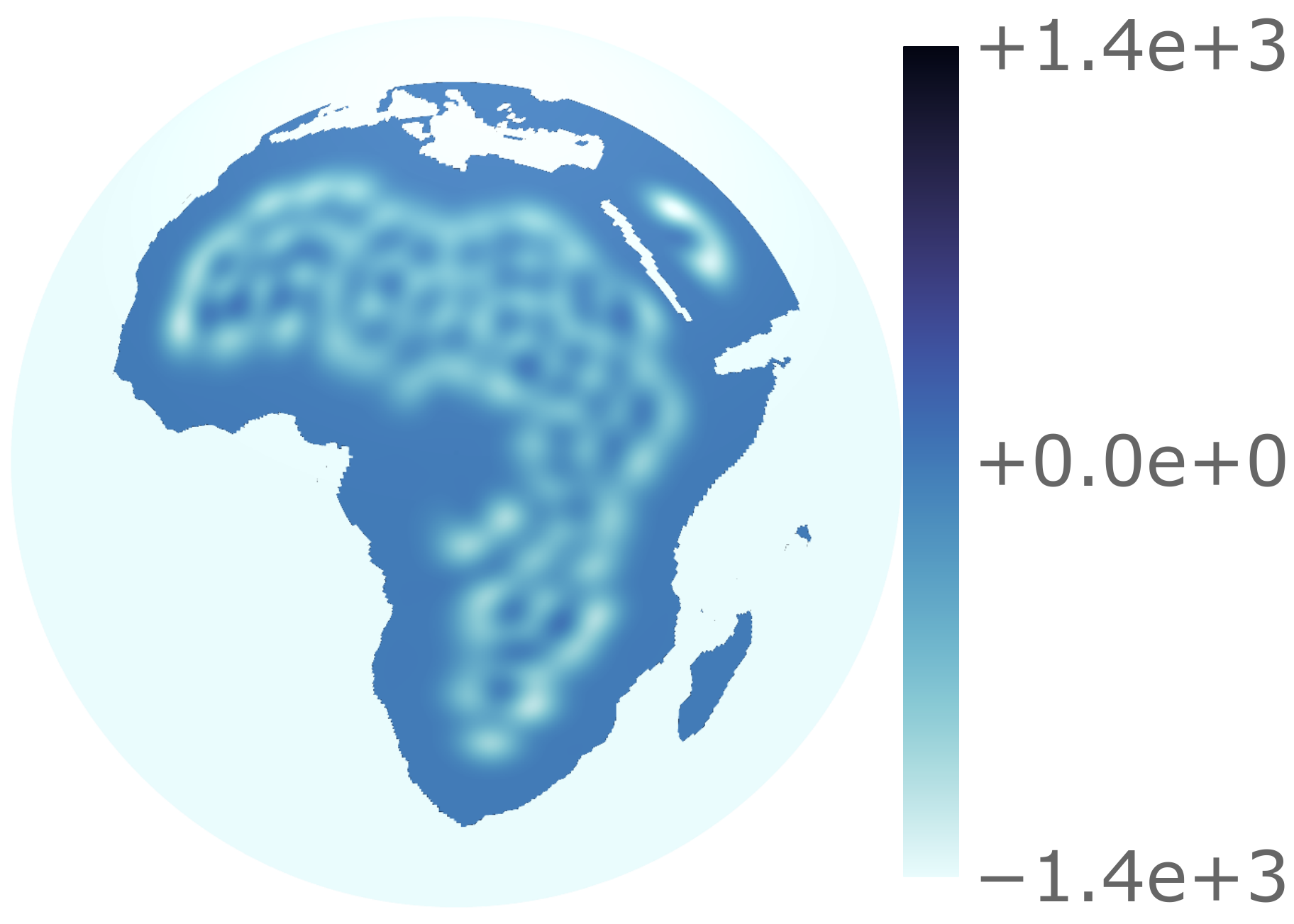}}
	\newline
	\subfloat[\(\Re\big\{\pixel{W^{\Psi^{5j}}}\big\}\)] 
	{\includegraphics[trim={4 7 3 6},clip,width=.5\columnwidth]{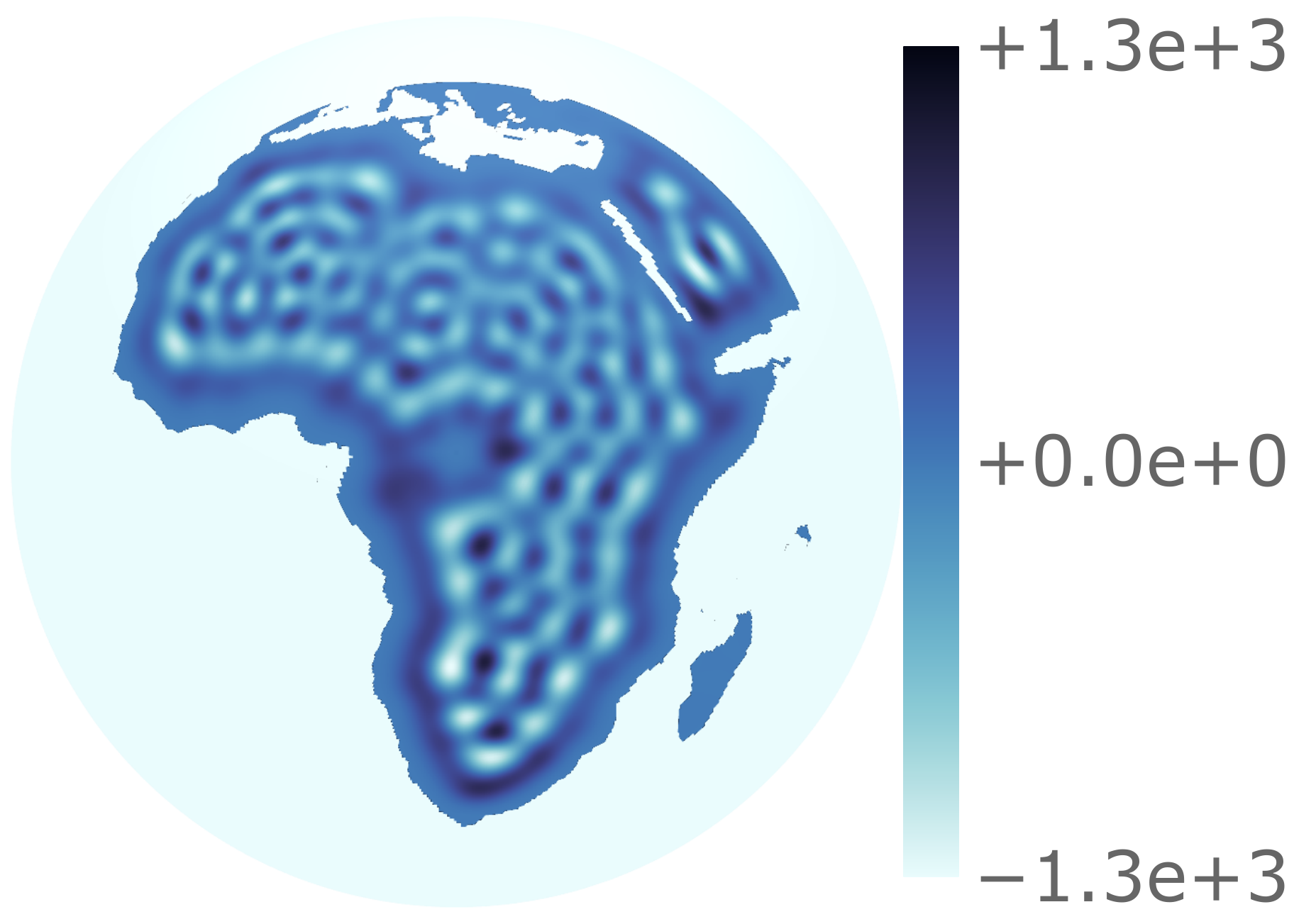}}
	\hfill
	\subfloat[\(\Re\big\{\pixel{W^{\Psi^{6j}}}\big\}\)] 
	{\includegraphics[trim={4 7 3 6},clip,width=.5\columnwidth]{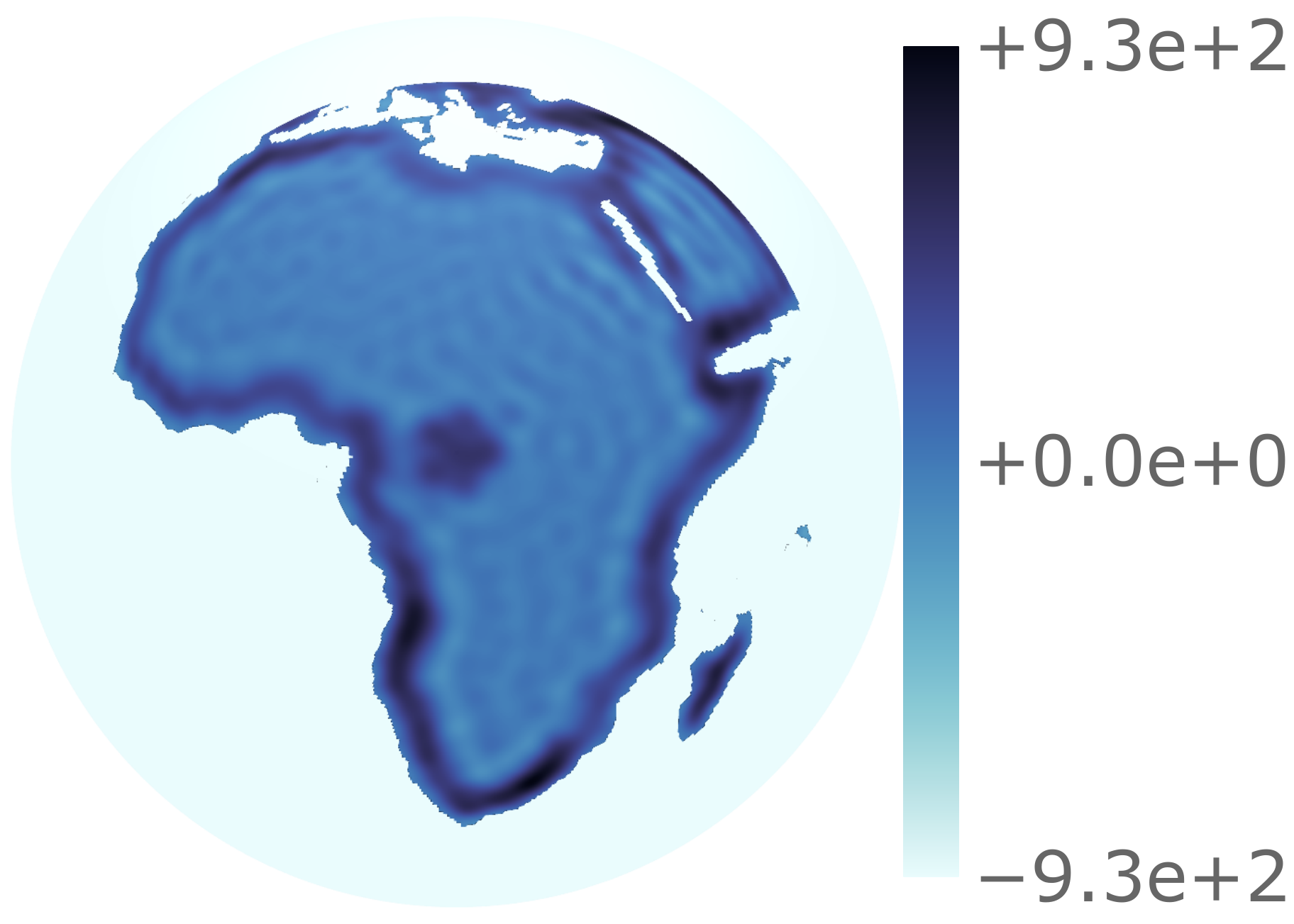}}
	\newline
	\subfloat[\(\Re\big\{\pixel{W^{\Psi^{7j}}}\big\}\)] 
	{\includegraphics[trim={4 7 3 6},clip,width=.5\columnwidth]{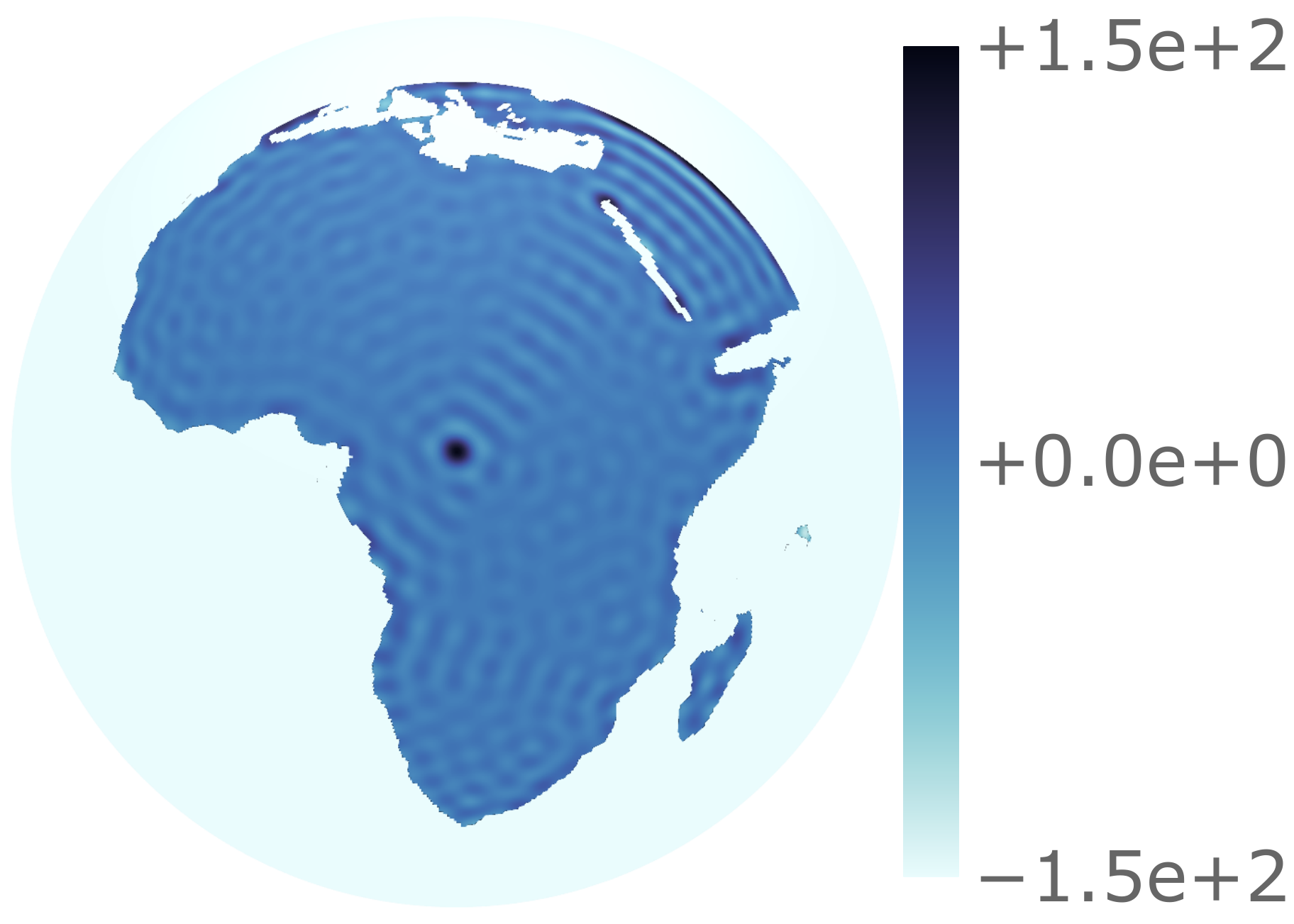}}
	\caption{
		The scale-discretised wavelet transform of the topographic map of Africa for parameters \(\lambda=3\), \(J_{0}=2\), and bandlimit \(\lmax=128\); \ie{} with the wavelets shown in \cref{fig:africa_slepian_wavelets}.
		Spatially localised, scale-dependent features of the bandlimited signal may be extracted by the wavelet coefficients given by the wavelet transform.
		The scaling coefficients are given in the top left plot, while the wavelet coefficients at scales \(j \in \set{2, 3, 4, 5, 6, 7}\) are shown left-to-right, top-to-bottom.
	}\label{fig:africa_slepian_wavelet_coefficients}
\end{figure}

\begin{figure*}
	\centering
	\subfloat[Initial Noisy Data \newline
		\(\snr{x} = \SI{1.78}{\dB}\)]
	{\includegraphics[trim={4 7 3 6},clip,width=.5\columnwidth]{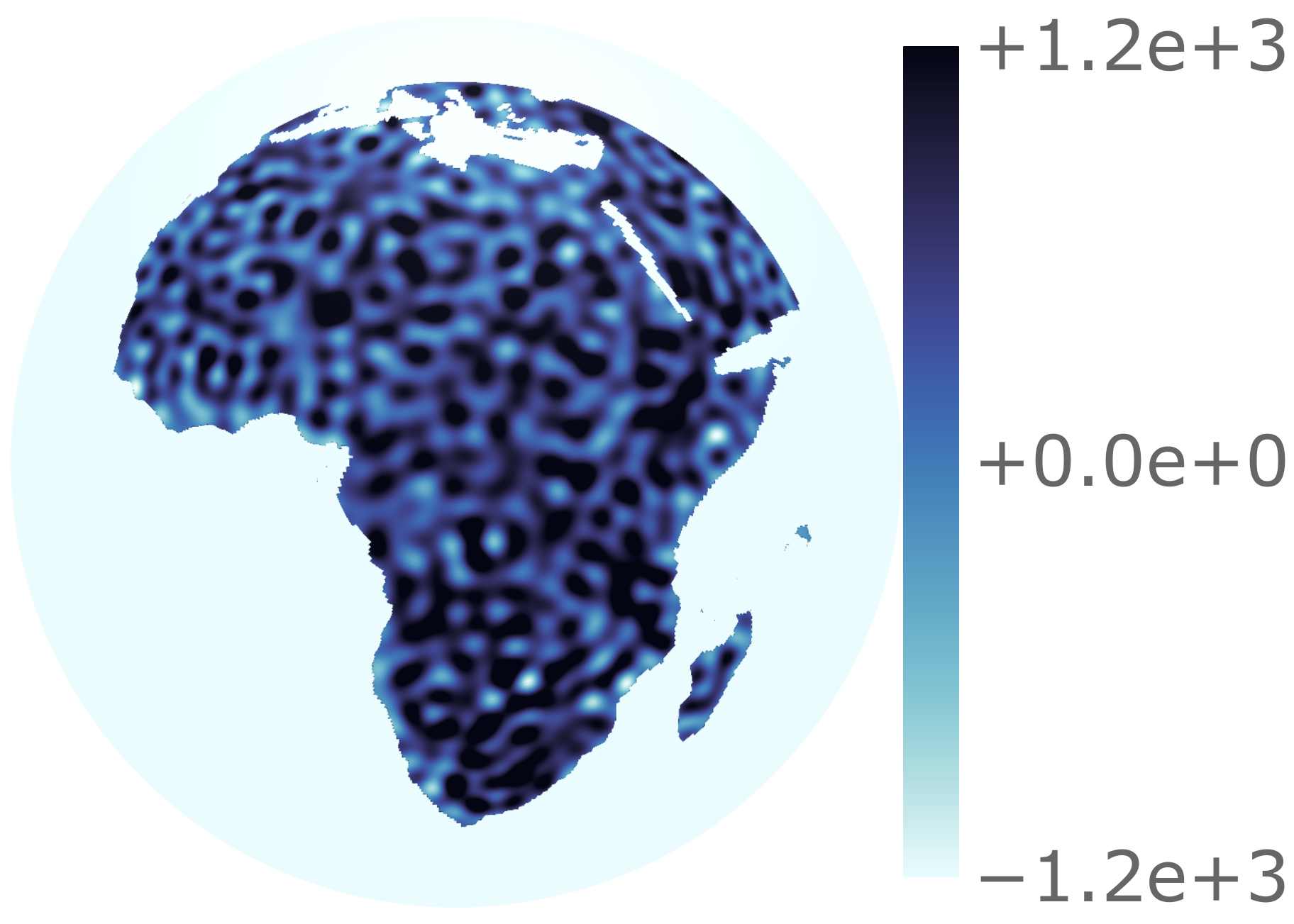}}
	\hfill
	\subfloat[Denoised \(N_{\sigma}=2\) \newline
		\(\snr{d} = \SI{3.95}{\dB}\)]
	{\includegraphics[trim={4 7 3 6},clip,width=.5\columnwidth]{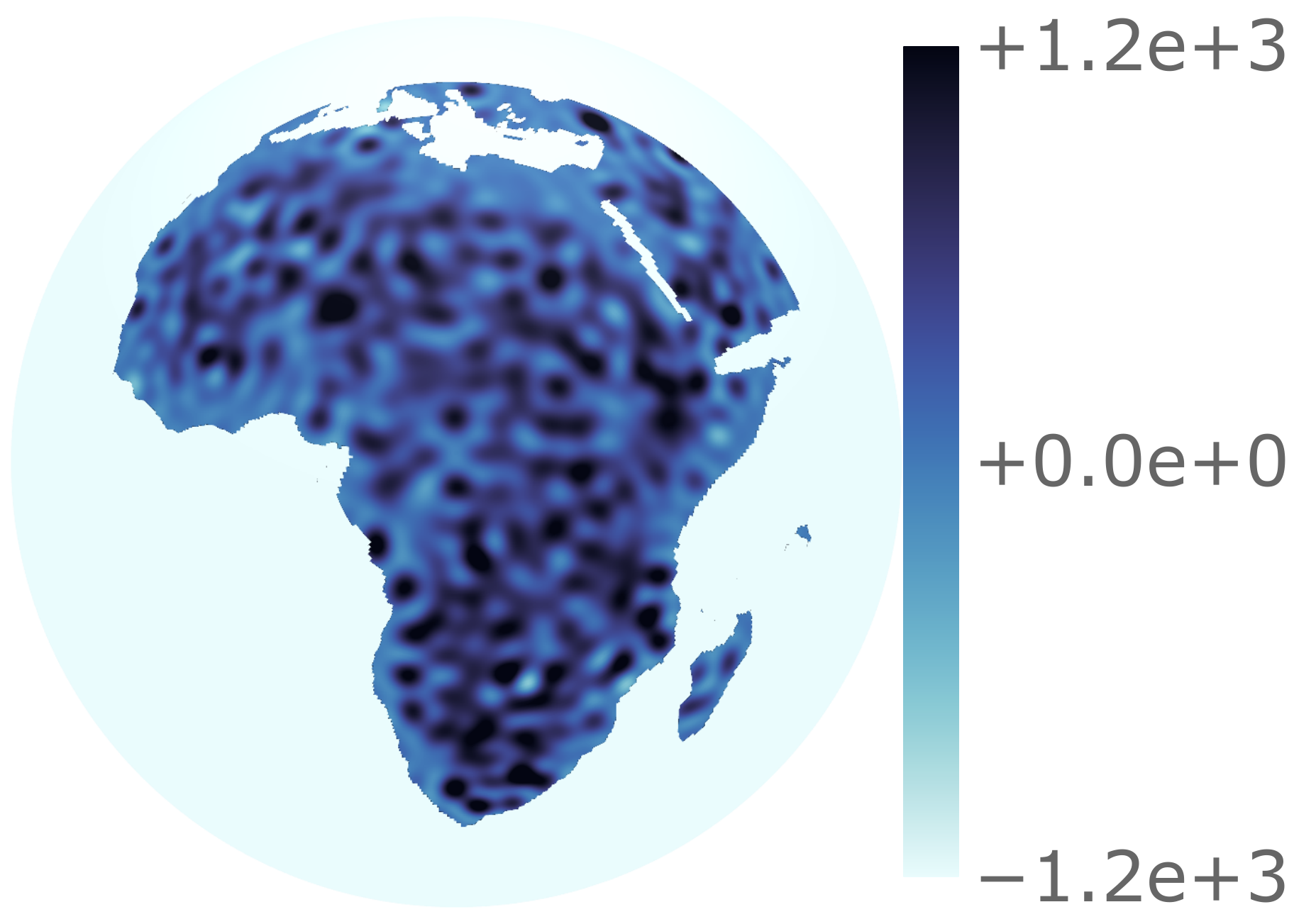}}
	\hfill
	\subfloat[Denoised \(N_{\sigma}=3\) \newline
		\(\snr{d} = \SI{2.93}{\dB}\)]
	{\includegraphics[trim={4 7 3 6},clip,width=.5\columnwidth]{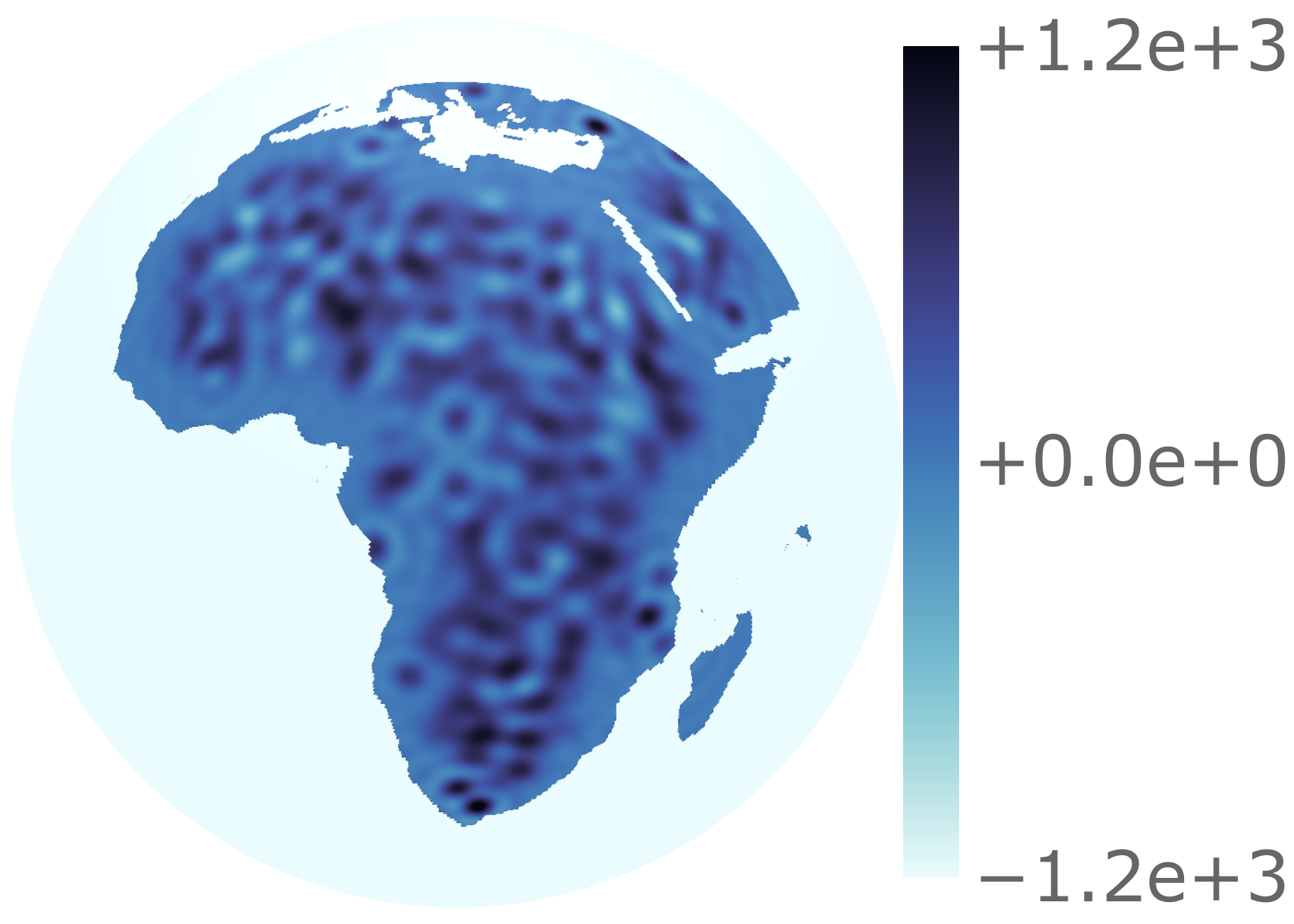}}
	\hfill
	\subfloat[Denoised \(N_{\sigma}=5\) \newline
		\(\snr{d} = \SI{0.55}{\dB}\)]
	{\includegraphics[trim={4 7 3 6},clip,width=.5\columnwidth]{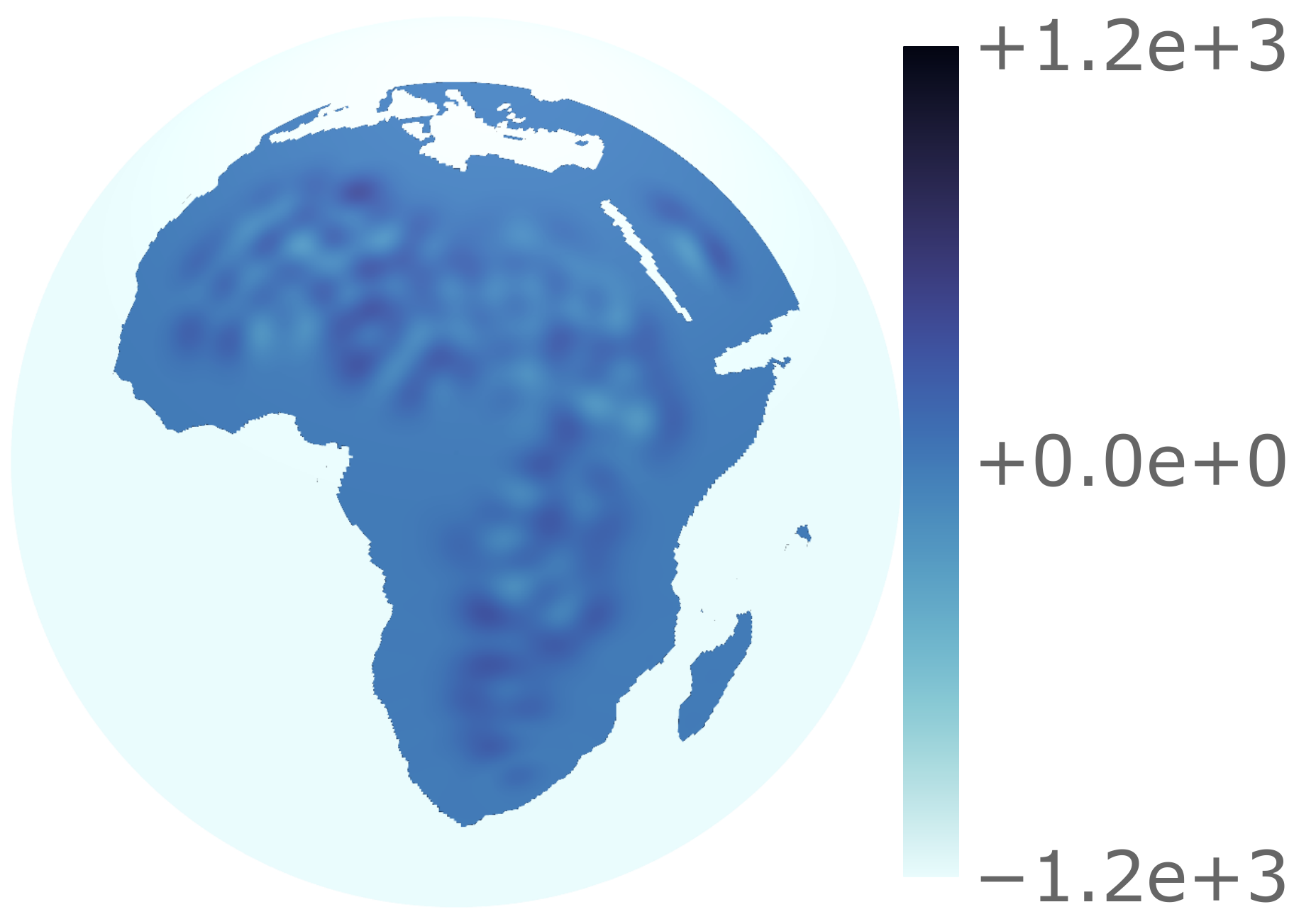}}
	\caption{
		Panel (a) shows the initial noisy signal of the Africa region with a signal-to-noise ratio of \(\SI{1.78}{\dB}\).
		The scaling and wavelet coefficients of the noisy signal are calculated and are then hard-thresholded for a few \(N_{\sigma}\) values.
		The corresponding denoised plots for \(N_{\sigma} \in \set{2, 3, 5}\) are shown in panels (b--d). 
		At \(N_{\sigma}=2\) the signal-to-noise ratio is boosted by \(\SI{2.17}{\dB}\) to \(\SI{3.95}{\dB}\).
		As more signal is removed the signal-to-noise ratio decreases to \(\SI{2.93}{\dB}\) at \(N_{\sigma}=3\), which is still higher than the initial noisy signal.
		At \(N_{\sigma}=5\) the signal-to-noise ratio is \(\SI{0.55}{\dB}\), where little signal remains.
	}\label{fig:africa_denoising}
\end{figure*}

Another region \(R'\) is constructed by centring a polar cap of angular opening \(\SI{41}{\degree}\) over Africa on the EGM2008 dataset similarly to \cref{sec:south_america_region} as shown in \cref{fig:africa_region}.
The eigenvalue plot is presented in \cref{fig:africa_eigenvalues}, with a selection of Slepian functions \(p \in \set{1, 10, 25, 50, 100, 200}\) given in \cref{fig:africa_eigenfunctions}.
\cref{fig:africa_slepian_wavelets} presents the corresponding scaling function and the wavelets for scales \(j \in \set{2, 3, 4, 5, 6, 7}\) for the Africa region.
The wavelet and scaling coefficients for the Earth topographic data of the Africa region are given in \cref{fig:africa_slepian_wavelet_coefficients}.
Lastly, the denoising procedure in \cref{sec:wavelet_denoising} is repeated for a noisy Africa signal with an initial signal-to-noise ratio of \(\SI{1.78}{\dB}\) as shown in \cref{fig:africa_denoising}.
This leads to a signal-to-noise ratio of \(\SI{3.95}{\dB}\), \(\SI{2.93}{\dB}\) and \(\SI{0.55}{\dB}\) respectively.

\bibliographystyle{IEEEtran}
\bibliography{library}

\end{document}